\documentclass{article}

\usepackage[left=1.25in,top=1.25in,right=1.25in,bottom=1.25in,head=1.25in]{geometry}
\usepackage{amsfonts,amsmath,amssymb,amsthm}
\usepackage{verbatim,float,url,dsfont}
\usepackage{graphicx,subfigure}
\usepackage{sectsty,etoc}

\usepackage{algorithm,algorithmic}
\usepackage[colon]{natbib}
\usepackage[colorlinks=true,citecolor=blue,urlcolor=blue,linkcolor=blue]{hyperref}

\theoremstyle{definition}

\newcommand{\argmin}{\operatornamewithlimits{argmin}}
\newcommand{\argmax}{\operatornamewithlimits{argmax}}
\newcommand{\minimize}{\operatorname*{minimize}}
\newcommand{\st}{\mathop{\mathrm{subject\,\,to}}}

\def\R{\mathbb{R}}
\def\E{\mathbb{E}}

\def\PVE{\mathrm{PVE}}
\def\SNR{\mathrm{SNR}}

\def\Var{\mathrm{Var}}

\def\hbeta{\hat{\beta}}

\title{Extended Comparisons of Best Subset Selection, Forward Stepwise
  Selection, and the Lasso \\ \bigskip
\Large Following ``Best Subset Selection from a Modern Optimization Lens'' by
Bertsimas, King, and Mazumder (2016)}   
\author{Trevor Hastie \and Robert Tibshirani \and Ryan J. Tibshirani} 
\date{}

\begin{document}
\maketitle

\begin{abstract}
In exciting new work, \citet{bertsimas2016best} showed that the
classical best subset selection problem in regression modeling can be
formulated as a mixed integer    
optimization (MIO) problem. Using recent
advances in MIO algorithms, they demonstrated that  best subset selection can
now be solved at much larger problem sizes that what was thought possible 
in the statistics community.  They presented
empirical comparisons of best subset selection with other popular  
variable selection procedures, in particular, the lasso and forward
stepwise selection. Surprisingly (to us), their simulations suggested
that best subset selection consistently outperformed both methods in
terms of prediction accuracy.  Here we present an expanded set of
simulations to shed more light on these comparisons.
The summary is roughly as follows: 
\begin{itemize}
\item neither best subset selection
nor the lasso uniformly dominate the other, with best subset
selection generally performing better in high signal-to-noise (SNR)
ratio regimes, and the lasso better in low SNR regimes; 
\item best subset selection and forward stepwise perform quite
  similarly throughout;
\item the relaxed lasso (actually, a simplified
version of the original relaxed estimator defined in
\citealp{meinshausen2007relaxed}) 
 is the overall winner, performing just about as well
as the lasso in low SNR scenarios, and as well as best subset selection
in high SNR scenarios. 
\end{itemize}
% For practical reasons, we limited the MIO
% solver for best subset selection in our simulations to 3 minutes per problem 
% instance per subset size (i.e., a budget of 150 minutes  
% for a path of 50 subset sizes).  We note that this may not be enough for the
% MIO  algorithm to find high-quality approximate solutions, especially in the 
% high-dimensional settings we considered.  
% Furthermore, we emphasize that these results are based on prediction 
% error as the metric of interest, and we note that different behaviors
% may occur for other metrics (e.g., a metric measuring the
% recovery of the correct variables in a sparse population linear 
% model may yield different results).  
\end{abstract}

\etocdepthtag.toc{main}
\section{Introduction}

Best subset selection, forward stepwise selection, and the lasso are
popular methods for selection and estimation of the parameters in a
linear model.  The first two are classical methods in statistics,
dating back to at least
\citet{beale1967discarding,hocking1967selection} for  
best subset selection and \citet{efroy1966stepwise,draper1966applied}
for forward selection; the lasso is (relatively speaking) more recent,
due to \citet{tibshirani1996regression,chen1998atomic}.

Given a response vector $Y \in \R^n$, predictor matrix 
$X \in \R^{n\times p}$, and a subset size $k$ between 0 and $\min\{n,p\}$, best     
subset selection finds the subset of $k$ predictors that produces the best fit 
in terms of squared error, solving the nonconvex problem
\begin{equation}
\label{eq:bs_prob}
\minimize_{\beta \in \R^p} \; \|Y-X\beta\|_2^2 \;\; \st \;\; 
\|\beta\|_0 \leq k,
\end{equation}
where \smash{$\|\beta\|_0=\sum_{i=1}^p 1\{\beta_i \not= 0\}$} is
the $\ell_0$ norm of $\beta$. (Here and throughout, for notational simplicity, 
we omit the intercept term from the regression model.) 

Forward stepwise selection is less ambitious: starting with the empty model,  
it iteratively adds the variable that best improves the fit.\footnote{Other ways
  of defining the variable $j_k$ that ``best improves the fit'' are possible,
  but the entry criterion is \eqref{eq:fs_crit} is the standard one in
  statistics.}  It hence yields a subset of each size $k=0,\ldots,\min\{n,p\}$, 
but none of these are generally globally optimal in the sense of
\eqref{eq:bs_prob}. Formally, the procedure starts with an empty active 
set $A_0=\{0\}$, and for $k=1,\ldots,\min\{n,p\}$, selects the variable
indexed by  
\begin{equation}
\label{eq:fs_crit}
j_k = \argmin_{j \notin A_{k-1}} \, \|Y - P_{A_{k-1} \cup \{j_k\}} Y
\|_2^2 = \argmax_{j \notin A_{k-1}} \, \frac{X_j^T P_{A_{k-1}}^\perp
  Y} {\|P_{A_{k-1}}^\perp X_j\|_2}
\end{equation}
that leads to the lowest squared error when added to $A_{k-1}$, or equivalently,
such that \smash{$X_{j_k}$}, achieves the maximum absolute correlation with $Y$, 
after we project out the contributions from \smash{$X_{A_{k-1}}$}. A note on
notation: here we write $X_S \in \R^{n\times |S|}$ for the submatrix of $X$
whose columns are indexed by a set $S$ (and when $S=\{j\}$, we 
simply use $X_j$). We also write $P_S$ for the projection matrix onto the column   
span of $X_S$, and $P_S^\perp=I-P_S$ for the projection matrix onto the 
orthocomplement.  At the end of step 
$k$ of the procedure, the active set is updated, \smash{$A_k=A_{k-1} \cup
  \{j_k\}$}, and the forward stepwise estimator of the 
regression coefficients is defined by the least squares fit onto
\smash{$X_{A_k}$}.      

% At step $k$, the forward stepwise estimator
% \smash{$\hbeta^{(k)}$} of the regression coefficients is
% \begin{equation}
% \label{eq:fs_coef}
% \begin{aligned}
% \hbeta^{(k)}_{A_k} &= (X_{A_k}^T X_{A_k})^{-1} X_{A_k}^T Y, \\
% \hbeta^{(k)}_{-A_k} &= 0,
% \end{aligned} 
% \end{equation}
% where we write $x_S \in \R^{|S|}$ to index the components of a vector
% $x \in \R^p$ in a set $S$, and we write $x_{-S}$ to index the
% components in $\{1,\ldots,p\}\setminus S$. 

The lasso solves a convex relaxation of \eqref{eq:bs_prob} where we
replace the $\ell_0$ norm by the $\ell_1$ norm, namely 
\begin{equation}
\label{eq:lasso_prob_cstr}
\minimize_{\beta \in \R^p} \; \|Y-X\beta\|_2^2  \;\; \st \;\; 
\|\beta\|_1 \leq t,
\end{equation}
where \smash{$\|\beta\|_1=\sum_{i=1}^p |\beta_i|$}, and $t \geq 0$ 
is a tuning parameter. By convex duality, the above problem
is equivalent to the more common (and more easily solveable) penalized
form   
\begin{equation}
\label{eq:lasso_prob}
\minimize_{\beta \in \R^p} \; \|Y-X\beta\|_2^2  + \lambda \|\beta\|_1 
\end{equation}
where now $\lambda \geq 0$ is a tuning parameter. This is the form 
that we focus on in this paper.

The lasso problem \eqref{eq:lasso_prob} is convex (and highly
structured) and there is by now a sizeable literature in statistics,
machine learning, and optimization dedicated to efficient algorithms
for this problem.  On the other hand, the best subset selection
problem \eqref{eq:bs_prob} is nonconvex and is known to be NP-hard
\citep{natarajan1995sparse}.  The accepted view in statistics for many
years has been that this problem is not solveable
beyond (say) $p$ in the 30s, this view being shaped by the
available software for best subset selection (e.g., in the R language,
the {\tt leaps} package implements a branch-and-bound algorithm for
best subset selection of \citealp{furnival1974regression}).  

For a much more detailed introduction to best subset selection, forward stepwise 
selection, and the lasso, see, e.g., Chapter 3 of \citet{hastie2009elements}. 

\subsection{An exciting new development}

Recently, \citet{bertsimas2016best} presented a mixed integer
optimization (MIO) formulation for the best subset selection problem
\eqref{eq:bs_prob}.  This allows one to use highly optimized MIO  
solvers, like Gurobi (based on branch-and-cut methods, 
hybrids of branch-and-bound and cutting plane algorithms), to solve 
\eqref{eq:bs_prob}. Using these MIO solvers, problems with $p$ in the
hundreds and even thousands are not out of reach, and this presents us
with exciting new ground on which to perform empirical 
comparisons. Simulation studies in 
\citet{bertsimas2016best} demonstrated that best subset selection
generally gives superior prediction accuracy compared to forward
stepwise selection and the lasso, over a variety of problem setups.  

In what follows, we replicate and expand these simulations to shed 
more light on such comparisons.
For convenience, we made an R package {\tt bestsubset} for   
optimizing the best subset selection problem using the Gurobi MIO
solver (after this problem has been translated into a mixed 
integer quadratic program as in \citealp{bertsimas2016best}). This
package, as well as R code for reproducing all of the results in this 
paper, are available at  
\url{https://github.com/ryantibs/best-subset/}. 

\section{Preliminary discussion}

\subsection{Is best subset selection the holy grail?}

Various researchers throughout the years have viewed best subset 
selection as the ``holy grail'' of estimators 
for sparse modeling in regression, suggesting (perhaps implicitly)
that it should be used whenever possible, and that other methods for
sparse regression---such as forward stepwise selection and the lasso---should  
be seen as approximations or heuristics, used only out of necessity when best  
subset selection is not computable.       
However, as we will demonstrate in the simulations that
follow, this is not the case.  Different procedures have different 
operating characteristics, i.e., give rise to different bias-variance 
tradeoffs as we vary their respective tuning parameters.  In fact, depending on 
the problem setting, the bias-variance tradeoff provided by best
subset selection may be more or less useful than the tradeoff
provided by the lasso. 

As a brief interlude, let us inspect the ``noiseless'' versions of the
best subset and lasso optimization problems, namely  
\begin{align}
\label{eq:bs_prob_noiseless}
\minimize_{\beta \in \R^p} \; \|\beta\|_0 \;\; \st \;\; X\beta=Y, \\
\label{eq:lasso_prob_noiseless}
\minimize_{\beta \in \R^p} \; \|\beta\|_1 \;\; \st \;\; X\beta=Y,
\end{align}
respectively.  Suppose that our goal is to find the sparsest
solution to the linear system $X\beta=Y$. Problem \eqref{eq:bs_prob_noiseless}, 
by definition of the $\ell_0$ norm, produces
it. Problem \eqref{eq:lasso_prob_noiseless}, in which the criterion has been  
convexified, does not generally give the sparsest solution and so 
in this sense we may rightly view it as a heuristic for the 
nonconvex problem \eqref{eq:bs_prob_noiseless}.  Indeed, much of the 
literature on compressed sensing (in which
\eqref{eq:bs_prob_noiseless}, \eqref{eq:lasso_prob_noiseless} have
been intensely studied) uses this language.
However, in the noiseless setting, there is no bias-variance tradeoff,
because (trivially) {\it there is no bias and no variance}; both of
the estimators defined by \eqref{eq:bs_prob_noiseless},
\eqref{eq:lasso_prob_noiseless} have zero mean squared error owing 
to the linear constraint $X\beta=Y$ (and the fact that $Y=\E(Y|X)$,
as there is no noise).  

\begin{figure}[htb]
\begin{minipage}[c]{0.425\textwidth}
\centering
\includegraphics[width=\textwidth]{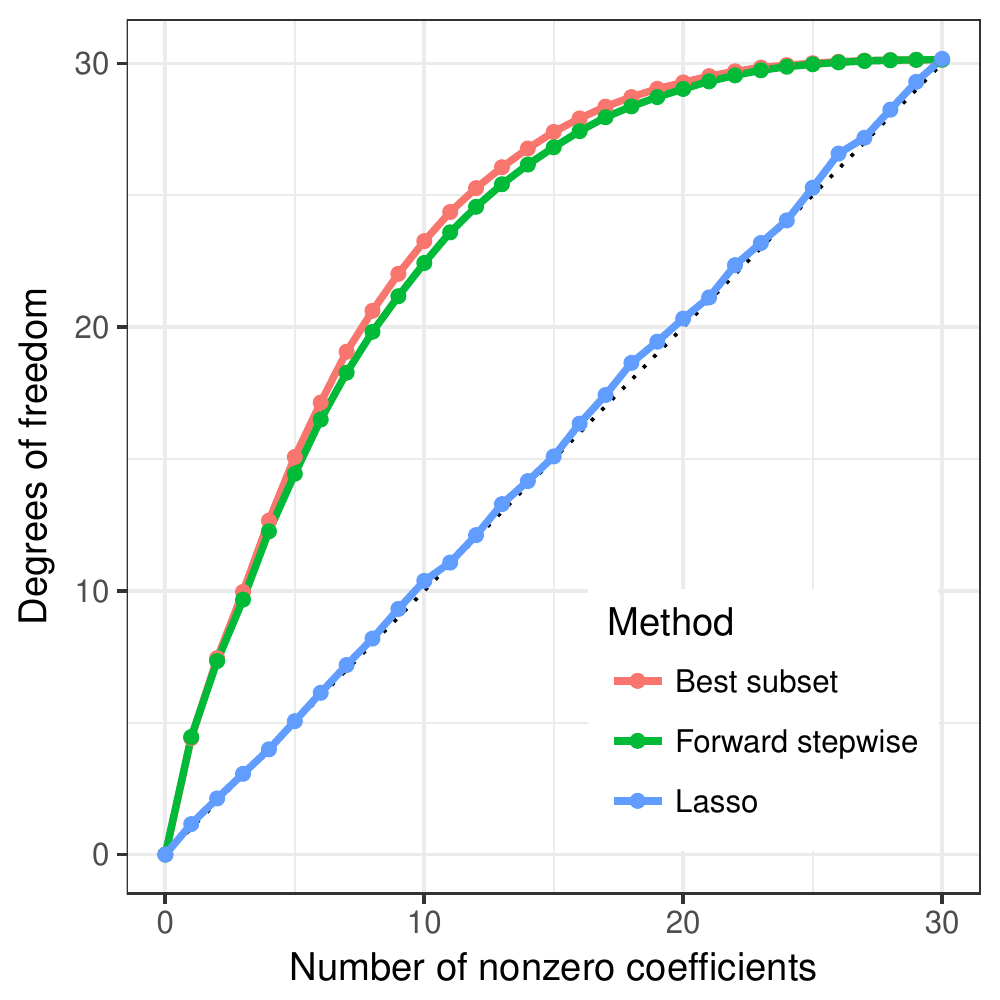}
\end{minipage}
\hspace{10pt}
\begin{minipage}[c]{0.525\textwidth}
\vspace{-25pt}
\caption{\it Effective degrees of freedom for the
  lasso, forward stepwise, and best subset selection, 
  in a problem setup with $n=70$ and $p=30$ (computed via Monte Carlo 
  evaluation of the covariance formula for degrees of freedom over 500  
  repetitions). The setup had an SNR of 0.7, predictor autocorrelation
  of 0.35, and the coefficients followed the beta-type 2 pattern 
  with $s=5$; see Section \ref{sec:setup} for details.  Note that the
  lasso degrees of freedom equals the (expected) number of nonzero coefficients,      
  whereas that of forward stepwise and best subset selection exceeds 
  the number of nonzero coefficients.}
\label{fig:df1}
\end{minipage}
\end{figure}

The noisy setting---which is the traditional and
most practical setting for statistical estimation, and that studied in 
this paper---is truly different. Here, it is no longer appropriate
to view the estimator defined by the $\ell_1$-regularized problem   
\eqref{eq:lasso_prob_cstr} 
as a heuristic for that defined by the $\ell_0$-regularized problem 
\eqref{eq:bs_prob} (or \eqref{eq:lasso_prob} as a heuristic for
\eqref{eq:bs_prob}). 
Generally speaking, the lasso and best subset selection 
differ in terms of their ``aggressiveness'' in selecting and estimating the 
coefficients in a linear model, with the lasso being less aggressive
than best subset selection; meanwhile, forward stepwise lands somewhere in the
middle, in terms of its aggressiveness. There are various ways to make this
vague but intuitive comparison more explicit.  For example:
\begin{itemize}
\item forward stepwise can be seen as a ``locally optimal''
  version of best subset selection, updating the active
  set by one variable at each step, instead of re-optimizing
  over all possible subsets of a given size; in turn, the lasso can
  be seen as a more ``democratic'' version of forward stepwise,
  updating the coefficients so as 
  maintain equal absolute correlation of all active variables with
  the residual \citep{efron2004least};
\item the lasso applies shrinkage to its nonzero estimated
  coefficients (e.g., see \eqref{eq:relaxed_lasso} with
  $\gamma=1$) but forward stepwise and best subset selection do
  not, and simply perform least squares on their respective active
  sets;  
\item thanks to such shrinkage, the fitted values from the lasso (for
  any fixed $\lambda \geq 0$) are continuous functions of $y$
  \citep{zou2007degrees,tibshirani2012degrees}, whereas the fitted
  values from forward stepwise and best subset selection
  (for fixed $k \geq 1$) jump discontinuously as $y$ moves across
  a decision boundary for the active set; 
\item again thanks to shrinkage, the effective degrees of freedom of
  the lasso (at any fixed $\lambda \geq 0$) is equal to the expected
  number of selected variables
  \citep{zou2007degrees,tibshirani2012degrees},  
  whereas the degrees of freedom of both forward stepwise and
  best subset selection can greatly exceed $k$ at any given step $k
  \geq 1$ \citep{kaufman2014when,janson2015effective}.  
\end{itemize}
Figure \ref{fig:df1} uses the latter perspective of effective degrees 
of freedom to contrast the aggressiveness of the three methods.  

\begin{figure}[htb]
\centering
\includegraphics[height=0.4\textwidth]{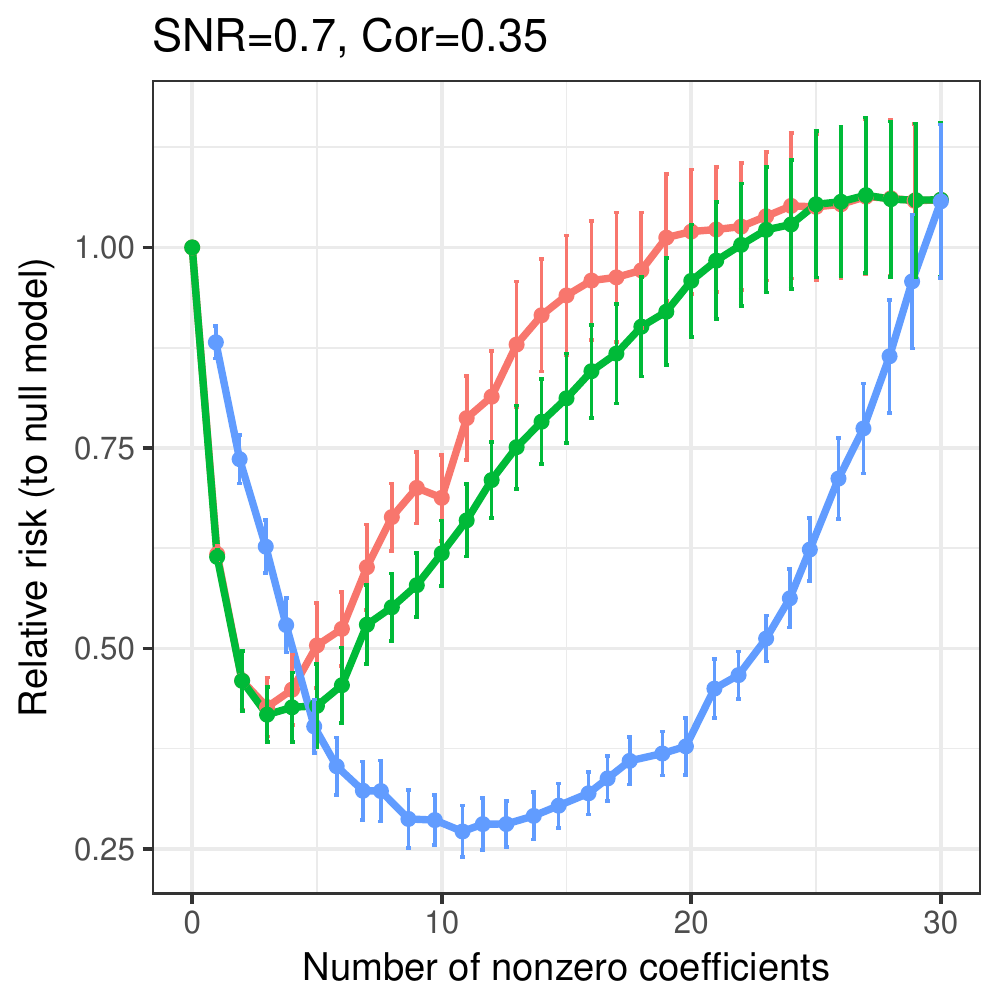} \hspace{2pt}
\includegraphics[height=0.4\textwidth]{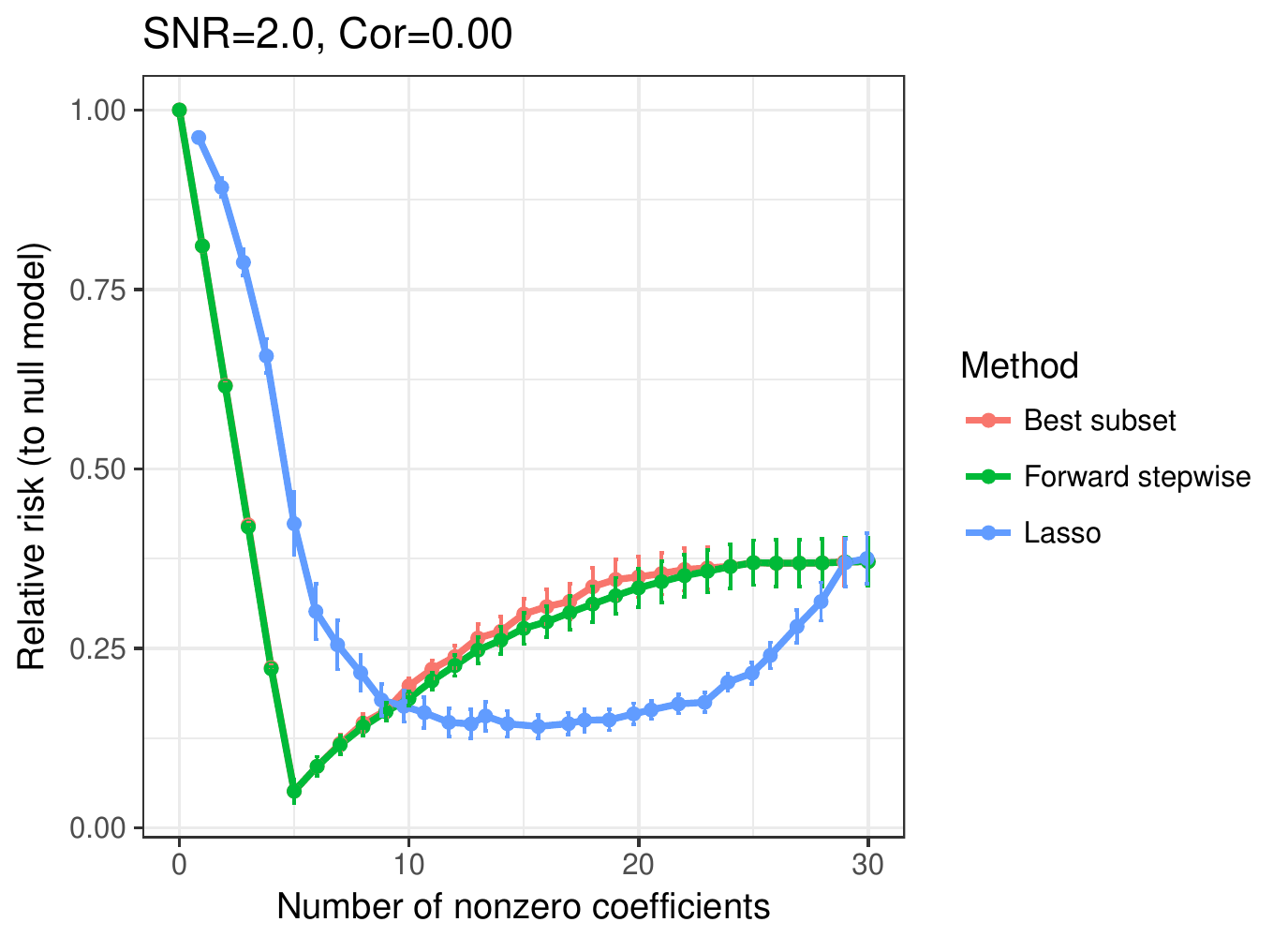}
\caption{\it Relative risk (risk divided by null risk) for the lasso, forward 
  stepwise selection, and best subset selection, for two different setups.  The
  results were averaged over 20 repetitions, and the bars denote one standard  
  errors. The setup for the left panel is identical to that used in Figure
  \ref{fig:df1}. The setup for the right panel used an SNR (signal-to-noise
  ratio) of 2 and zero predictor autocorrelation.  Note that in the left
  panel, the lasso is more accurate than forward stepwise selection and best
  subset selection, and in the right panel, the opposite is true.}
\label{fig:snr}
\end{figure}

When the signal-to-noise ratio (SNR) is low, and also depending  
on other factors like the correlations between predictor variables,  
the more aggressive best subset and  
forward stepwise methods can already have quite high
variance at the start of their model paths (i.e., for small
step numbers $k$). Even after optimizing over the tuning 
parameter $k$ (using say, an external validation set or an
oracle which reveals the true risk), we can arrive at an
estimator with unwanted variance and worse accuracy than a 
properly tuned lasso estimator. On the other hand, for high SNR
values, and other configurations for the correlations between
predictors, etc., the story can be completely flipped and the
shrinkage applied by the lasso estimator can result in unwanted bias  
and worse accuracy than best subset selection and forward stepwise
selection. See Figure \ref{fig:snr} for empirical evidence.   

This is a simple point, but is worth emphasizing.  To convey
the idea once more:
\begin{quote}\it
Different procedures bring us from the high bias to the high variance
ends of the tradeoff along different model paths; and these paths are 
affected by aspects of the problem setting, like the SNR and predictor
correlations, in different ways. For some classes of problems, some 
procedures admit more fruitful paths, and for other classes, other
procedures admit more fruitful paths.  For example, neither best
subset selection nor the lasso dominates the other, across
all problem settings.   
\end{quote}

\subsection{What is a realistic signal-to-noise ratio?}
\label{sec:snr}

In their simulation studies, \citet{bertsimas2016best} considered 
SNRs in the range of about 2 to 8 in their low-dimensional cases, and
about 3 to 10 in their high-dimensional cases. Is this a realistic
range that one encounters in practice? % Of course, the answer to this
% depends on the specific domain of application; for some domains, these 
% range seem to be quite optimistic.  In any case,
In our view, inspecting the proportion of variance explained (PVE) 
can help to answer this question.

Let $(x_0,y_0) \in \R^p \times \R$ be a pair of predictor and response    
variables, and define $f(x_0)=\E(y_0|x_0)$ and $\epsilon_0=y_0-f(x_0)$, so that  
we may express the relationship between $x_0,y_0$ as:
$$
y_0 = f(x_0) + \epsilon_0.
$$
The signal-to-noise ratio (SNR) in this model is defined as 
$$
\SNR = \frac{\Var(f(x_0))}{\Var(\epsilon_0)}.
$$
For a given prediction fuction $g$---e.g., one trained on $n$ samples
$(x_i,y_i)$, $i=1,\ldots,n$ that are i.i.d.\ to $(x_0,y_0)$---its associated
proportion of variance explained (PVE) is defined as 
$$
\PVE(g) = 1 - \frac{\E(y_0-g(x_0))^2}{\Var(y_0)}.
$$
Of course, this is maximized when we take $g$ to be the mean function
$f$ itself, in which case
$$
\PVE(f) = 1 - \frac{\Var(\epsilon_0)}{\Var(y_0)} = \frac{\SNR}{1+\SNR}.
$$
In the second equality we have assumed independence of $x_0,
\epsilon_0$, so $\Var(y_0)=\Var(f(x_0))+\Var(\epsilon_0)$.
As the optimal prediction function is $f$, it sets the gold-standard of
$\SNR/(1+\SNR)$ for the PVE, so we should always expect to see the attained PVE
be less than $\SNR/(1+\SNR)$ and greater than 0 (otherwise we could simply
replace our prediction function by $g=0$.) 

\begin{figure}[htb]
\begin{minipage}[c]{0.55\textwidth}
\centering
\includegraphics[width=\textwidth]{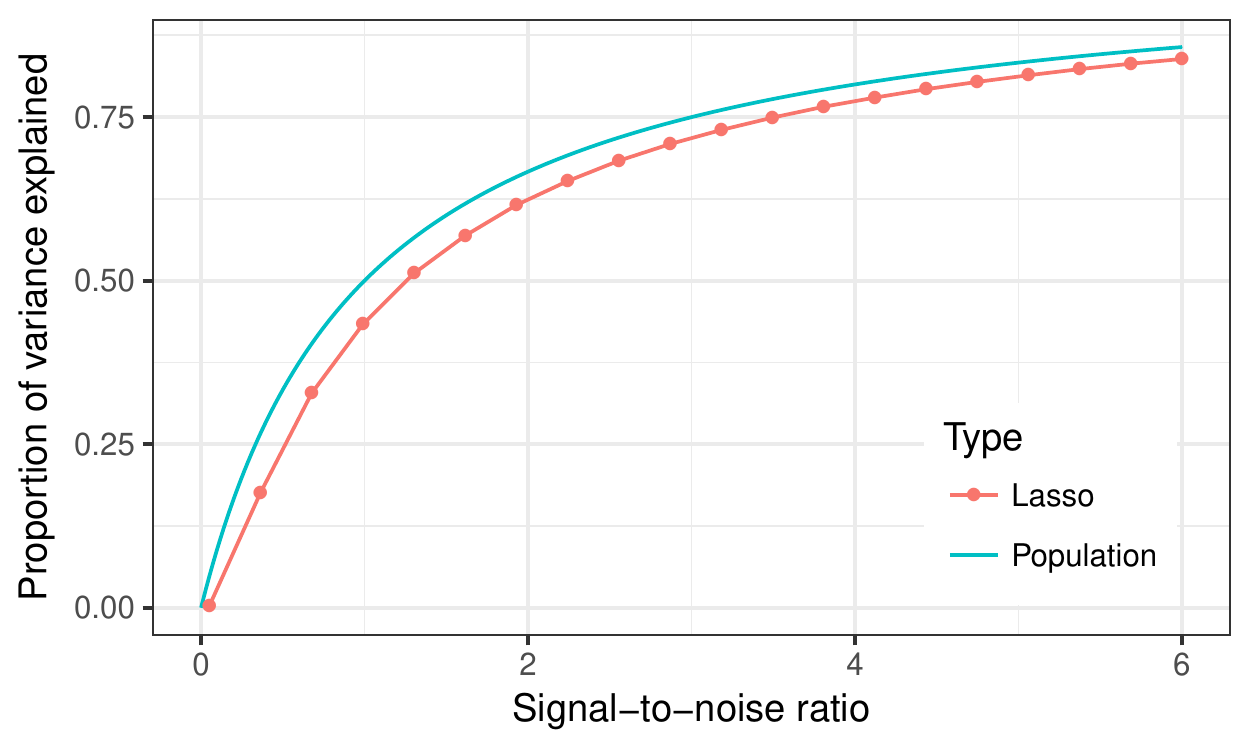}
\end{minipage}
\hspace{10pt}
\begin{minipage}[c]{0.4\textwidth}
\vspace{-25pt}
\caption{\it PVE (proportion of variance explained) of the lasso in a  
  simulation setup with $n=200$ and $p=100$, as 
  the SNR varies from 0.05 to 6 (more details are provided in the text). 
  The red curve is the population PVE, the maximum achievable PVE at 
  any given SNR value.  We see that SNRs above 2 give PVEs roughly above  
  0.6, which seems to us to be rare in many practical applications.}
\label{fig:pve}
\end{minipage}
\end{figure}

We illustrate using a simulation with $n=200$ and $p=100$. The 
predictor autocorrelation was zero and the coefficients followed the beta-type 2 
pattern with $s=5$; see Section \ref{sec:setup} for details.  We varied the SNR 
in the simulation from 0.05 to 6 in 20 equally spaced values. We 
computed the lasso over 50 values of the tuning parameter $\lambda$, and
selected the tuning 
parameter by optimizing prediction error on a separate validation set of size 
$n$. Figure \ref{fig:pve} shows the PVE of the tuned lasso estimator, 
averaged over 20 repetitions from this simulation setup.  Also shown is the
population PVE, i.e., the maximum possible PVE at any given SNR level, of 
$\SNR/(1+\SNR)$.   
We see that an SNR of 1.0 corresponds to a PVE of about 0.45 (with a maximum of 
0.5), while an SNR as low as 0.25 yields a PVE of 0.1 (with a maximum of 0.2).   
In our experience, a
PVE of 0.5 is rare for noisy observational data, and 0.2 may be more
typical. A PVE of 0.86, corresponding to an SNR of 6, is unheard of!  With
financial returns data, explaining even 2\% of the variance (PVE 
of 0.02) would be considered huge, and the corresponding prediction function
could lead to considerable profits if used in a trading scheme.  Therefore,
based on these observations, we examine a wider lower range of SNRs in our 
simulations, compared to the SNRs studied in \citet{bertsimas2016best}.   

\subsection{A (simplified) relaxed lasso}

In addition to the lasso estimator, we consider a simplified version
of the relaxed lasso estimator as originally defined by
\citet{meinshausen2007relaxed}.
Let \smash{$\hbeta^{\mathrm{lasso}}(\lambda)$} denote the solution in
problem \eqref{eq:lasso_prob}, i.e., the lasso estimator at the tuning 
parameter value $\lambda \geq 0$. Let $A_\lambda$ denote its active set, and let 
\smash{$\hbeta^{\mathrm{LS}}_{A_\lambda}$} denote
the least squares coefficients from regressing of $Y$ on $X_{A_\lambda}$, the 
submatrix of active predictors. Finally, let
\smash{$\hbeta^{\mathrm{LS}}(\lambda)$} be the full-sized ($p$-dimensional)
version of the least squares coefficients, padded with zeros in the
appropriately.  We consider the estimator
\smash{$\hbeta^{\mathrm{relax}}(\lambda,\gamma)$} defined by 
\begin{equation}
\label{eq:relaxed_lasso}
\hbeta^{\mathrm{relax}}(\lambda,\gamma) =
\gamma \hbeta^{\mathrm{lasso}}(\lambda) + (1-\gamma)
\hbeta^{\mathrm{LS}}(\lambda) 
\end{equation}
with respect to the pair of tuning parameter values $\lambda \geq 0$ 
and $\gamma \in [0,1]$. Recall 
\citep{tibshirani2013lasso} that when the columns of $X$ are in
general position (a weak condition occurring almost surely for
continuously distributed pedictors, regardless of $n,p$), it holds
that:  
\begin{itemize}
\item the lasso solution is unique;
\item the submatrix $X_{A_\lambda}$ of active predictors has full column rank,
thus \smash{$\hbeta_{A_\lambda}^{\mathrm{LS}}=(X_{A_\lambda}^T
  X_{A_\lambda})^{-1} X_{A_\lambda}^T Y$}
is well-defined; 
\item the lasso solution can be written (over its active set) as
\smash{$\hbeta^{\mathrm{lasso}}_{A_\lambda}(\lambda)=(X_{A_\lambda}^T
  X_{A_\lambda})^{-1}(X_{A_\lambda}^T Y -  \lambda s)$}, where 
\smash{$s \in \{-1,1\}^{|{A_\lambda}|}$} contains the signs of the active lasso
coefficients.  
\end{itemize}
Thus, under the general position assumption on $X$, the simplified
relaxed lasso can be rewritten as 
\begin{equation}
\label{eq:relaxed_lasso_2}
\begin{aligned}
\hbeta^{\mathrm{relax}}_{A_\lambda}(\lambda,\gamma) &= (X_{A_\lambda}^T X_{A_\lambda})^{-1} X_{A_\lambda}^T
Y - \gamma \lambda (X_{A_\lambda}^T X_{A_\lambda})^{-1} s \\
\hbeta^{\mathrm{relax}}_{-{A_\lambda}} (\lambda,\gamma)&= 0,
\end{aligned}
\end{equation}
so we see that $\gamma \in [0,1]$ acts as a multiplicative factor
applied directly to the ``extra'' shrinkage term apparent in the lasso
coefficients.  Henceforth, we will drop the word ``simplified'' and 
will just refer to this estimator as the relaxed lasso.

\begin{figure}[htb]
\begin{minipage}[c]{0.425\textwidth}
\centering
\includegraphics[width=\textwidth]{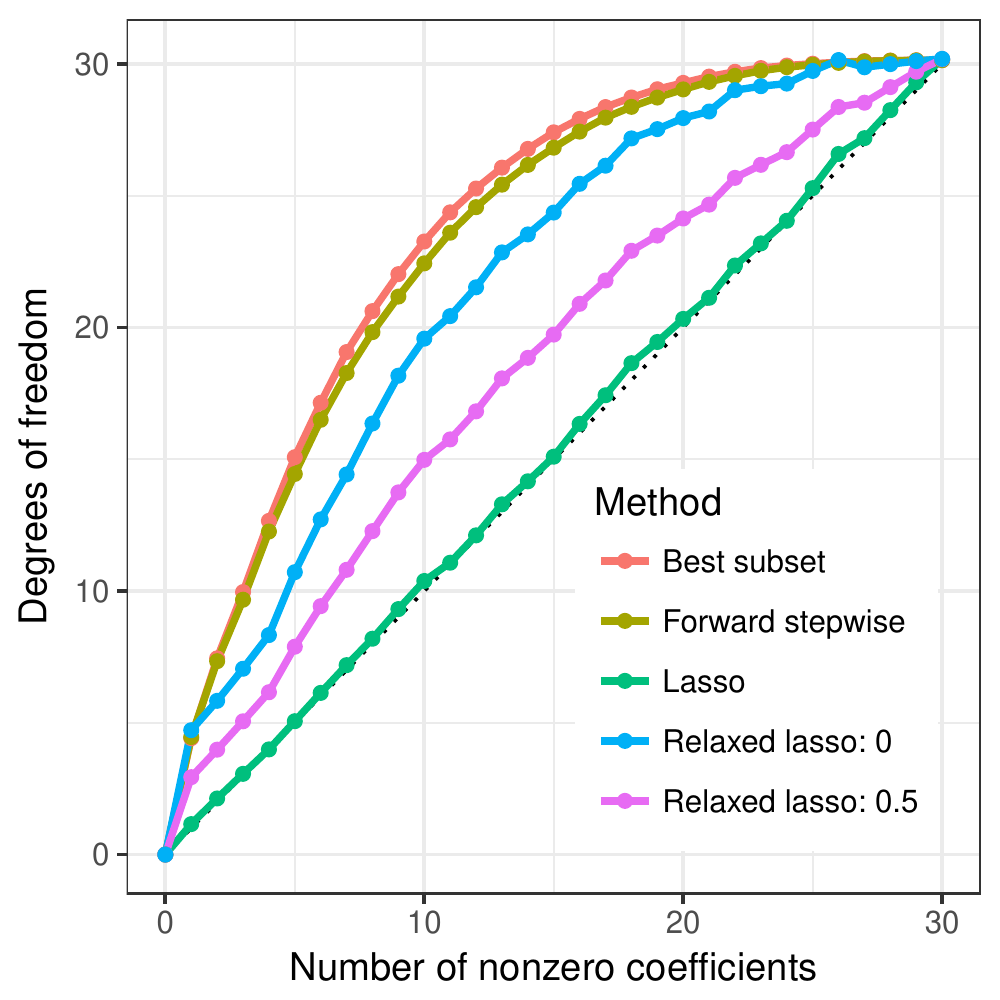}
\end{minipage}
\hspace{10pt}
\begin{minipage}[c]{0.525\textwidth}
\vspace{-25pt}
\caption{\it Degrees of freedom for the lasso, forward stepwise, best subset
  selection, and the relaxed lasso with   
  $\gamma=0.5$ and $\gamma=0$. The problem setup is the same as that in the 
  left panel of Figure \ref{fig:df1}.  Note that the relaxed lasso has an inflated 
  degrees of freedom compared to the lasso and generally has a larger degrees of
  freedom than the expected number of nonzero coefficients.  But, even when  
  $\gamma=0$, its degrees of freedom is smaller than that of forward stepwise
  and best subset selection throughout their model paths.}
\label{fig:df2}
\end{minipage}
\end{figure}

The relaxed lasso tries to undo the shrinkage inherent in the lasso estimator,
to a varying degree, depending on $\gamma$.  In this sense, we would expect it
to be more aggressive than the lasso, and have a larger effective degrees of
freedom.  However, even in its most aggressive mode, $\gamma=0$, it 
is typically less aggressive than both forward stepwise selection
and best subset selection, in that it often has a smaller degrees of
freedom than these two.  See Figure \ref{fig:df2} for an example.  

\subsection{Other estimators}

Many other sparse estimators for regression could be
considered, for example, $\ell_1$-penalized alternatives to 
the lasso, like the Dantzig selector
\citep{candes2007dantzig} and square-root lasso
\citep{belloni2011square}; greedy alternatives to forward stepwise
algorithm, like matching pursuit \citep{mallat1993matching} and
orthogonal matching pursuit \citep{davis1994adaptive};
nonconvex-penalized methods, such as SCAD \citep{fan2001variable},
MC+ \citep{zhang2010nearly}, and SparseNet \citep{mazumder2011sparsenet};  
hybrid lasso/stepwise approaches like FLASH \citep{radchenko2011improved}; 
and many others.   

It would be interesting to include all of these estimators in our
comparisons, though that would make for a huge simulation suite and
would dilute 
the comparisons between best subset selection, forward stepwise, and  
the lasso that we would like to highlight. Roughly speaking, we would
expect the Dantzig selector and square-root lasso to perform similarly 
to the lasso; the matching pursuit variants to perform similarly
to forward stepwise; and the nonconvex-penalized methods to perform
somewhere in between the lasso and best subset selection.  (It is
worth noting that our R package is structured in such a way to
make further simulations and comparisons straightforward. We
invite interested readers to use it to perform comparisons to other
methods.) 

\subsection{Brief discussion of computational costs}
\label{sec:comp}

Computation of the lasso solution in 
\eqref{eq:lasso_prob} has been a
popular topic of research, and there are by now many efficient lasso  
algorithms.  In our simulations, we use coordinate descent with warm starts
over a sequence of tuning parameter values $\lambda_1 > \cdots  
> \lambda_m > 0$, as implemented in the {\tt glmnet} R package 
\citep{friedman2007pathwise,friedman2010regularization}.  The base
code for this is 
written in Fortran, and warm starts---plus additional tricks like 
active set optimization and screening rules
\citep{tibshirani2012strong}---make this implementation highly
efficient.  For example, for a problem with $n=500$ observations and
$p=100$ variables, {\tt glmnet} delivers the lasso solutions across 
100 values of $\lambda$ in less than 0.01 seconds, on a standard
laptop computer.  The relaxed lasso in \eqref{eq:relaxed_lasso} comes
at only a slight increase in computational cost, seeing as we must
only additionally compute the least squares coefficients on each
active set.  We provide an implementation in the {\tt bestsubset} R
package accompanying this paper, which
just uses an R wrapper around {\tt glmnet}. 
For the same example with $n=500$ and $p=100$, computing the relaxed
lasso path over 100 values of $\lambda$ and 10 values of $\gamma$ 
again took less than 0.01 seconds.

For forward stepwise selection, we implemented our own version 
in the {\tt bestsubset} R package.  The core matrix
manipulations for this method are written in C, and the rest 
is written in R.  The forward stepwise path is highly
structured and this greatly aids its computation: at step $k$, we have 
$k-1$ active variables included in the model, and we seek the variable
among the remaining $p-k+1$ that---once orthogonalized with respect
to the current active set of variables---achieves the greatest
absolute correlation with $Y$, as in \eqref{eq:fs_crit}. Suppose that we 
have maintained a QR decomposition of the active submatrix
\smash{$X_{A_{k-1}}$} of predictors, as well as the orthogonalization of the
remaining $p-k-1$ predictors with respect to \smash{$X_{A_{k-1}}$}.
We can compute the necessary
correlations in $O(n(p-k+1))$ operations, update the  QR  
factorization of \smash{$X_{A_k}$} in constant time, and orthogonalize
the remaining predictors with respect to the one just
included in $O(n(p-k))$  
operations (refer to the modified Gram-Schmidt algorithm in
\citealt{golub1996matrix}).  Hence, the forward stepwise 
selection path can be seen as a certain guided QR decomposition for 
computing the least squares coefficients on all $p$ variables (or, on some 
subset of $n$ variables when $p>n$).  For the same
example with $n=500$ and $p=100$, our implementation computes the
forward stepwise path in less than 0.5 seconds.

Best subset selection \eqref{eq:bs_prob} is the most
computationally challenging, by a large margin.
\citet{bertsimas2016best} 
describe two reformulations of \eqref{eq:bs_prob} as a mixed integer
quadratic program, one that is preferred when $n \geq p$, and the
other when $p>n$, and recommend using the Gurobi commercial MIO
solver (which is free for academic use).  They also describe a
proximal gradient descent method for computing approximate solutions
in \eqref{eq:bs_prob}, and recommend using the best output from this
algorithm over many randomly-initialized runs to warm start the
Gurobi solver.  See \citet{bertsimas2016best} for details.
We have implemented the method of these authors\footnote{We thank  
  the third author Rahul Mazumder for his help and guidance.}---which 
transforms the best subset selection problem into one of two MIO formulations
depending on the relative sizes of $n$ and $p$, uses proximal gradient
to compute a warm start, and then calls Gurobi through its R
interface---in our accompanying R package {\tt bestsubset}.  

Gurobi uses branch-and-cut techniques 
(a combination of branch-and-bound and cutting plane methods),
along with many other sophisticated optimization tools, for MIO
problems.  Compared to the pure branch-and-bound method from 
the {\tt leaps} R package, 
its speed can be impressive: for example, in one run with $n=500$ and 
$p=100$, it returned the best subset selection solution of size $k=8$
in about 3 minutes (brute-force search for this problem would need to
have looked at about 186 billion candidates!).  But for most problems
of this size ($n=500$ and $p=100$) it has been our experience that
Gurobi typically requires 1 hour or longer to complete its optimization.    
The third author Rahul Mazumder of
\citet{bertsimas2016best} suggested to us that for these problem
sizes, it is often the case that Gurobi has found the solution in less
than 3 minutes, though it takes much longer to certify its optimality. For
our simulations in the next section, we used a time limit of 3 minutes for
Gurobi to optimize the best subset selection problem \eqref{eq:bs_prob} at any
particular value of the subset size $k$ (once the time limit has been reached,
the solver returns its best iterate).  For more discussion on this
choice and its implications, see Section \ref{sec:gurobi}.  We note that this is
already quite a steep computational cost for ``regular'' practical usage: at 3
minutes per value of $k$, if we wanted to use 10-fold cross-validation to choose
between the subset sizes $k=0,\ldots,50$, then we are already facing 25 hours of   
computation time.  

\section{Simulations}

\subsection{Setup}
\label{sec:setup}

We present simulations, basically following the simulation setup of  
\citet{bertsimas2016best}, except that we consider a wider range of SNR 
values. Given $n,p$ (problem dimensions), $s$
(sparsity level), beta-type (pattern of sparsity), $\rho$ 
(predictor autocorrelation level), and $\nu$ (SNR level), our process can be
described as follows:    
\begin{enumerate}
\item[i.] we define coefficients $\beta_0 \in \R^p$ according to $s$
  and the beta-type, as described below;
\item[ii.] we draw the rows of the predictor matrix $X \in 
  \R^{n \times p}$ i.i.d.\ from $N_p(0,\Sigma)$, where $\Sigma \in 
  \R^{p\times p}$ has entry $(i,j)$ equal to $\rho^{|i-j|}$; 
\item[iii.] we draw the response vector $Y \in \R^n$ from 
$N_n(X\beta_0, \sigma^2 I)$, with $\sigma^2$ defined to meet the   
  desired SNR level, i.e., $\sigma^2=\beta_0^T \Sigma \beta_0 /\nu$;  
\item[iv.] we run the lasso, relaxed lasso, forward stepwise
  selection, and best subset selection on the data $X,Y$, each over a 
  wide range of tuning parameter values; for each method we choose
  the tuning parameter by minimizing prediction error on  
  validation set \smash{$\widetilde{X} \in \R^{n\times p}, 
    \widetilde{Y} \in \R^n$} that is generated independently of and  
  identically to $X,Y$, as in steps ii--iii above;
\item[v.] we record several metrics of interest, as specified below; 
\item[vi.] we repeat steps ii-v a total of 10 times, and average the
  results. 
\end{enumerate}
Below we describe some aspects of the simulation process in more
detail. 

\paragraph{Coefficients.}  We considered four settings for the
coefficients $\beta_0 \in \R^p$: 
\begin{itemize}
\item {\bf beta-type 1:} $\beta_0$ has $s$ components equal to 1, 
  occurring at (roughly) equally-spaced indices between 1 and $p$,
  and the rest equal to 0; 
\item {\bf beta-type 2:} $\beta_0$ has its first $s$ components equal
  to 1, and the rest equal to 0;
\item {\bf beta-type 3:} $\beta_0$ has its first $s$ components
  taking nonzero values equally-spaced between 10 and 0.5, and the  
  rest equal to 0; 
\item {\bf beta-type 5:} $\beta_0$ has its first $s$ components equal
  to 1, and the rest decaying exponentially to 0, 
  specifically, $\beta_{0i}=0.5^{i-s}$, for $i=s+1,\ldots,p$. 
\end{itemize}
The first three types were studied in
\citet{bertsimas2016best}.  They also defined a fourth type 
that we did not include here, as we found it yielded basically the 
same results as beta-type 3.  The last type above is new: we
included it to investigate the effects of weak sparsity and call it 
beta-type 5, to avoid confusion.

\paragraph{Evaluation metrics.} Let $x_0 \in \R^p$ denote test  
predictor values drawn from $N_p(0,\Sigma)$ (as in the rows of the 
training predictor matrix $X$) and let $y_0 \in \R$ denote its
associated response value drawn from $N(x_0^T \beta_0,
\sigma^2)$. Also let \smash{$\hbeta$} denote estimated coefficients
from one of the regression procedures.  We considered the following
evaluation metrics:  
\begin{itemize}
\item {\bf Relative risk:} this is the accuracy metric studied in
  \cite{bertsimas2016best}\footnote{Actually, these authors used
    an ``in-sample'' version of this metric defined as
    \smash{$\|X\hbeta-X\beta_0\|_2^2/\|X\hbeta\|_2^2$}, whereas our 
    definition is ``out-of-sample'', with an expectation over the new
    test predictor value $x_0$ taking the place of the sample average 
    over the training values $x_i$, $i=1,\ldots,n$.}, defined as
$$
\mathrm{RR}(\hbeta) = 
\frac{\E(x_0^T\hbeta-x_0^T\beta_0)^2}{\E(x_0^T\beta_0)^2} = 
\frac{(\hbeta-\beta_0)^T\Sigma(\hbeta-\beta_0)}
{\beta_0^T\Sigma\beta_0}.
$$
The expectations here and below are taken over the test point $(x_0,y_0)$, 
with all training data and validation data (and thus \smash{$\hbeta$}) held
fixed. A perfect score is 0 (if \smash{$\hbeta=\beta_0$})
and the null score is 1 (if \smash{$\hbeta=0$}).

% \item {\bf Relative test error:} this is the expected test error relative
%   to that of best subset selection,
% $$
% \mathrm{RTE}(\hbeta) = \frac{\E(y_0-x_0^T\hbeta)^2}{\E(y_0-x_0^T
%   \hbeta^\mathrm{sub})} = 
% \frac{(\hbeta-\beta_0)^T\Sigma(\hbeta-\beta_0) + \sigma^2}
% {(\hbeta^\mathrm{sub}-\beta_0)^T \Sigma
% (\hbeta^\mathrm{sub}-\beta_0) + \sigma^2},
% $$
% where \smash{$\hbeta^\mathrm{sub}$} denotes a solution in \eqref{eq:bs_prob}. 
% % (or the result of running the Gurobi optimizer until the time limit has been
% % reached). 
% A perfect score is
% \smash{$\sigma^2/
% ((\hbeta^\mathrm{sub}-\beta_0)^T \Sigma
% (\hbeta^\mathrm{sub}-\beta_0) + \sigma^2)$}.
% % and the null score is 
% % \smash{$(\beta_0^T \Sigma \beta_0 + \sigma^2)/
% % ((\hbeta^\mathrm{sub}-\beta_0)^T \Sigma
% % (\hbeta^\mathrm{sub}-\beta_0) + \sigma^2)$}.

% This metric allows for the most direct comparison to best subset selection. 
% While one could also consider the risk relative to that of best subset
% selection, we prefer the relative test error defined above.  The reason is that
% the former metric can very high variance in the high SNR regime, as the
% denominator can become quite close to zero; the latter metric does not suffer
% from this problem, as the denominator is always at least $\sigma^2$.

\item {\bf Relative test error:} this measures the expected test error  
  relative to the Bayes error rate, 
$$
\mathrm{RTE}(\hbeta) =
\frac{\E(y_0-x_0^T\hbeta)^2}{\sigma^2} =
\frac{(\hbeta-\beta_0)^T\Sigma(\hbeta-\beta_0) 
  +\sigma^2}{\sigma^2}.     
$$
A perfect score is 1 and the null score is
\smash{$(\beta_0^T\Sigma\beta_0+\sigma^2)/\sigma^2=\SNR+1$}.  

\item {\bf Proportion of variance explained:} 
as defined in Section \ref{sec:snr}, this is
$$
\mathrm{PVE}(\hat\beta)
=1-\frac{\E(y_0-x_0^T \hbeta)^2}{\Var(y_0)} 
=1-\frac{(\hbeta-\beta_0)^T\Sigma(\hbeta-\beta_0)+\sigma^2}
{\beta_0^T\Sigma\beta_0+\sigma^2}.
$$
A perfect score is $\SNR/(1+\SNR)$ and the null score is 0. 

\item {\bf Number of nonzeros:} unlike the last three metrics
  which measure predictive accuracy, this metric simply records the
  number of nonzero estimated coefficients, 
  \smash{$\|\hbeta\|_0 = \sum_{i=1}^p 1\{\hbeta_i\not=0\}$}.
\end{itemize}

It is worth noting that, in addition to metrics based on
predictive accuracy, it would be useful to consider a metric
that measures proper variable recovery, i.e., the extent to which the 
sparsity pattern in the estimated \smash{$\hbeta$} matches that in
$\beta_0$. We briefly touch on this in the discussion. Here we mention
one advantage to studying predictive accuracy: any of the metrics
defined above are still relevant when $\E(y|x)$ is
no longer assumed to be linear, making the predictive angle more
broadly practically relevant than a study of proper variable recovery 
(which necessarily requires linearity of the true mean). 

\paragraph{Configurations.}  We considered the following
four problem settings:
\begin{itemize} 
\item {\bf low:} $n=100$, $p=10$, $s=5$;
\item {\bf medium:} $n=500$, $p=100$, $s=5$; 
\item {\bf high-5:} $n=50$, $p=1000$, $s=5$;
\item {\bf high-10:} $n=100$, $p=1000$, $s=10$.
\end{itemize}
In each setting, we considered ten values for the SNR ranging
from 0.05 to 6 on a log scale, namely
$$
\begin{array}{c|cccccccccc}
  \mbox{SNR}&0.05&
                    0.09&0.14&0.25&0.42&0.71&1.22&2.07&3.52&6.00 \\[7pt]
  \hline\\[-5pt]
  \mbox{PVE}& 0.05& 0.08& 0.12& 0.20& 0.30& 0.42& 0.55& 0.67& 0.78& 0.86
\end{array}
$$
(For convenience we provide the corresponding population PVE as well.)
In each setting, we also considered three values for the
predictor autocorrelation $\rho$, namely 0, 0.35, and 0.7.

\paragraph{Tuning of procedures.}
In the low setting, the lasso was tuned over 50       
values of $\lambda$ ranging from $\lambda_{\max}=\|X^T Y\|_\infty$
to a small fraction of $\lambda_{\max}$ on a log scale, as per the 
default in {\tt glmnet}, and the relaxed lasso was tuned over the same 50 
values of $\lambda$, and 10 values of $\gamma$ equally spaced from 1 to
0 (hence a total of 500 tuning parameter values).  Also in the low setting,  
forward stepwise and best subset selection were tuned over steps
$k=0,\ldots,10$.  In all other problem settings (medium, high-5, and high-10),
the lasso was tuned over 100 values of $\lambda$, the relaxed lasso was tuned
over the same 100 values of $\lambda$ and 10 values of $\gamma$ (hence 1000
tuning parameter values total), and forward stepwise and best subset selection
were tuned over steps $k=0,\ldots,50$. In all cases, tuning was performed by  
by minimizing prediction error on an external validation set of size $n$, which
we note approximately matches the precision of leave-one-out cross-validation.  

\subsection{Time budget for Gurobi}
\label{sec:gurobi}

As mentioned in Section \ref{sec:comp}, for each problem instance
and subset size $k$, we used a time limit of 3
minutes for Gurobi to optimize the best subset selection problem.  In
comparison, \citet{bertsimas2016best} 
used much larger time budgets: 15 minutes (per problem per $k$) for
problems with $p=100$ as in our medium setup, and 66 minutes (per
problem per $k$) for problems with $p \geq 2000$ as in our high-5 and high-10
setups.  Their simulations however were not as extensive, as they
looked at fewer combinations of beta-types, SNR levels, and correlation levels.  
Another important difference worth mentioning: the Gurobi optimizer,
when run through its Python/Matlab interface,
 as in \citet{bertsimas2016best}, 
automatically takes advantage of multithreading capabilities; this does not
appear to be the case when run through its R interface, as in our simulations. 

The third author of \citet{bertsimas2016best}, Rahul Mazumder, suggested 
in personal communication that the MIO solver in the medium setting will often
arrive at the best subset selection solution in less than 3 minutes, but it can
take much longer to certify its optimality\footnote{Gurobi
  constructs a sequence of lower and upper bounds on the criterion in  
  \eqref{eq:bs_prob}; typically the lower bounds come from convex 
  relaxations and the upper bounds from the current iterates, and it
  is the lower bounds that take so long to converge.} (usually over 1 hour, in
absence of extra speedup tricks as described in \citealp{bertsimas2016best}).  
Meanwhile, in the high-5 and high-10 settings, this author also pointed out that
3 minutes may no longer be enough.  For practical reasons, we have kept 
the 3 minute budget per problem instance per subset size.  
Note that this amounts to 150 minutes per path of 50 solutions, 1500 minutes or
25 hours per set of 10 repetitions, and in total 750 hours or 31.25 days for
any given setting, once we go through the 10 SNR levels and 3 correlation levels.    

\subsection{Results: computation time}

In Table \ref{tab:timings}, we report the time in seconds taken by each 
method to compute one path of solutions, averaged over 10 repetitions and all
SNR and predictor correlation levels in the given setting.  All
timings were recorded on a Linux cluster.
As explained above, the lasso path consisted of 50 tuning parameter values in
the low setting and 100 in all other settings, the relaxed lasso path consisted
of 500 tuning parameter values in the low setting and 1000 in all other
settings, and the forward stepwise and best subset selection paths each
consisted of $\min\{p,50\}$ tuning parameter values.  

\begin{table}[hbtp]
\centering
\begin{tabular}{l|rrrr}
Setting & BS & FS & Lasso & RLasso \\ 
\hline
{\bf low} ($n=100$, $p=10$, $s=5$)
& 3.43 & 0.006 & 0.002 & 0.002 \\  
{\bf medium} ($n=500$, $p=100$, $s=5$)
& {\it  $\approx 120$ min} & 0.818 & 0.009 & 0.009 \\   
{\bf high-5} ($n=50$, $p=1000$, $s=5$)
& {\it $\approx 126$ min} & 0.137 & 0.011 & 0.011 \\
{\bf high-10} ($n=100$, $p=1000$, $s=10$) 
& {\it $\approx 144$ min} & 0.277 & 0.019 & 0.021 \\   
%{\bf big} (n=1000, p=1000) & ? & {\bf 3.5 min} 
%& 0.630 & 0.632 \\ 
\end{tabular}
\caption{\it Time in seconds for one path of solutions 
  for best subset selection (BS), forward stepwise selection (FS), the lasso,
  and relaxed lasso (RLasso).  The times were averaged over 20
  repetitions, and all SNR and predictor correlation levels in the given
  setting.} 
\label{tab:timings}
\end{table}

We can see that the lasso and relaxed lasso are very fast,
requiring less than 25 milliseconds in every case.
Forward stepwise selection is also fast, though not quite as fast as the lasso
(some of the differences here might be due to the fact that 
our forward stepwise  
algorithm is implemented partly in R).  Moreover, it should be noted that 
when $n$ and $p$ is large, and one wants to explore models with a sizeable
number of variables (we limited our search to models of size 50), forward
stepwise has to plod through its path one variable at a time, but the lasso 
can make jumps over subset sizes bigger than one by varying $\lambda$ and
leveraging warm starts.

Recall, the MIO solver for best subset selection was allowed 3 minutes per
subset size $k$, or 150 minutes for a path of 50 subset sizes.
As the times in Table \ref{tab:timings} suggest, the maximum allotted time was
not reached in all instances, and the MIO solver managed to verify optimality
of some solutions along the path. In the medium setting, on average 
17.55 of the 50 solutions were verified as being optimal. In the
high-5 and high-10 settings, only 1.61 of the 50 were verified on average (note
this count includes the subset of size 1, which is trivial).  These measures may 
be pessismistic, as Gurobi may have found high-quality approximate solutions or 
even exact solutions but was just not able to verify them in time, see the
discussion in the above subsection.

\subsection{Results: accuracy metrics}

Here we display a slice of the accuracy results, focusing for concreteness on
the case in which the predictor autocorrelation is $\rho=0.35$, and the 
population coefficients follow the beta-type 2 pattern.  In a supplementary
document, we display the full set of results, over the whole simulation
design.  

\begin{figure}[p]
\centering
{\bf Low setting: $n=100$, $p=10$, $s=5$ \\
Correlation $\rho=0.35$, beta-type 2} 
\bigskip 

\begin{tabular}{ll}
\includegraphics[height=0.45\textwidth]{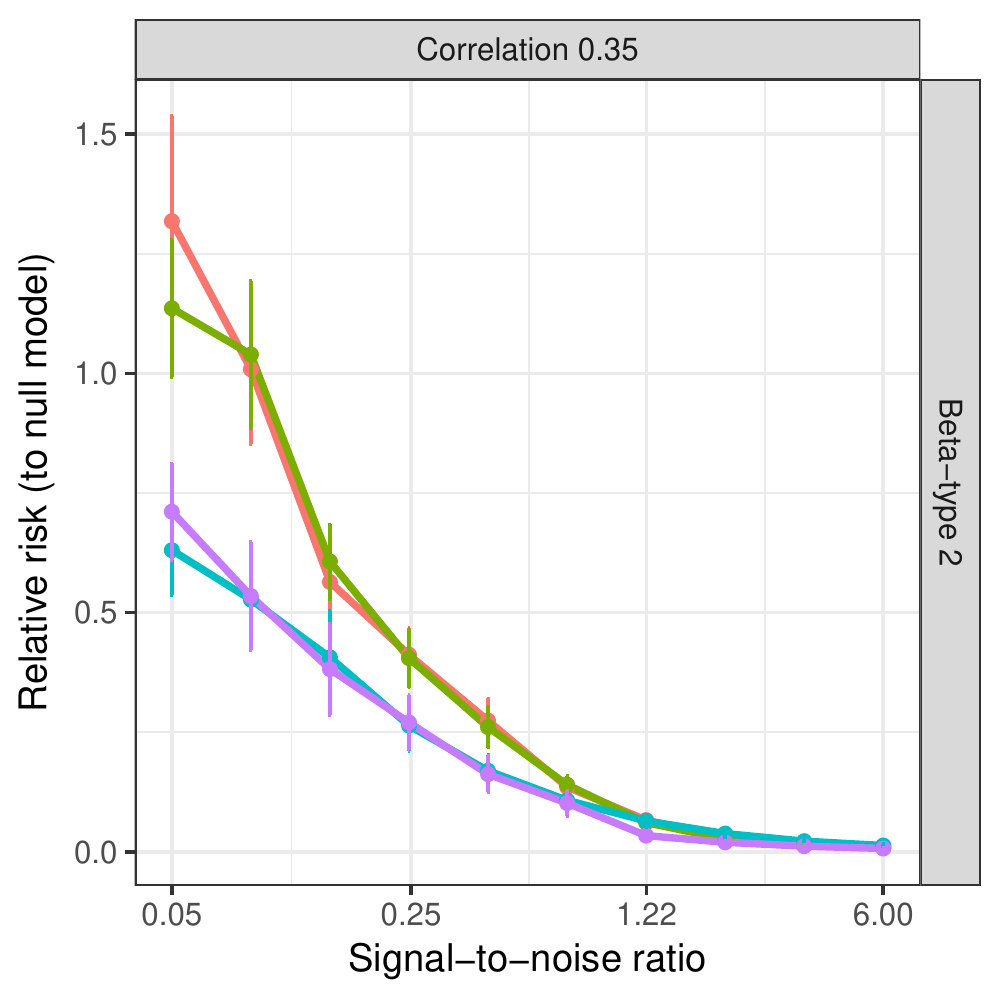} &
\includegraphics[height=0.45\textwidth]{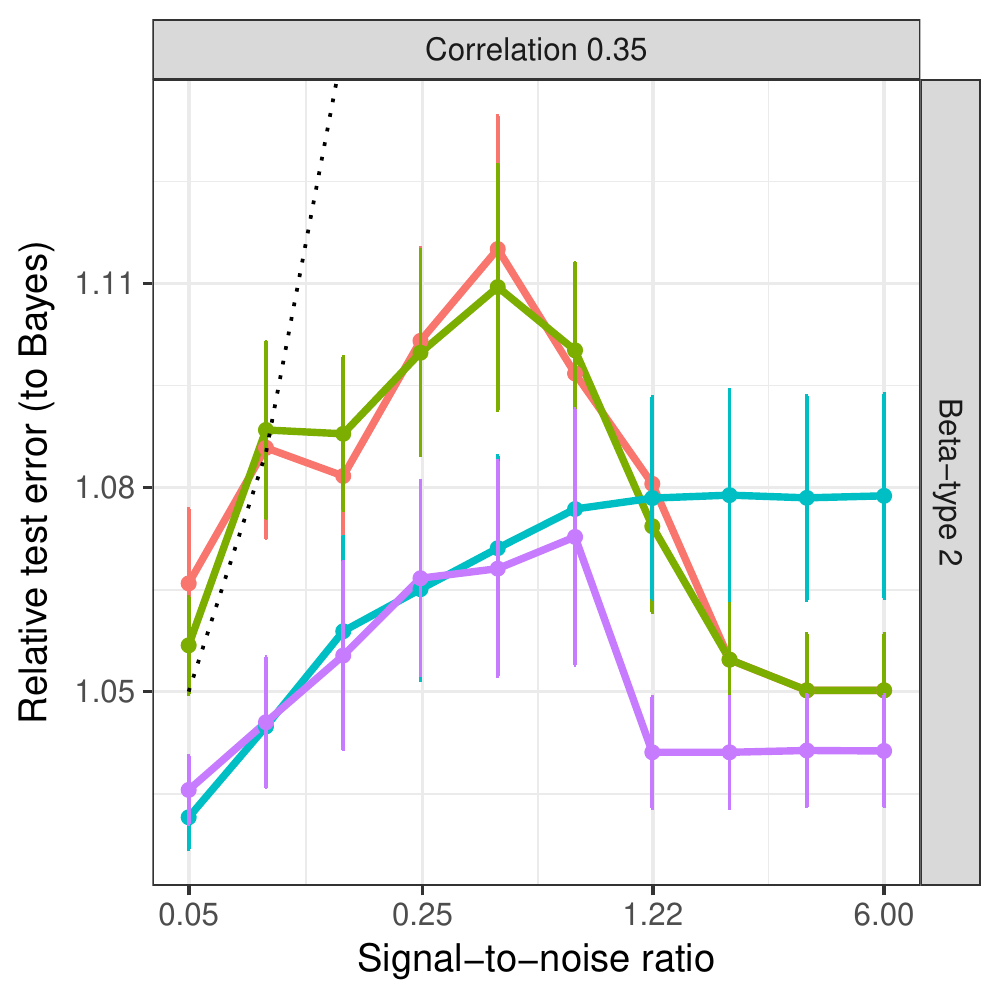} \\
\includegraphics[height=0.45\textwidth]{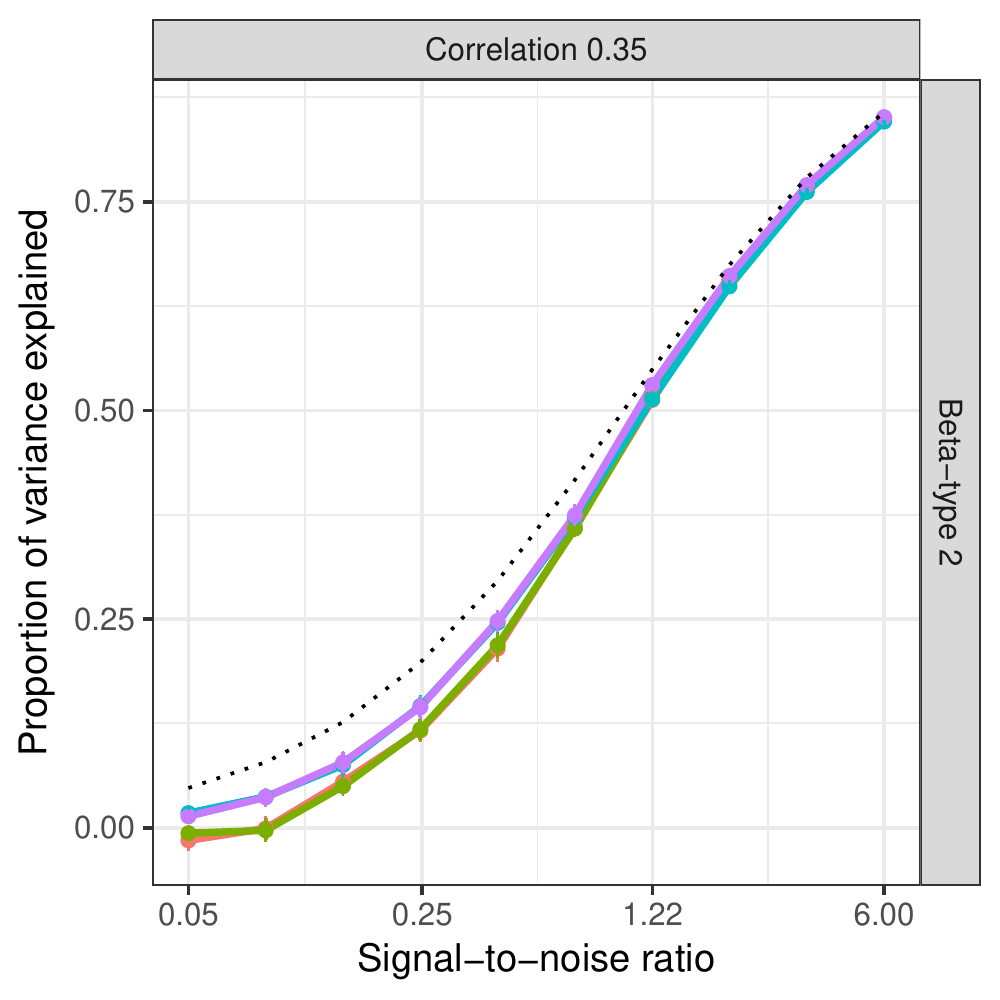} &
\includegraphics[height=0.45\textwidth]{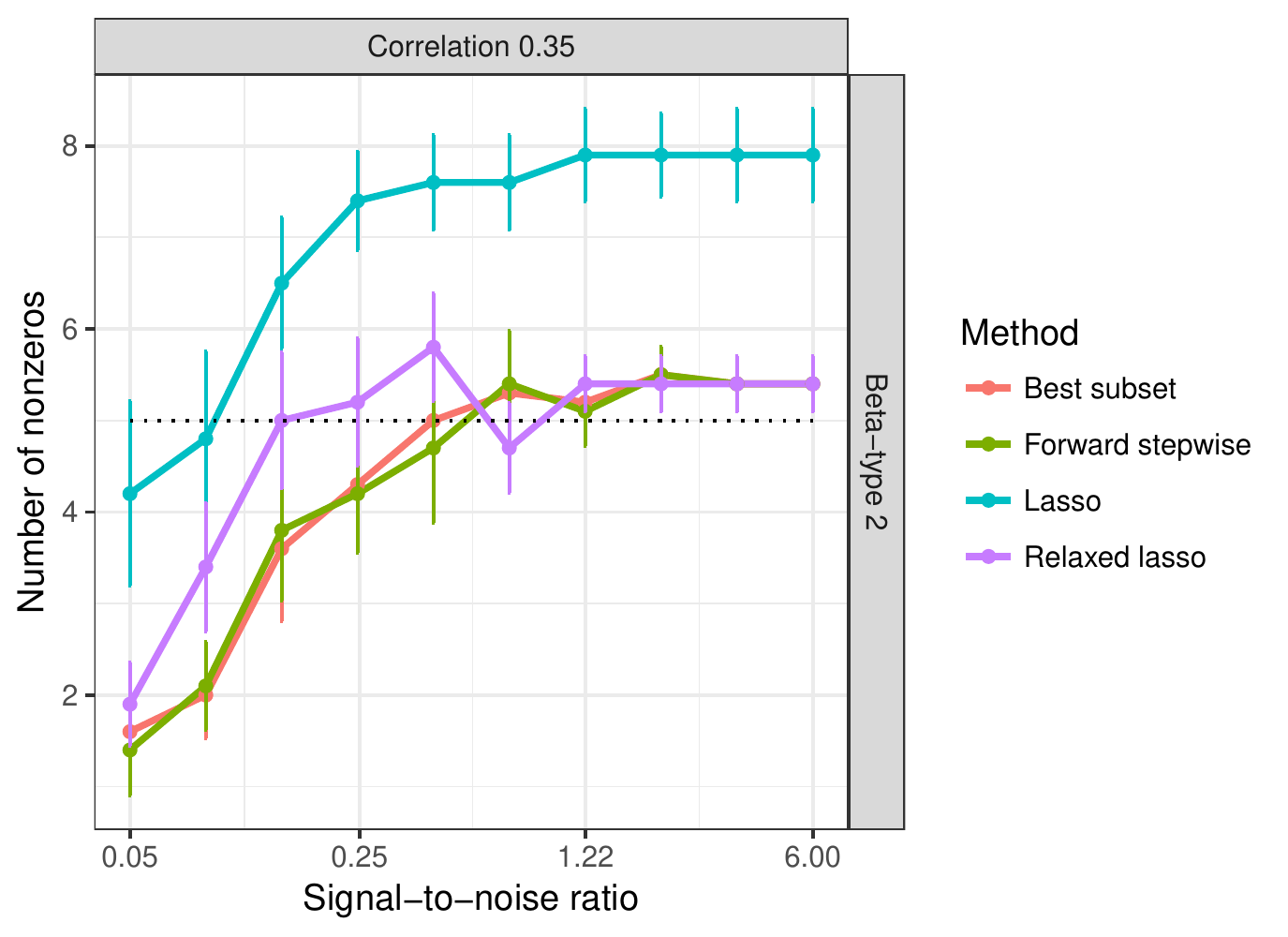}
\end{tabular}
\caption{\it Relative risk, relative test error, PVE, and number of nonzero
  curves as functions of SNR, in the low setting with $n=100$, $p=10$, and
  $s=5$.} 
\label{fig:lo}
\end{figure}

\begin{figure}[p]
\centering
{\bf Medium setting: $n=500$, $p=100$, $s=5$ \\
Correlation $\rho=0.35$, beta-type 2} 
\bigskip 

\begin{tabular}{ll}
\includegraphics[height=0.45\textwidth]{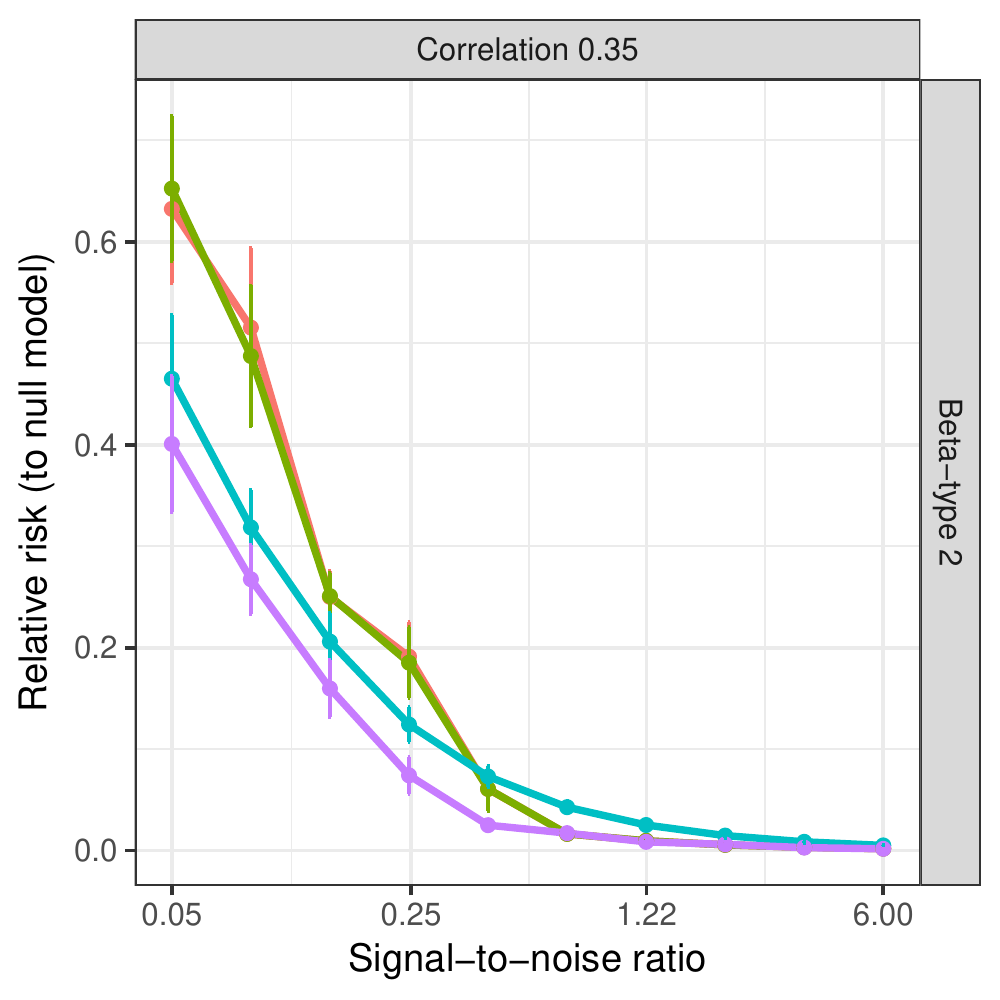} &
\includegraphics[height=0.45\textwidth]{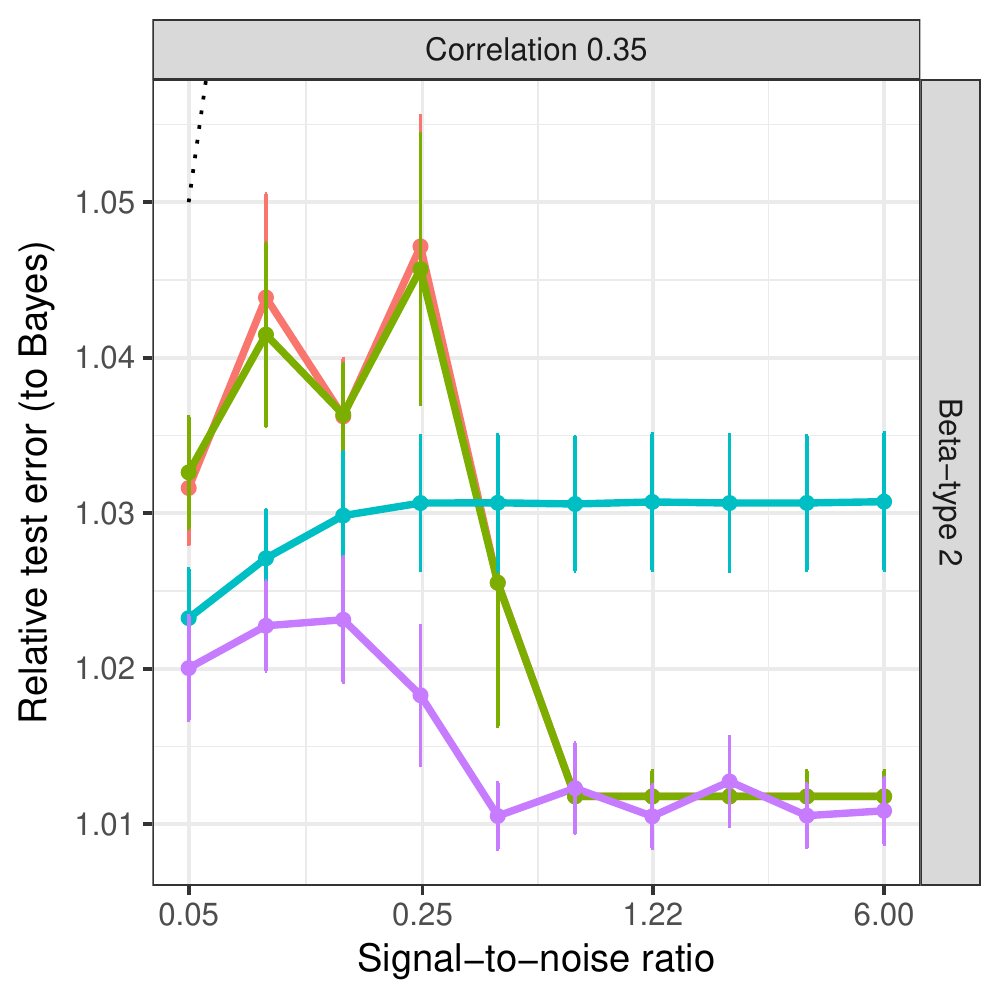} \\
\includegraphics[height=0.45\textwidth]{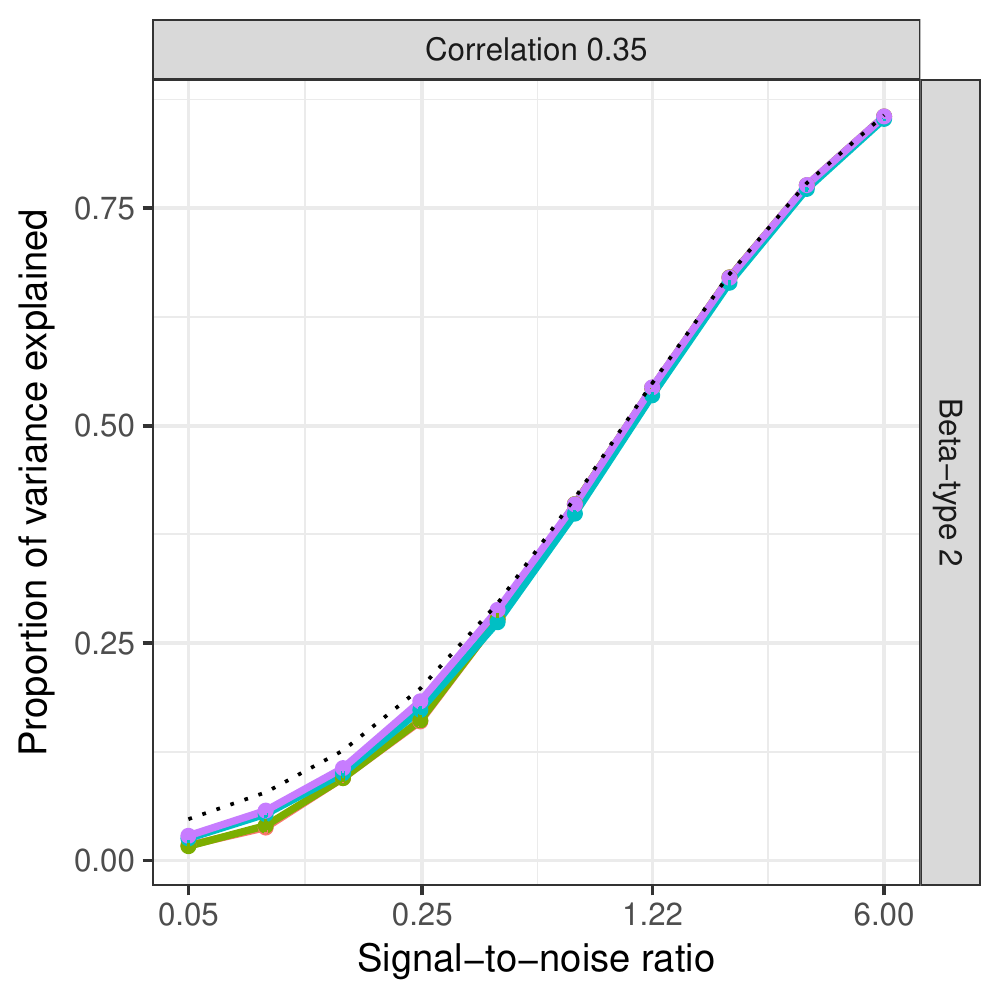} &
\includegraphics[height=0.45\textwidth]{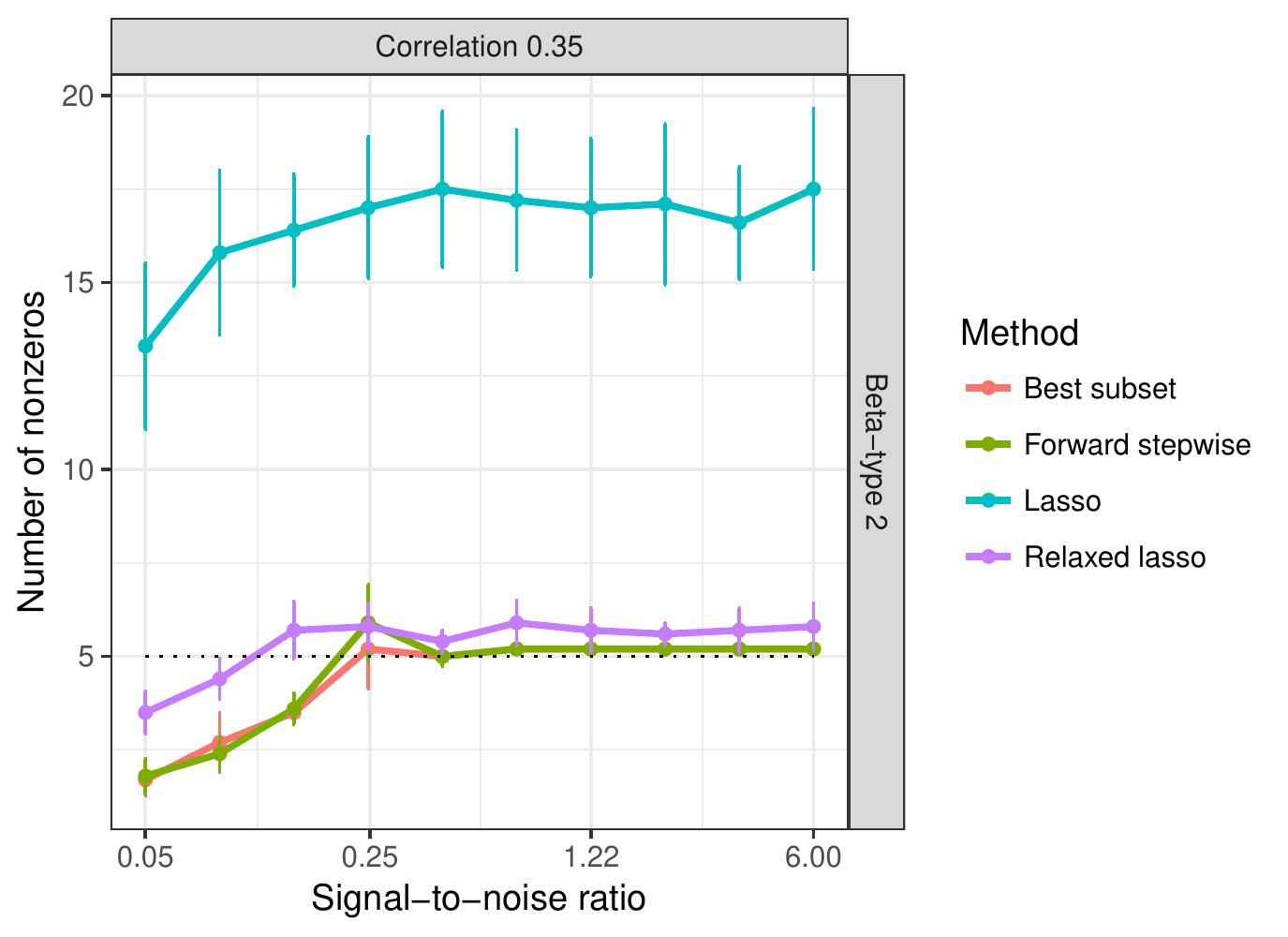}
\end{tabular}
\caption{\it Relative risk, relative test error, PVE, and number of nonzero
  curves as functions of SNR, in the medium setting with $n=500$, $p=100$, and  
  $s=5$.} 
\label{fig:med}
\end{figure}

\begin{figure}[p]
\centering
{\bf High-5 setting: $n=50$, $p=1000$, $s=5$ \\
Correlation $\rho=0.35$, beta-type 2} 
\bigskip 

\begin{tabular}{ll}
\includegraphics[height=0.45\textwidth]{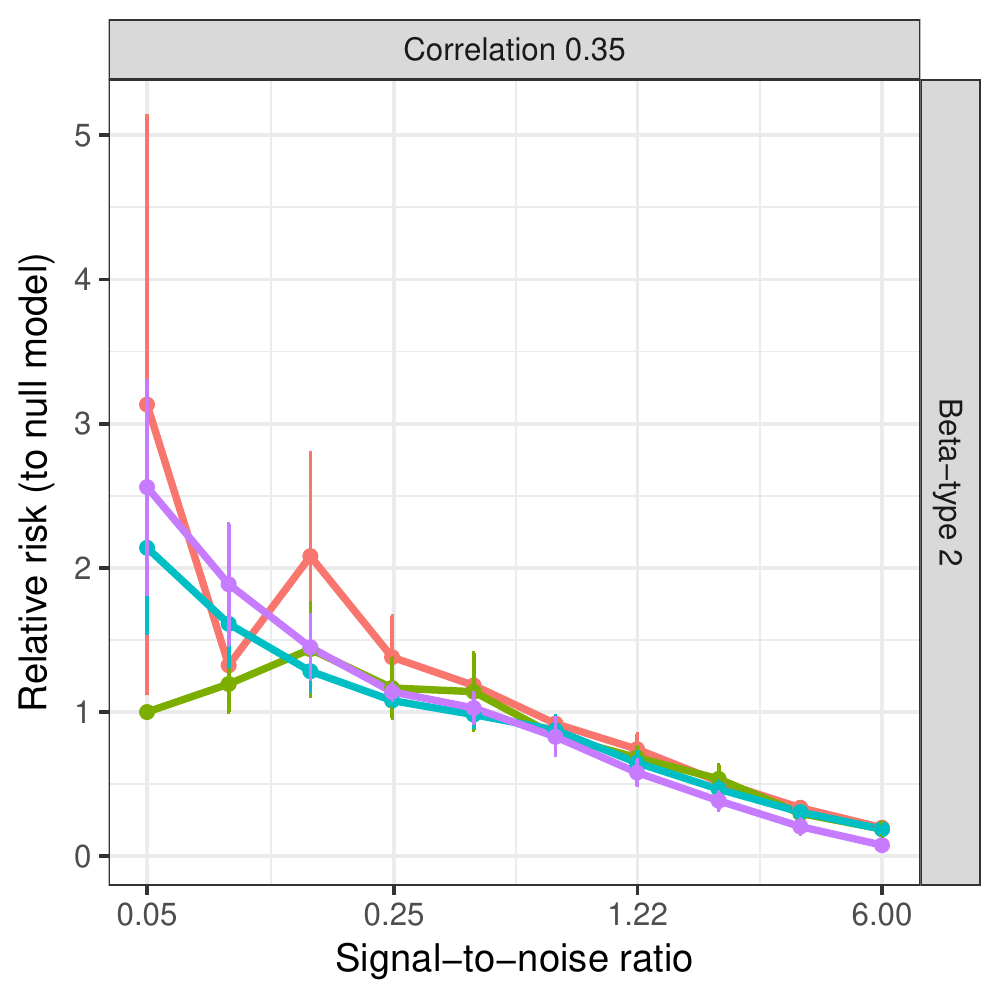} &
\includegraphics[height=0.45\textwidth]{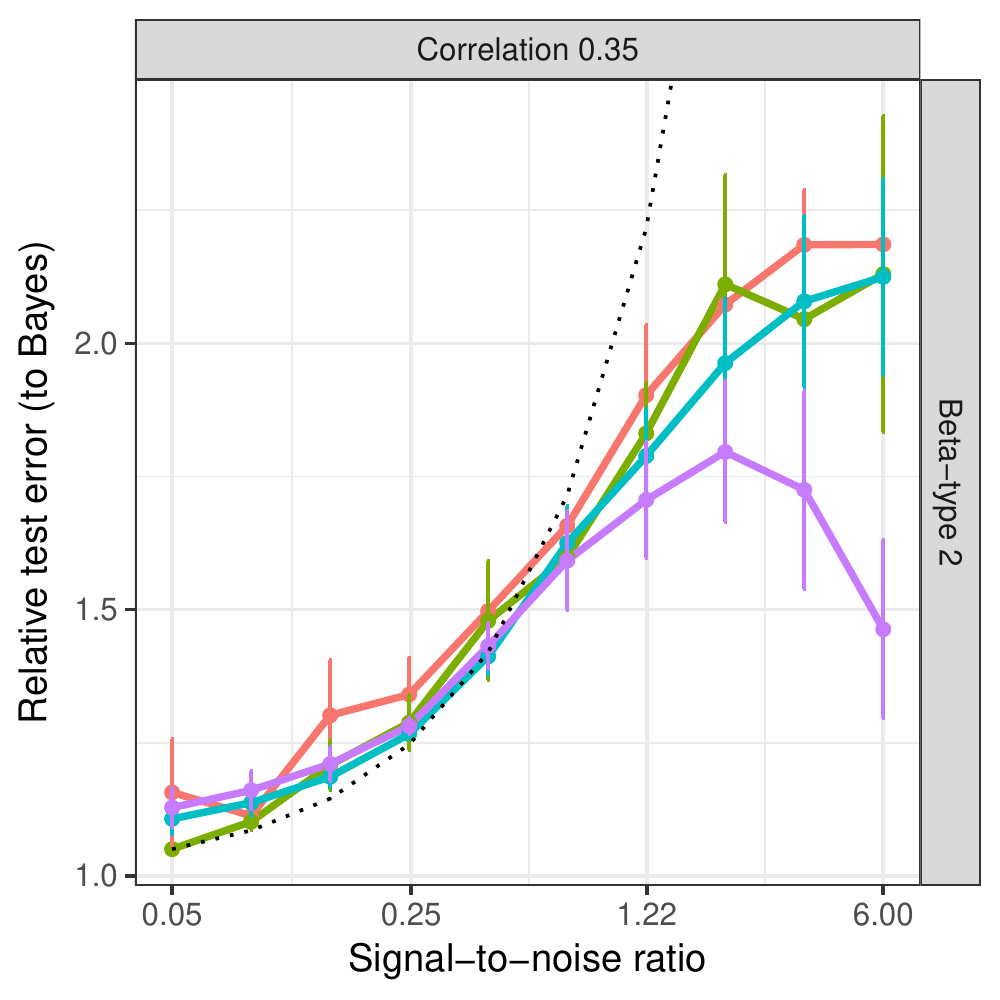} \\
\includegraphics[height=0.45\textwidth]{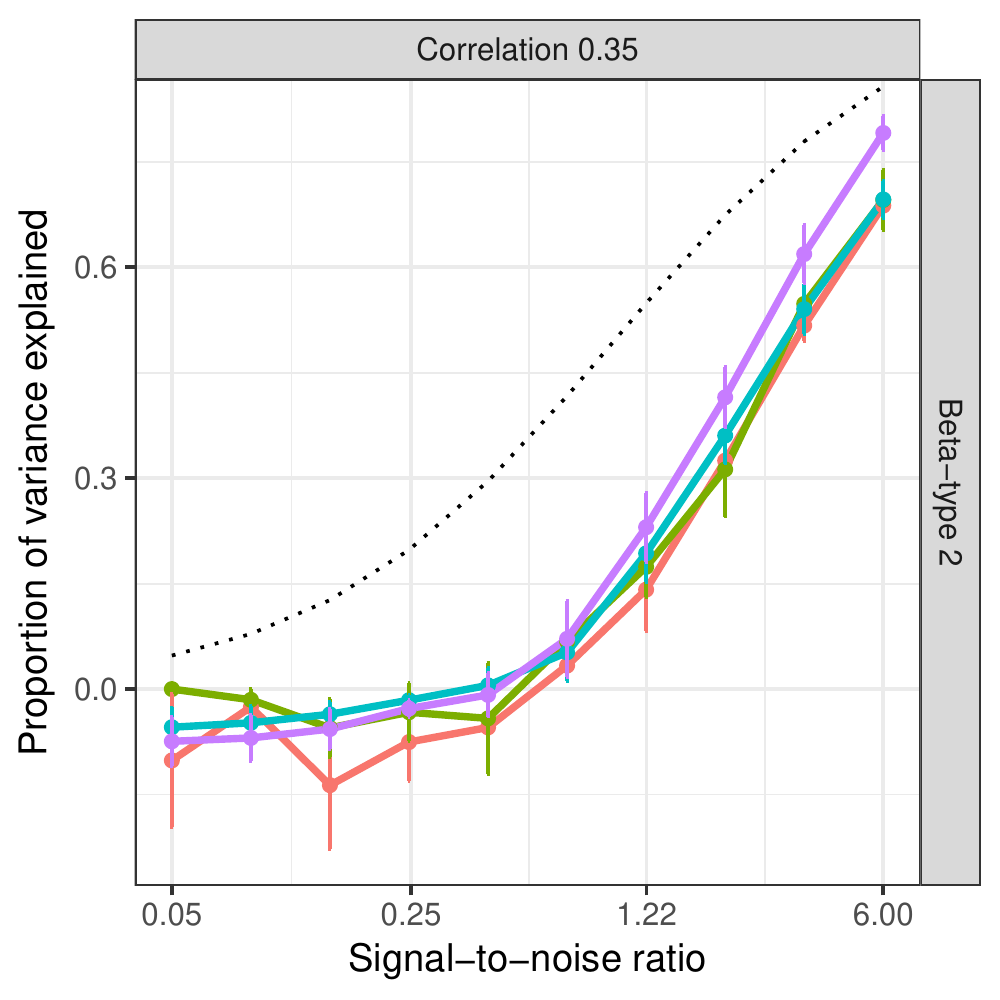} &
\includegraphics[height=0.45\textwidth]{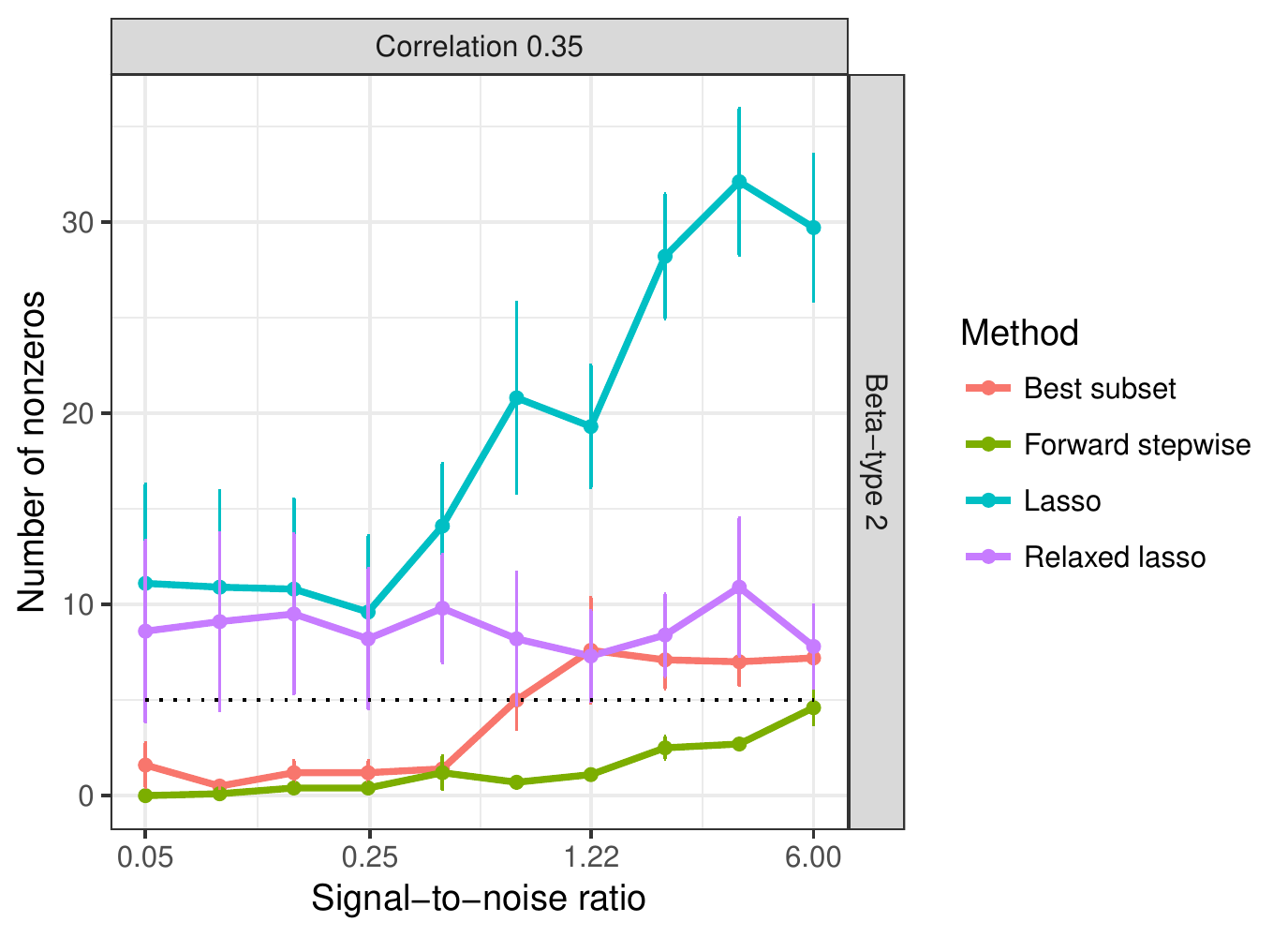}
\end{tabular}
\caption{\it Relative risk, relative test error, PVE, and number of nonzero
  curves as functions of SNR, in the high-5 setting with $n=50$, $p=1000$, and  
  $s=5$.} 
\label{fig:hi5}
\end{figure}

\begin{figure}[p]
\centering
{\bf High-10 setting: $n=100$, $p=1000$, $s=10$ \\
Correlation $\rho=0.35$, beta-type 2} 
\bigskip 

\begin{tabular}{ll}
\includegraphics[height=0.45\textwidth]{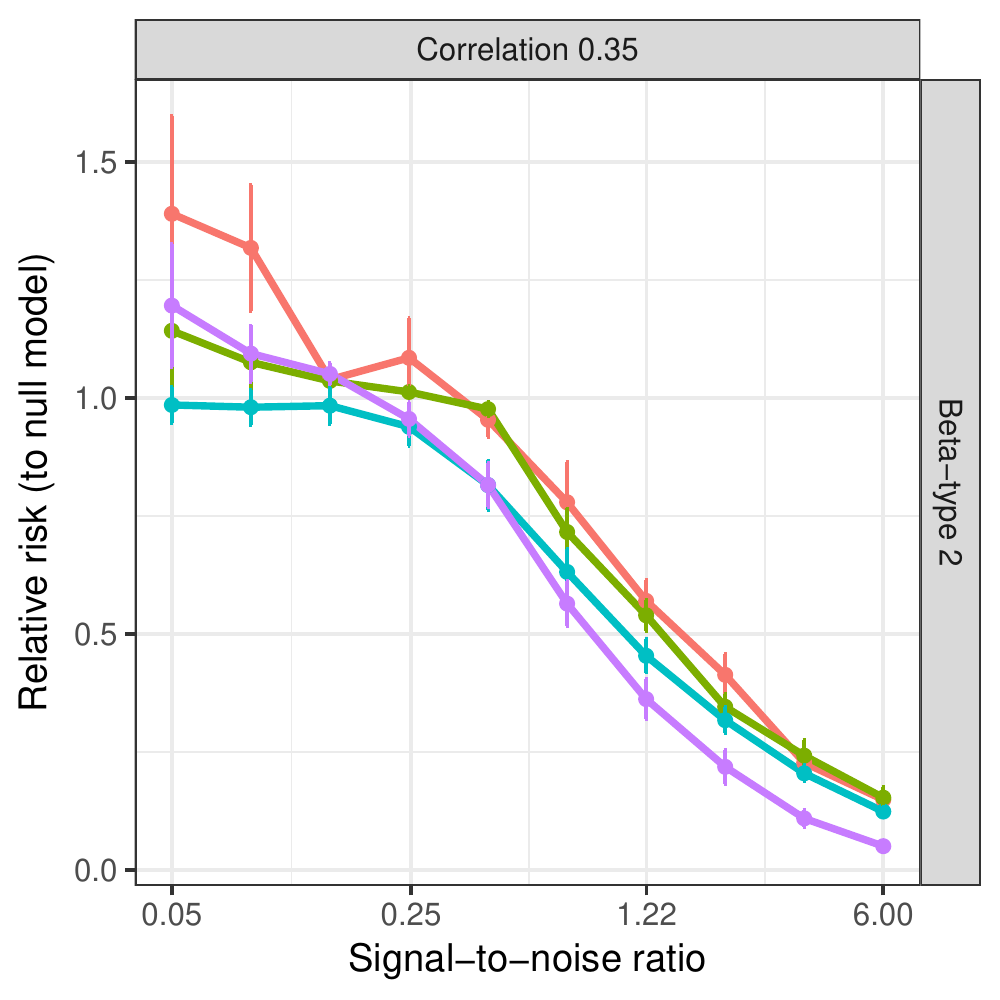} &
\includegraphics[height=0.45\textwidth]{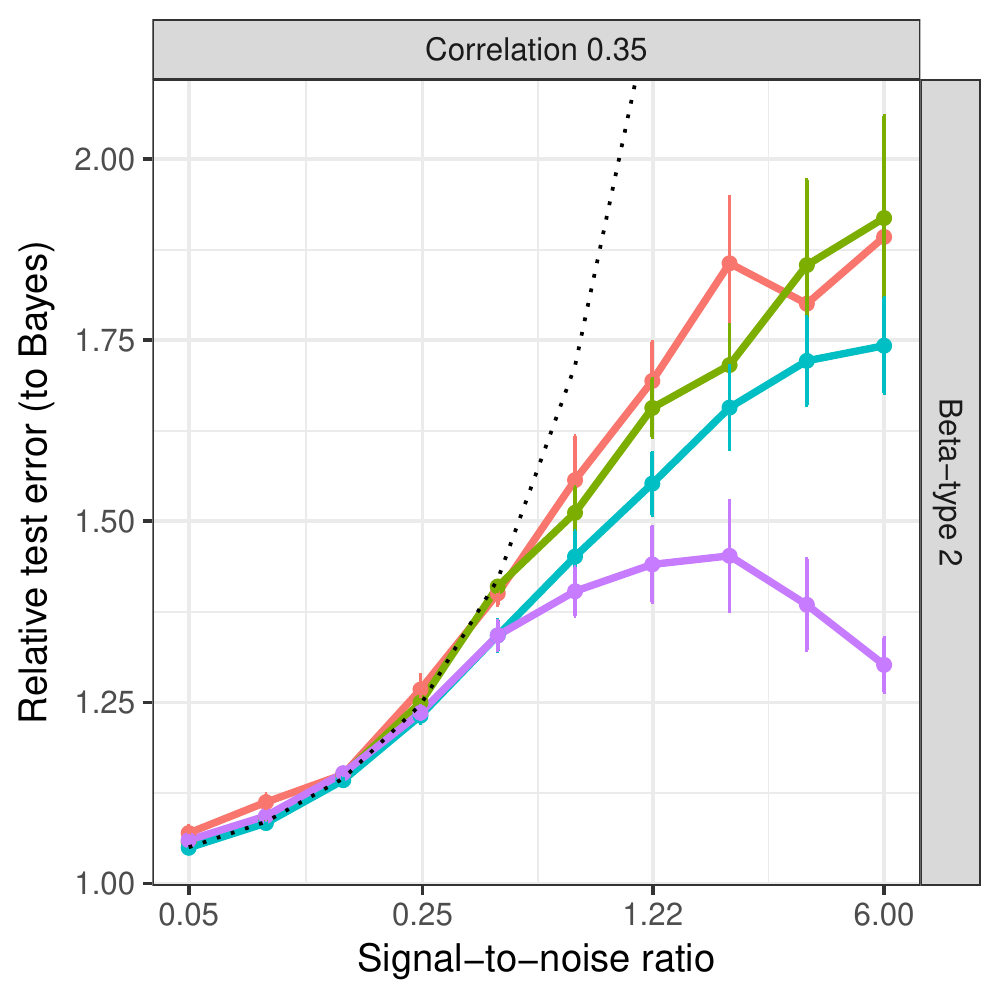} \\
\includegraphics[height=0.45\textwidth]{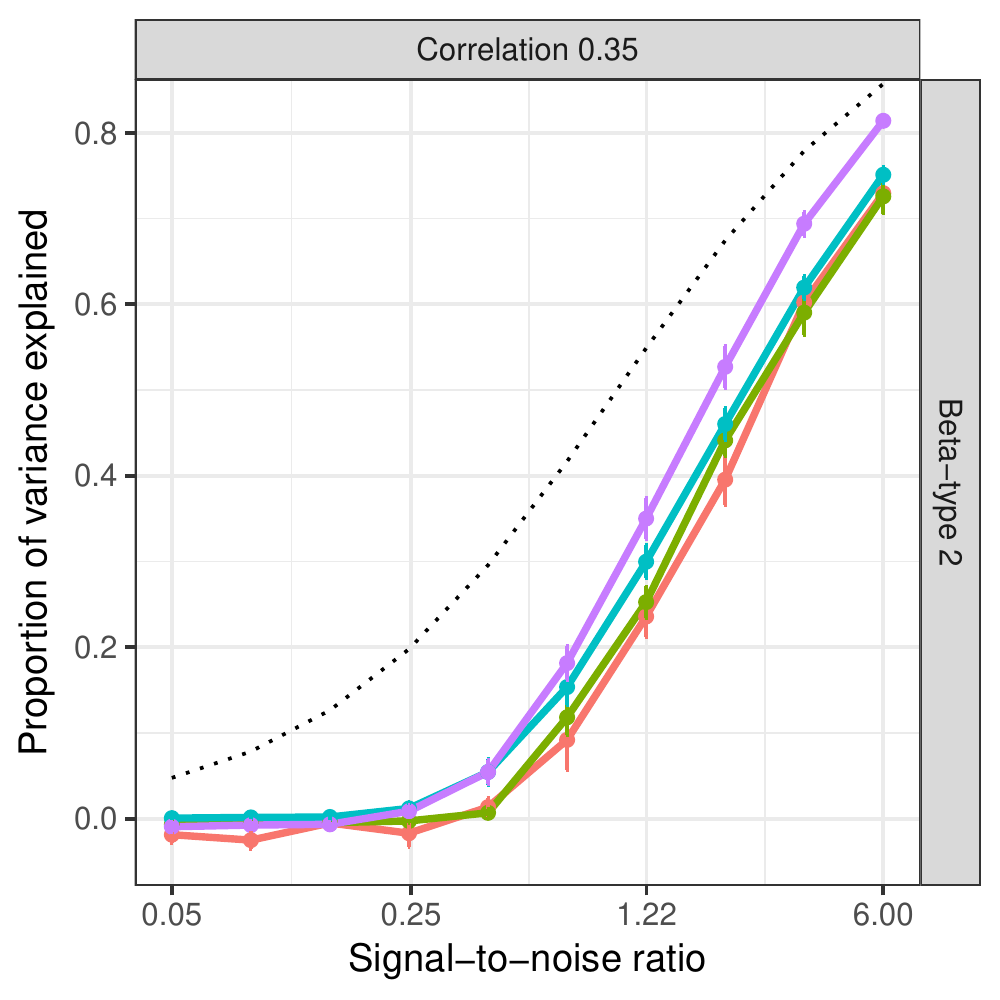} &
\includegraphics[height=0.45\textwidth]{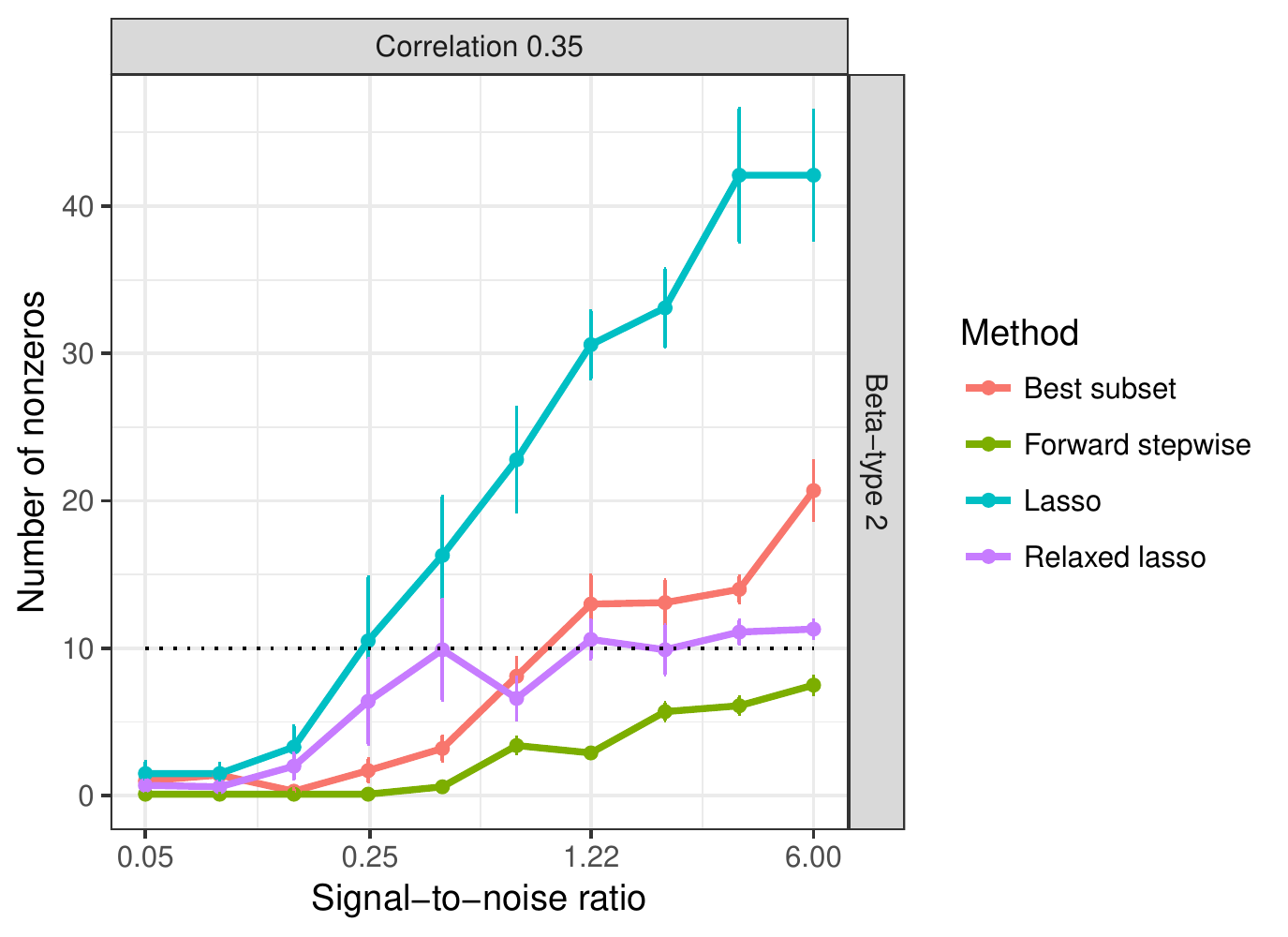}
\end{tabular}
\caption{\it Relative risk, relative test error, PVE, and number of nonzero
  curves as functions of SNR, in the high-10 setting with $n=100$, $p=1000$, and    
  $s=10$.} 
\label{fig:hi10}
\end{figure}

Figure \ref{fig:lo} plots the relative risk, relative test error, PVE, and
number of nonzero coefficients as functions of 
the SNR level, for the low setting.  Figures \ref{fig:med}, \ref{fig:hi5},
and \ref{fig:hi10} show the same for the medium, high-5, and high-10 settings,  
respectively.  Each panel in the figures displays the average of
a given metric over 10 repetitions, for the four methods in question, and
vertical bars denote one standard 
error.  In the relative test error plots, the dotted curve denotes
the performance of the null model (null score); in the PVE plots, it denotes 
the performance of the true model (perfect score); in the number of 
nonzero plots, it marks the true support size $s$. 

The low and medium settings, Figures \ref{fig:lo} and \ref{fig:med}, yield
somewhat similar results.  In the relative 
risk and PVE plots (top left and bottom left panels), we see that best subset
and forward stepwise selection lag behind the lasso and relaxed lasso in terms
of accuracy for low SNR levels, and as the SNR increases, we see that all
four methods converge to nearly perfect accuracy.  The relative test error plot
(top right panel) magnifies the differences between the methods.
For low SNR levels, 
we see that the lasso outperforms the more aggressive best subset and forward
stepwise methods, but for high SNR levels, it is outperformed by the
latter two methods.  The critical transition point---the SNR value at which  
their relative test error curves cross---is different for the low and medium 
settings: for the low setting, it is around 1.22, and for the medium setting,
it is earlier, around 0.42.  The relaxed lasso, meanwhile, is competitive across
all SNR levels: at low SNR levels it matches the performance of the lasso at low
SNR levels, and at high SNR levels it matches that of best subset and forward
stepwise selection. It is able to do so by properly tuning the amount of
shrinkage (via its parameter $\gamma$) on the validation set.  Lastly, the
number of nonzero estimated coefficients from the four methods (bottom right
panel) is also revealing.  The lasso consistently delivers much denser
solutions; essentially, to optimize prediction error on the validation set, it
is forced to do so, as the sparser solutions along its path feature too
much shrinkage.  The relaxed lasso does not suffer from this issue, again thanks
to its ability to unshrink (move $\gamma$ away from 1); it delivers
solutions that are just as sparse as those from best subset and forward stepwise
selection, except at the low SNR range.

The high-5 and high-10 settings, Figures \ref{fig:hi5} and \ref{fig:hi10},
behave quite differently.  The high-5 setting (smaller $n$ and 
smaller $s$) is more dire: the PVEs delivered by all methods---especially best  
subset selection---are {\it negative} for low SNR values, due to poor tuning 
on the validation set (had we chosen the null model for each 
method, the PVE would have been zero).  In both the high-5 and high-10
settings, we see that there is generally no reason, based on relative risk, 
relative test error, or PVE, to favor best subset selection or forward stepwise 
selection over the lasso.  At low SNR levels, best subset and forward stepwise
selection often have 
worse accuracy metrics (and certainly more erratic metrics); at high SNR
levels, these procedures do not show much of an advantage.  For best subset 
selection, it is quite possible that its performance at the high SNR range
would improve if we gave Gurobi a greater budget (than 3 minutes per problem
instance per subset size).  The relaxed lasso again performs the best overall,
with a noticeable gap in performance at the high SNR levels.  As is confirmed by
the number of nonzero  
coefficients plots, the lasso and best subset/forward stepwise selection achieve
similar accuracy in the high SNR range using two opposite strategies:
the former uses high-bias and low-variance estimates, and the latter uses
low-bias and high-variance estimates.  The relaxed lasso 
is most accurate by striking a favorable balance between these two polar
regimes.  

\subsection{Summary of results}

As mentioned above, the results from our entire simulation suite can be found in  
a supplementary document.  Here is a high-level summary.  

\begin{itemize}
\item An important caveat to emphasize upfront is that the Gurobi MIO algorithm
  for best subset selection was given 3 minutes per problem
  instance per subset size.  This practical restriction may have caused
  best subset selection to underperform, in particular, at the high SNR levels
  in the high-5 and high-10 settings. 

\item Forward stepwise selection and best subset selection perform quite
  similarly throughout (with the former being much faster). This does not agree
  with the results for forward stepwise in  \citet{bertsimas2016best}, where
  it performed quite poorly in comparison.  In talking with the third author,
  Rahul Mazumder, we have learned that this was due to the fact that forward
  stepwise in their study was tuned using AIC, rather than a separate
  validation set. So, when put on equal footing and allowed to select its
  tuning  parameter using validation data just as the other methods, we see that
  it performs quite comparably. 

\item The lasso gives better accuracy results than best subset selection in the
  low SNR range and worse accuracy than best subset in the high SNR range.  The
  transition point---the SNR level past which best subset outperforms the
  lasso---varies depending on the problem dimensions ($n,p$) predictor
  autocorrelation ($\rho$), and beta-type (1 through 5).  For the medium 
  setting, the transition point comes earlier than in the low setting.  For the
  high-5 and high-10 settings, the transition point often does not come at all
  (before an SNR of 6, which is the maximum value we considered).  As the
  predictor autocorrelation level increases, the transition point typically
  appears later (again, in some cases it does not come at all, e.g., for beta-type
  5 and autocorrelation $\rho=0.7$).

\item The relaxed lasso provides altogether the top accuracy results.  In 
  nearly all cases (across all SNR levels, and in all problem configurations) we
  considered, it performs as well as or better than all other methods.  We 
  conclude that it is able to use its auxiliary shrinkage parameter ($\gamma$)
  to get the ``best of both worlds'': it accepts the heavy shrinkage from the
  lasso when such shrinkage is helpful, and reverses it when it is not. 

\item The proportion of variance explained plots remind us that, despite what
  may seem like large relative differences, the four methods under consideration
  do not have very different absolute performances in this intuitive
  and important metric.  It thus makes sense overall to favor the methods that
  are easy to compute. 
\end{itemize}

\section{Discussion}

The recent work of \citet{bertsimas2016best}  has enabled the first large-scale
empirical examinations of best subset selection.  In this paper, we have
expanded and refined the simulations in their work, comparing best subset 
selection to forward stepwise selection, the lasso, and the relaxed lasso.  We
have found: (a) forward stepwise selection and best subset selection perform
similarly throughout; (b) best subset selection often loses to the lasso except
in the high SNR range; (c) the relaxed lasso achieves ``the best of both
worlds'' and performs on par with the best method in each scenario. 
We note that these comparisons are based on (various measures of)
out-of-sample prediction accuracy.  A different target, e.g., a measure of
support recovery, may yield different results.

Our R package {\tt bestsubset}, designed to easily replicate all of   
the simulations in this work, or forge new comparisons, is available at 
\url{https://github.com/ryantibs/best-subset/}. 

\bibliographystyle{agsm}
\bibliography{bestsubset}

\newpage
\sectionfont{\fontsize{16}{0}\selectfont}
\subsectionfont{\fontsize{14}{0}\selectfont}
\subsubsectionfont{\fontsize{12}{0}\selectfont}

\appendix
\begin{center}
\LARGE
Supplement to ``Extended Comparisons of Best Subset Selection, 
  Forward Stepwise Selection, and the Lasso''
\end{center}

\bigskip
This supplementary document contains plots from the simulation suite
described in the paper ``Extended Comparisons of Best Subset Selection,  Forward 
Stepwise Selection, and the Lasso''. 
The plots in Section 1 precisely follow the simulation format described in the
paper. Those in Section 2 follow an analogous format, except that the
tuning has been done using an ``oracle'', rather than a validation set as in
Section 1. Specifically, the tuning parameter for each method in each scenario
is chosen to minimize the average risk over all of the repetitions. 

\newpage
\etocdepthtag.toc{app}
\etocsettagdepth{main}{none}
\etocsettagdepth{app}{subsubsection}
\tableofcontents‎‎
\newpage

\section{Validation tuning}\vspace{-3pt}
\subsection{Low setting: $n=100$, $p=10$, $s=5$} 
\subsubsection{Relative risk (to null model)} 
\begin{figure}[!h]
\centering
\includegraphics[width=0.99\textwidth]{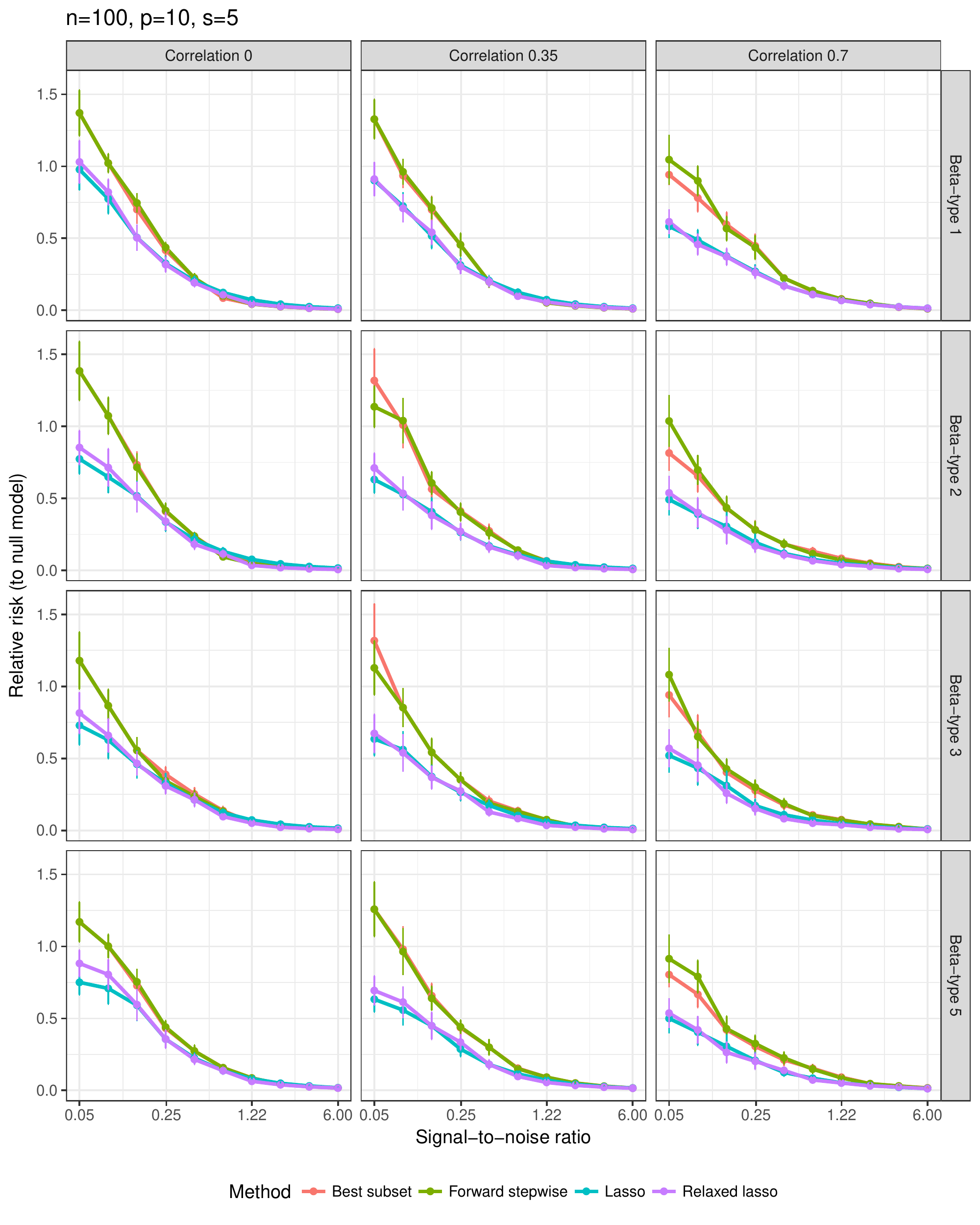}
\end{figure}
\newpage

\subsubsection{Relative test error (to Bayes)}
\begin{figure}[!h]
\centering
\includegraphics[width=0.99\textwidth]{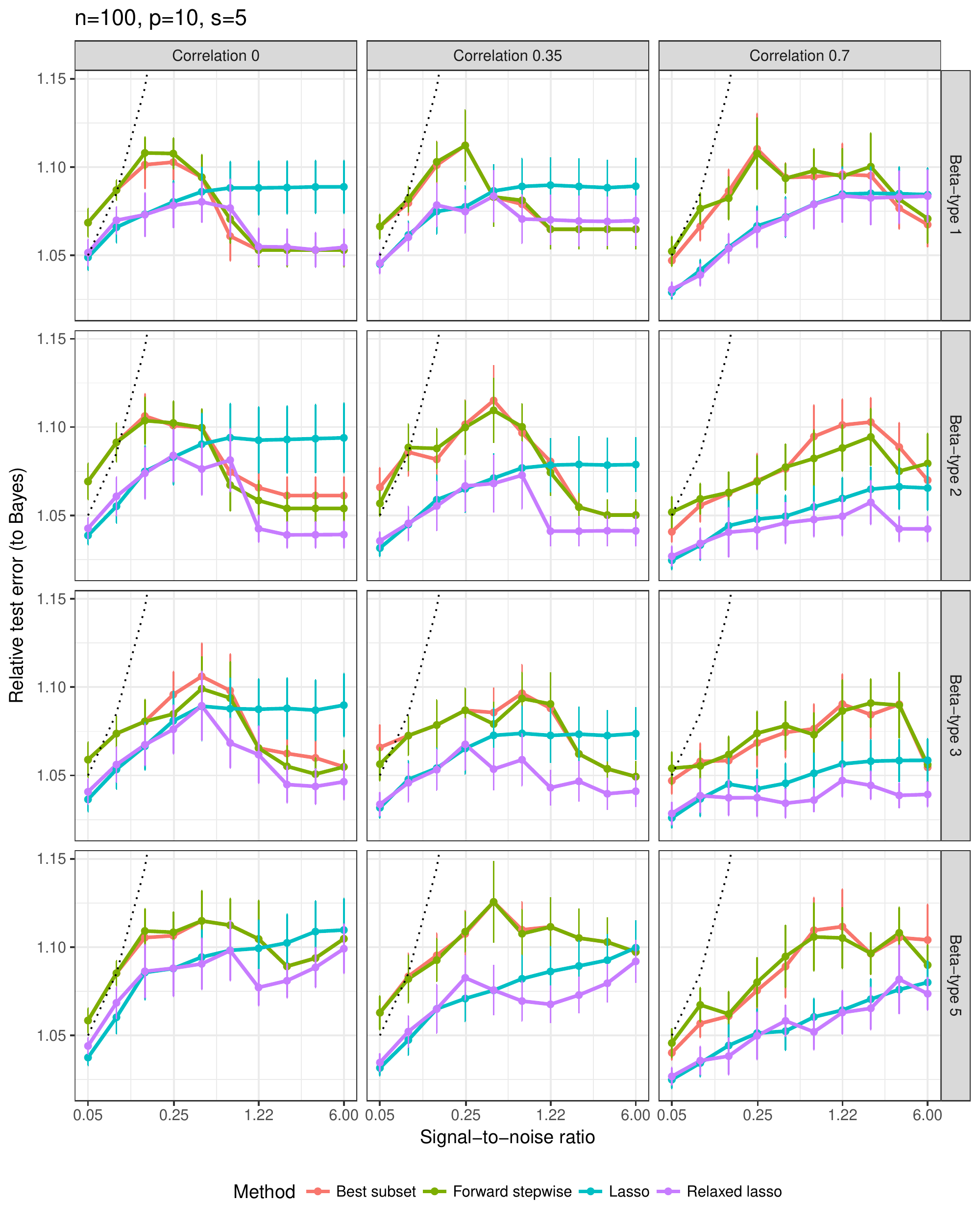}
\end{figure}
\newpage

\subsubsection{Proportion of variance explained}
\begin{figure}[!h]
\centering
\includegraphics[width=0.99\textwidth]{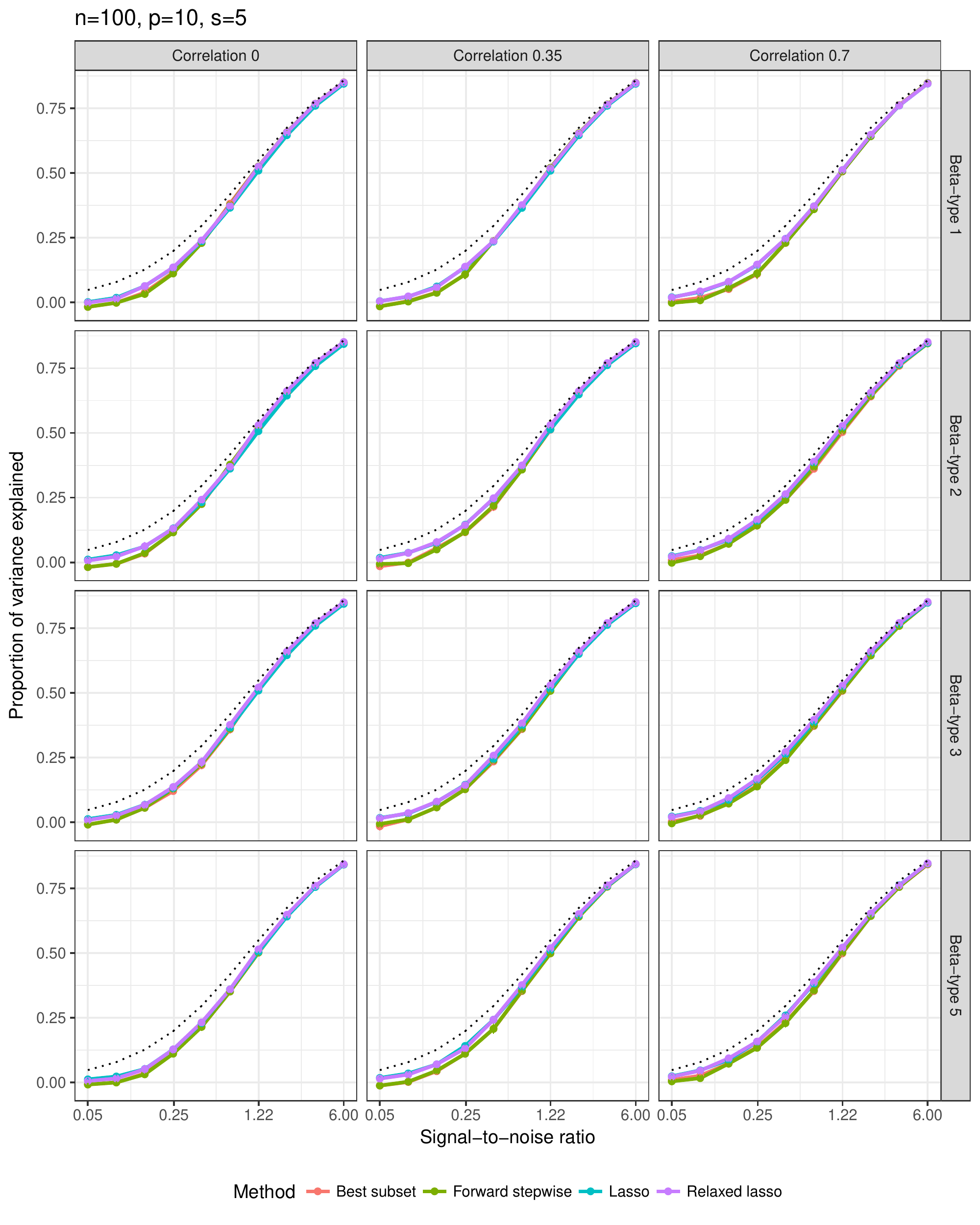}
\end{figure}
\newpage

\subsubsection{Number of nonzero coefficients}
\begin{figure}[!h]
\centering
\includegraphics[width=0.99\textwidth]{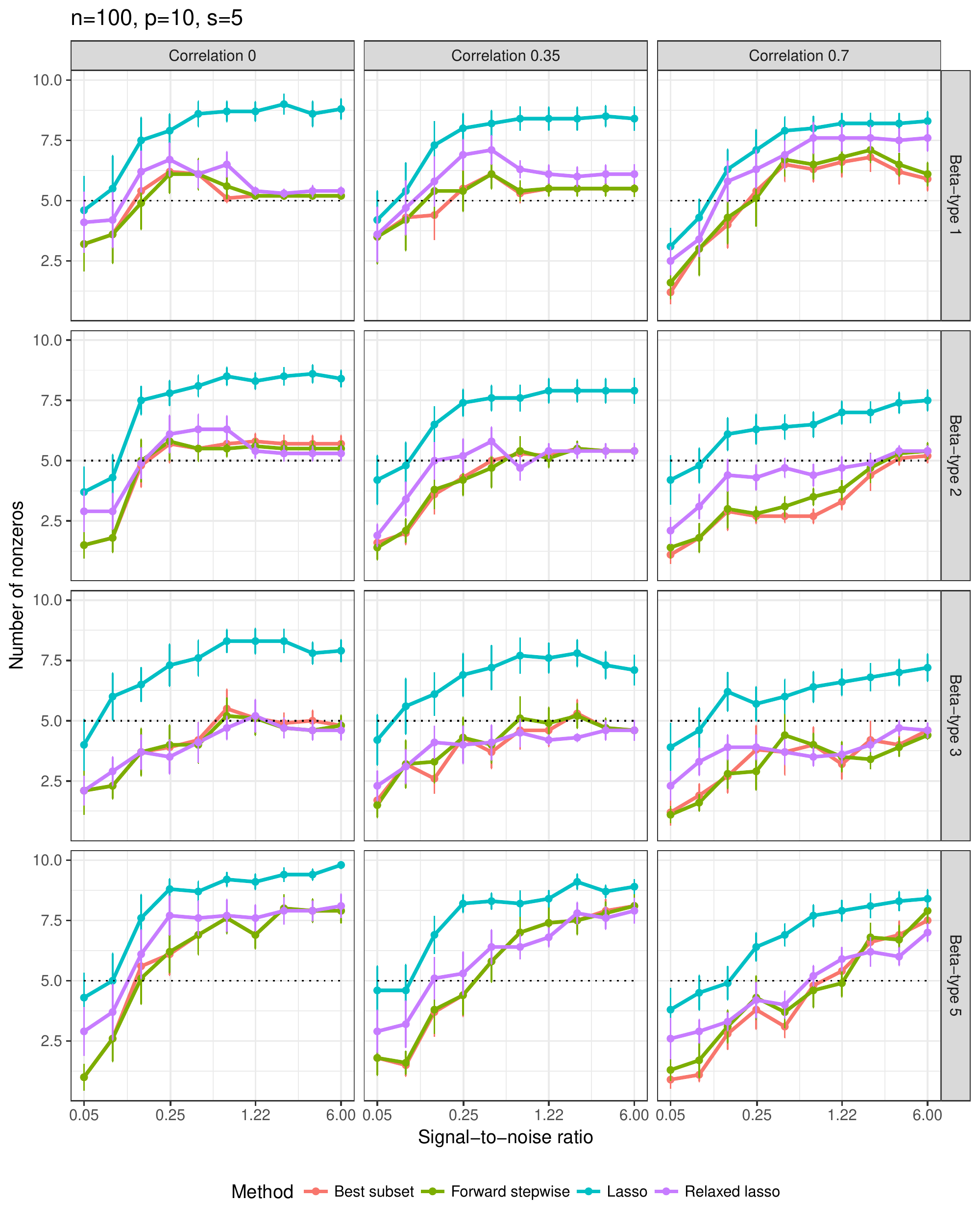}
\end{figure}
\newpage

\subsection{Medium setting: $n=500$, $p=100$, $s=5$} 
\subsubsection{Relative risk (to null model)} 
\begin{figure}[!h]
\centering
\includegraphics[width=0.99\textwidth]{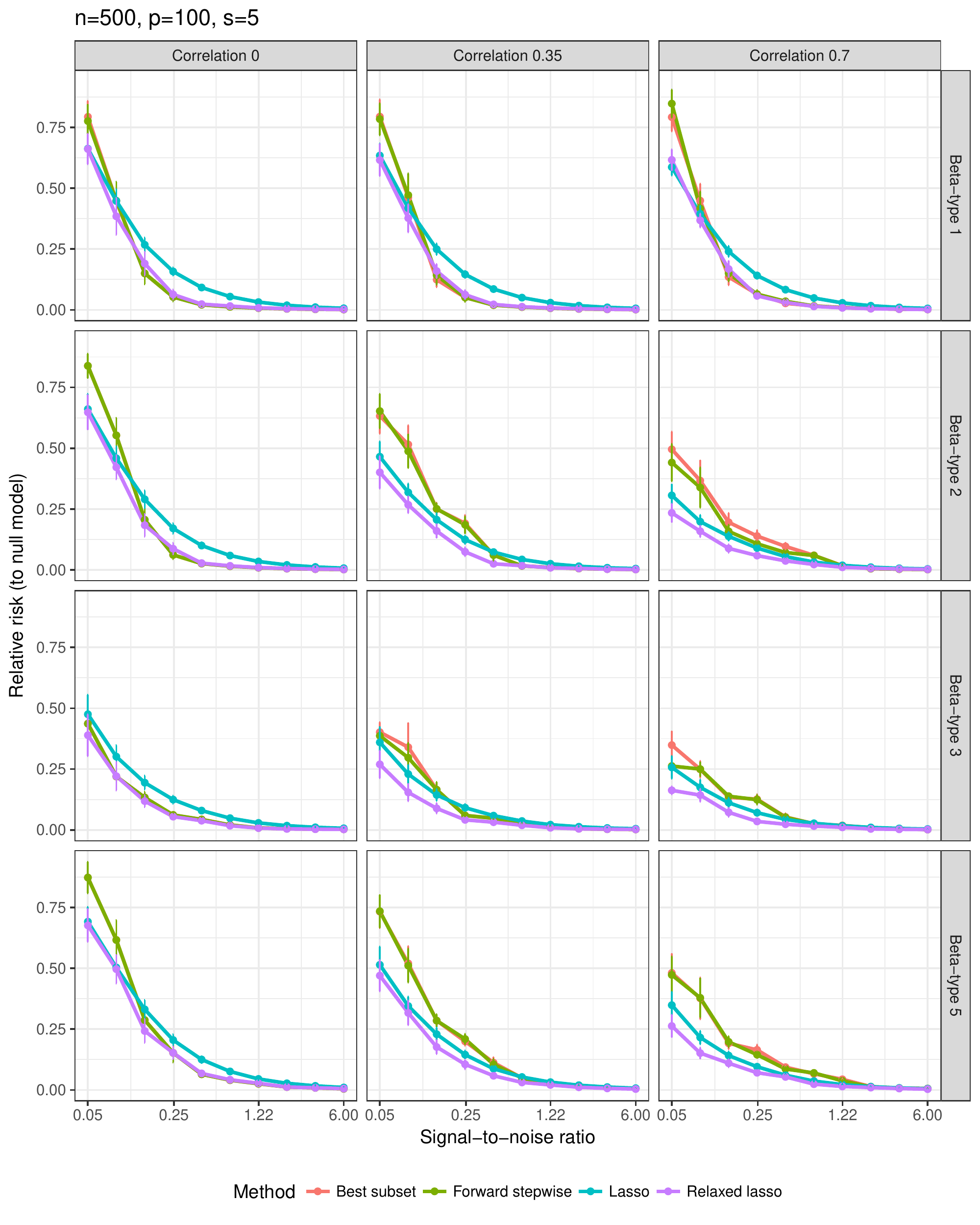}
\end{figure}
\newpage

\subsubsection{Relative test error (to Bayes)}
\begin{figure}[!h]
\centering
\includegraphics[width=0.99\textwidth]{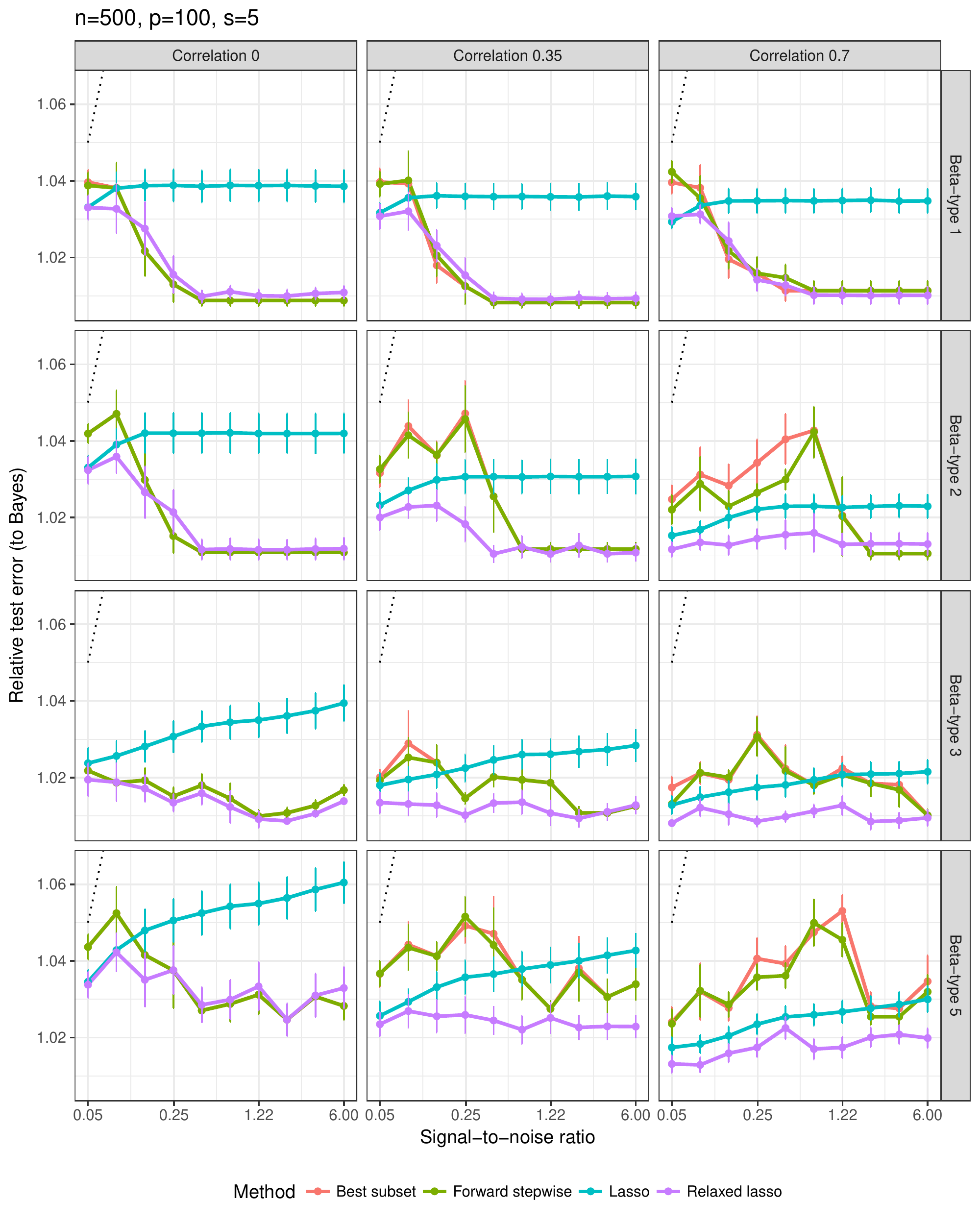}
\end{figure}
\newpage

\subsubsection{Proportion of variance explained}
\begin{figure}[!h]
\centering
\includegraphics[width=0.99\textwidth]{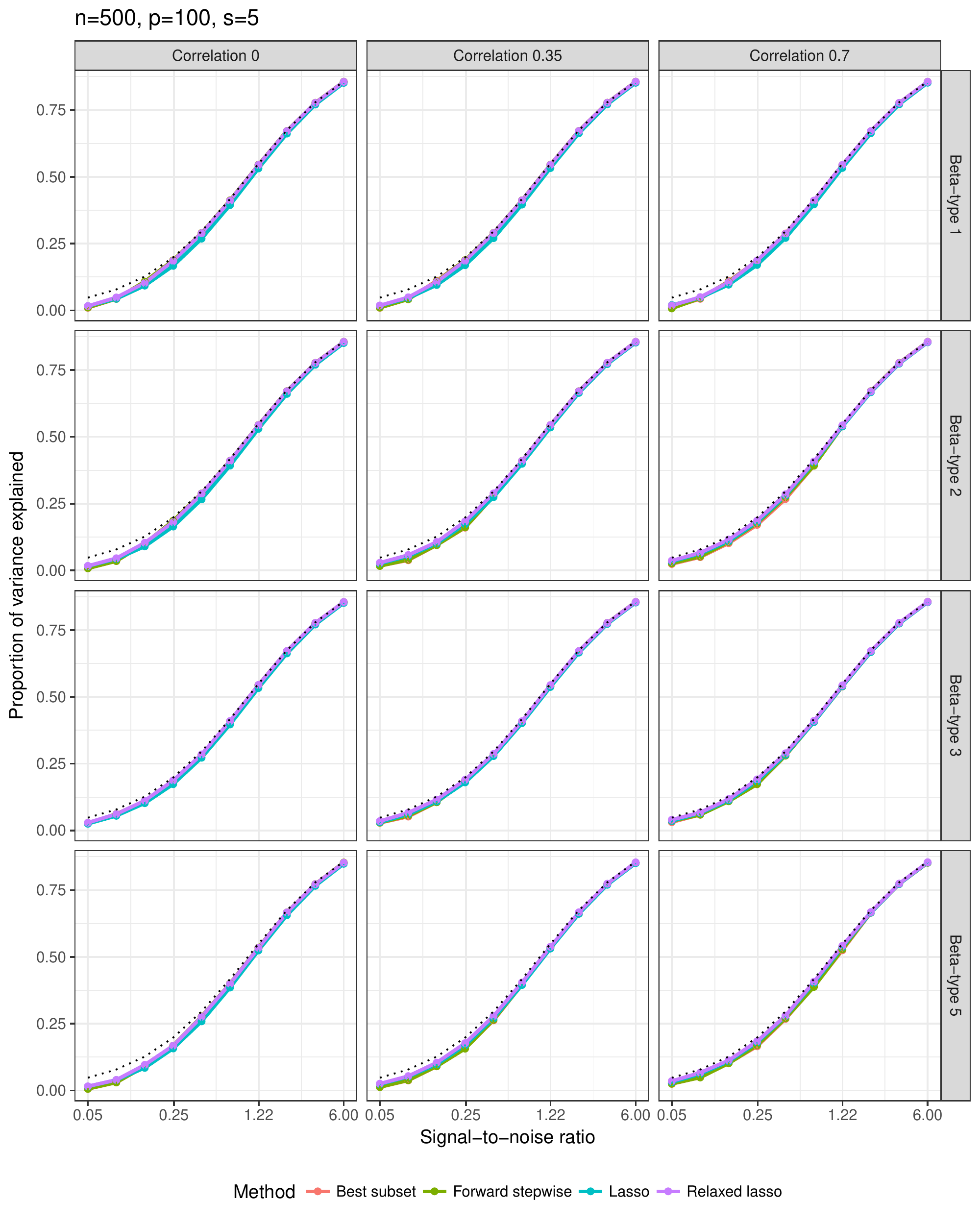}
\end{figure}
\newpage

\subsubsection{Number of nonzero coefficients}
\begin{figure}[!h]
\centering
\includegraphics[width=0.99\textwidth]{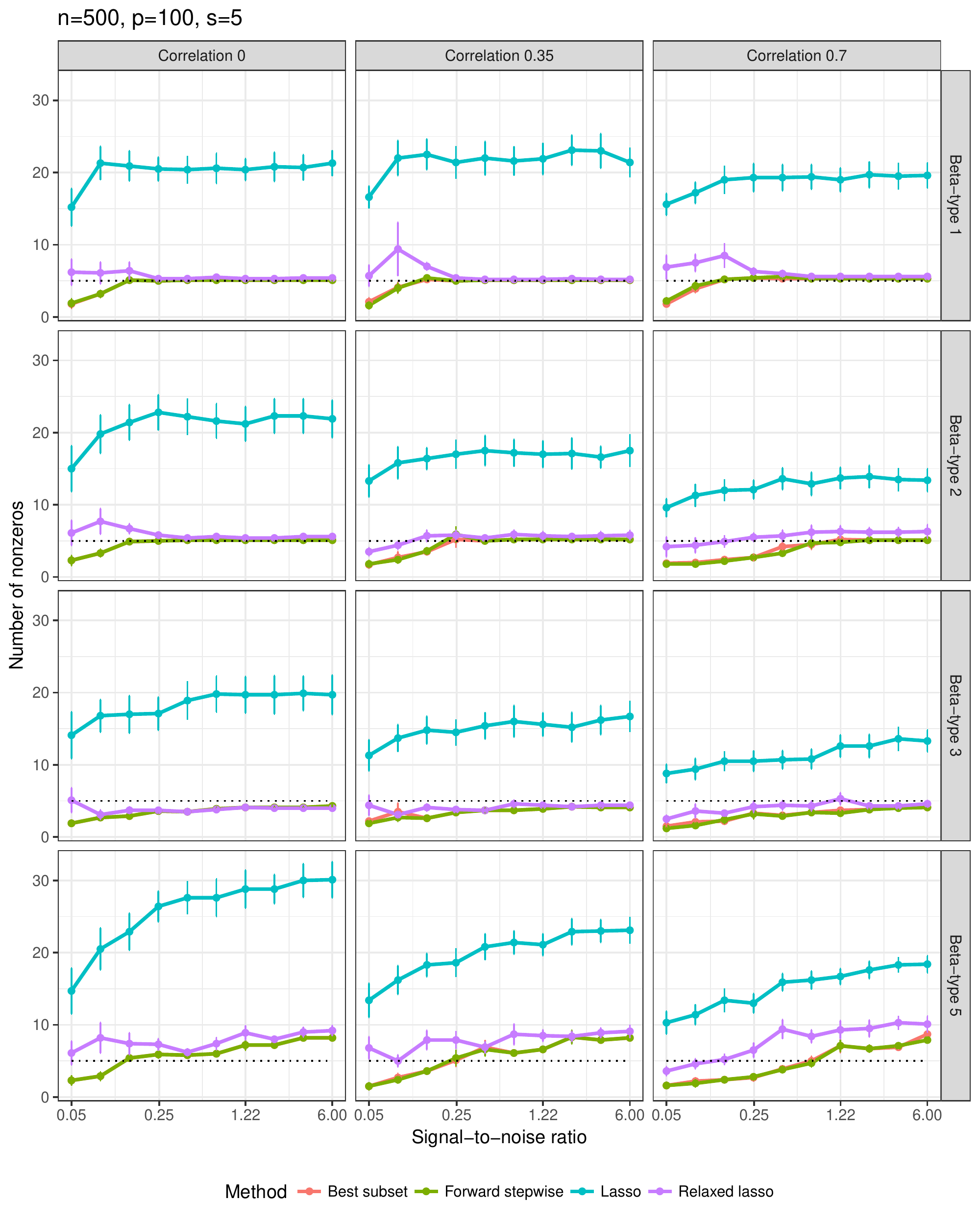}
\end{figure}
\newpage

\subsection{High-5 setting: $n=50$, $p=1000$, $s=5$} 
\subsubsection{Relative risk (to null model)} 
\begin{figure}[!h]
\centering
\includegraphics[width=0.99\textwidth]{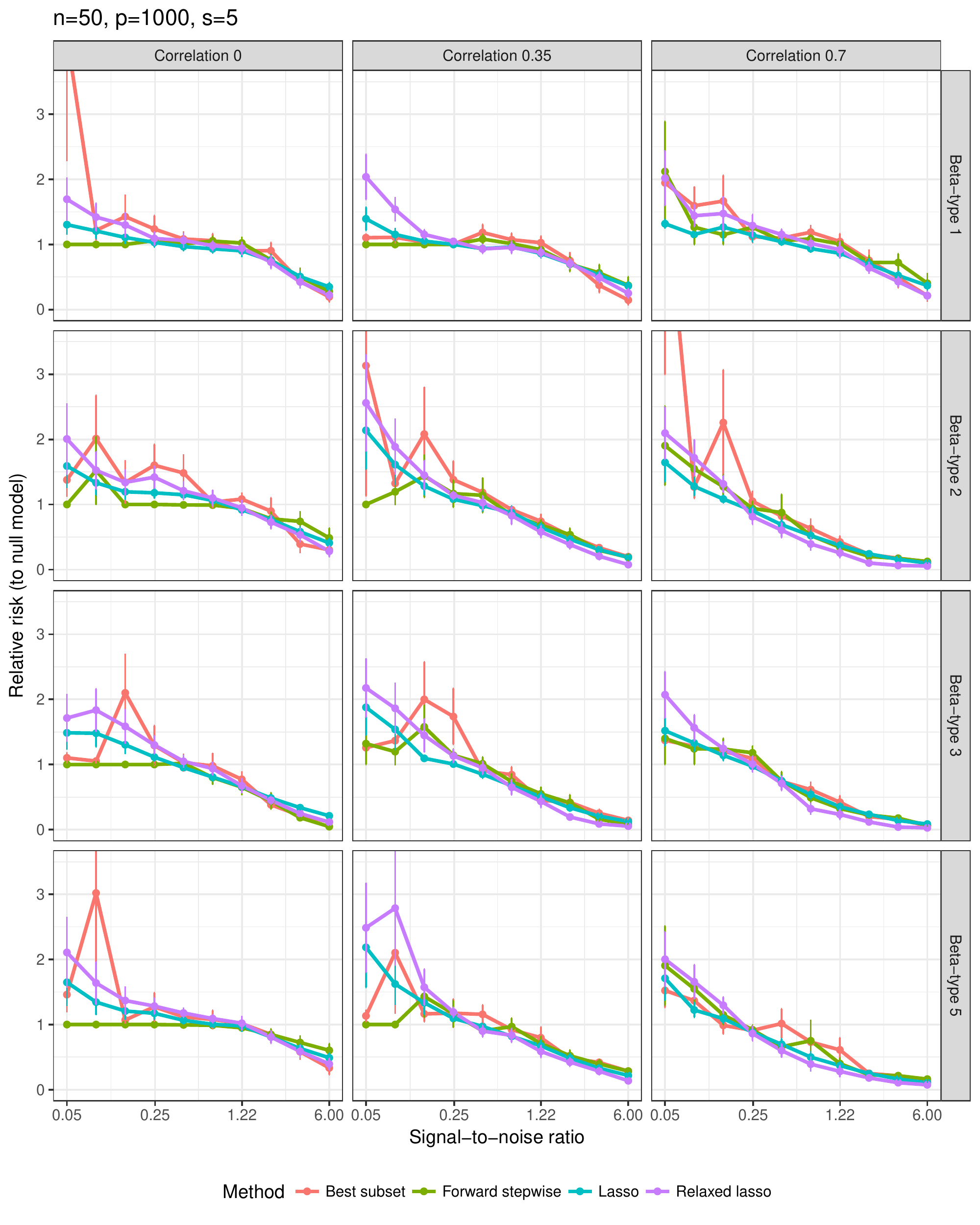}
\end{figure}
\newpage

\subsubsection{Relative test error (to Bayes)}
\begin{figure}[!h]
\centering
\includegraphics[width=0.99\textwidth]{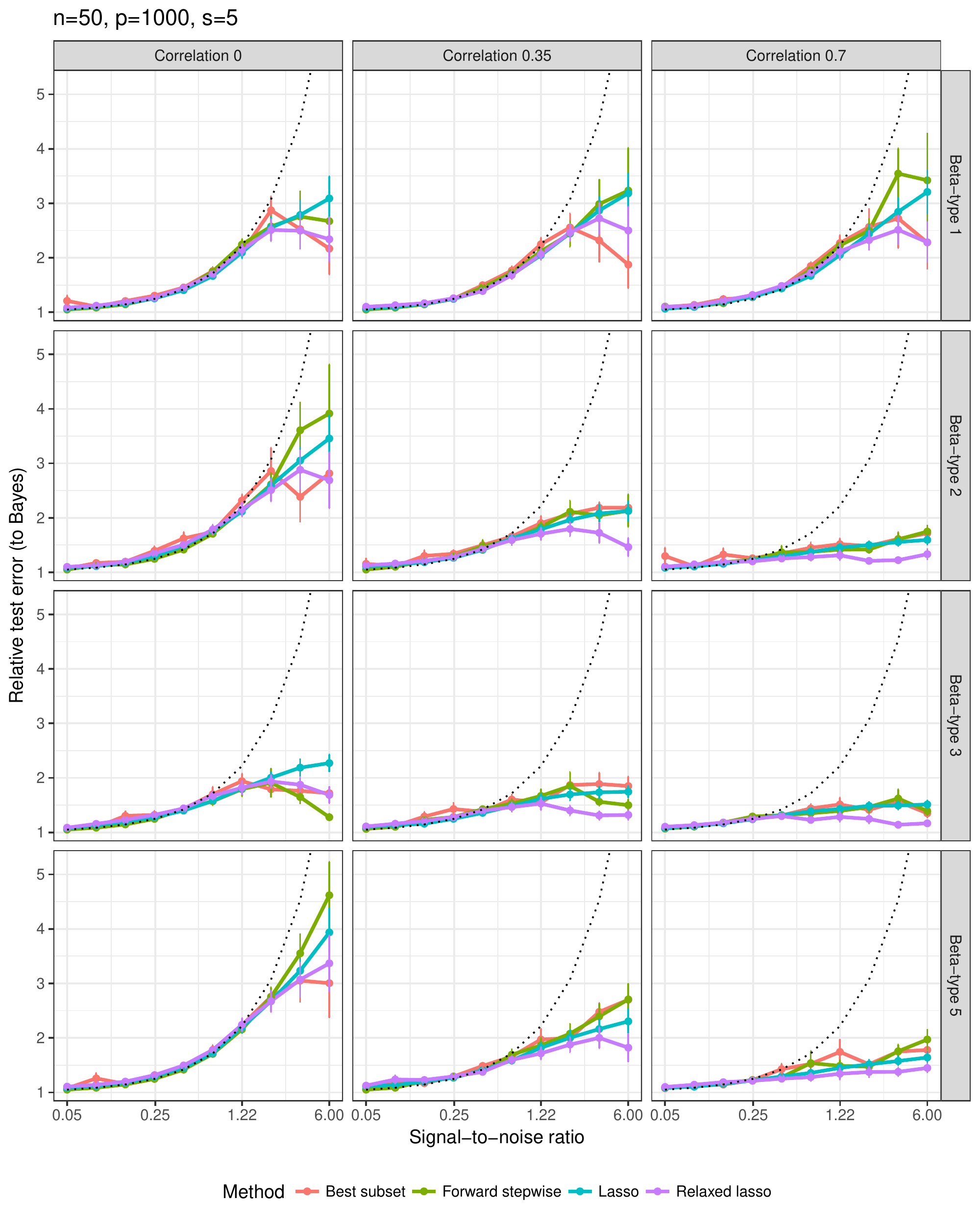}
\end{figure}
\newpage

\subsubsection{Proportion of variance explained}
\begin{figure}[!h]
\centering
\includegraphics[width=0.99\textwidth]{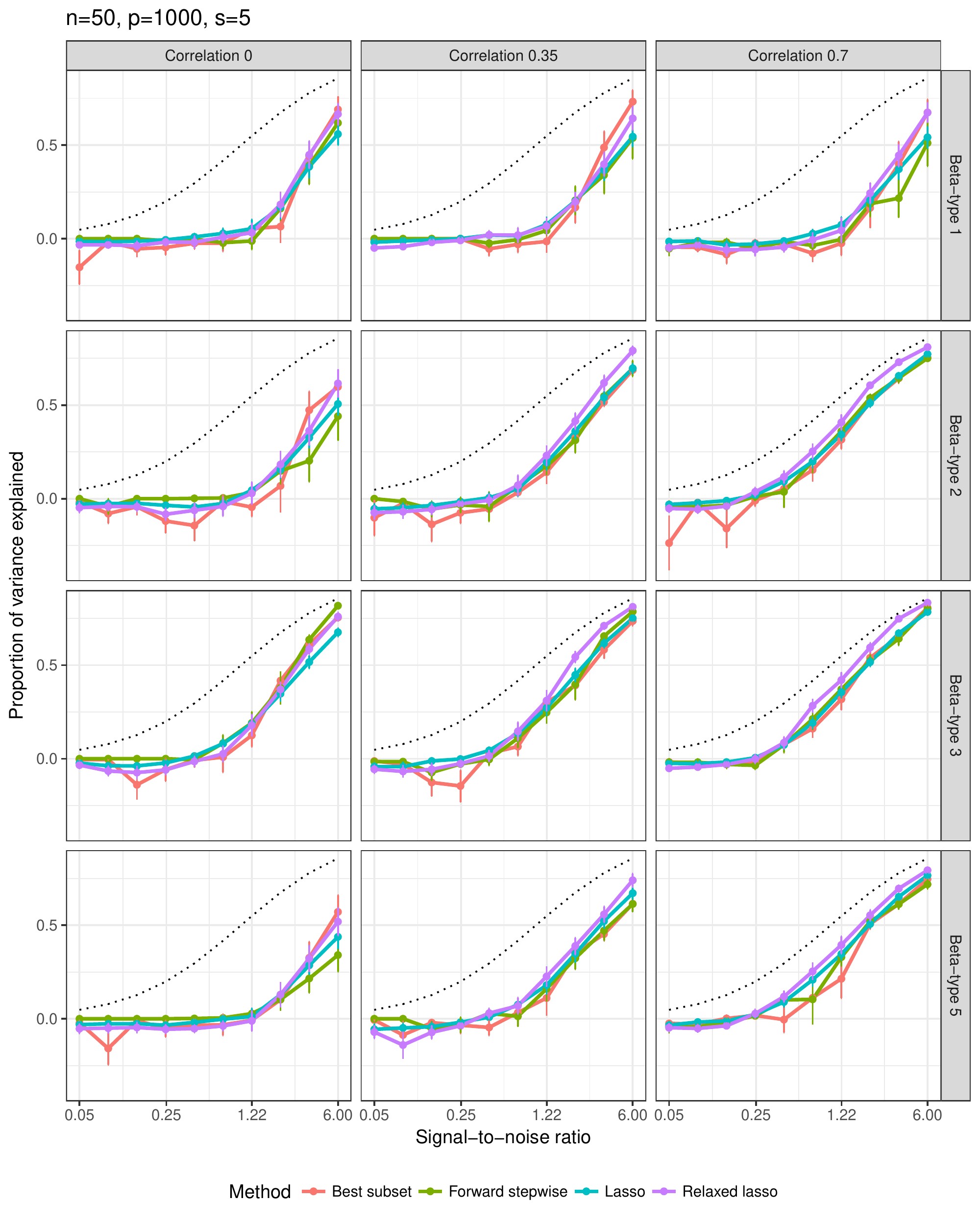}
\end{figure}
\newpage

\subsubsection{Number of nonzero coefficients}
\begin{figure}[!h]
\centering
\includegraphics[width=0.99\textwidth]{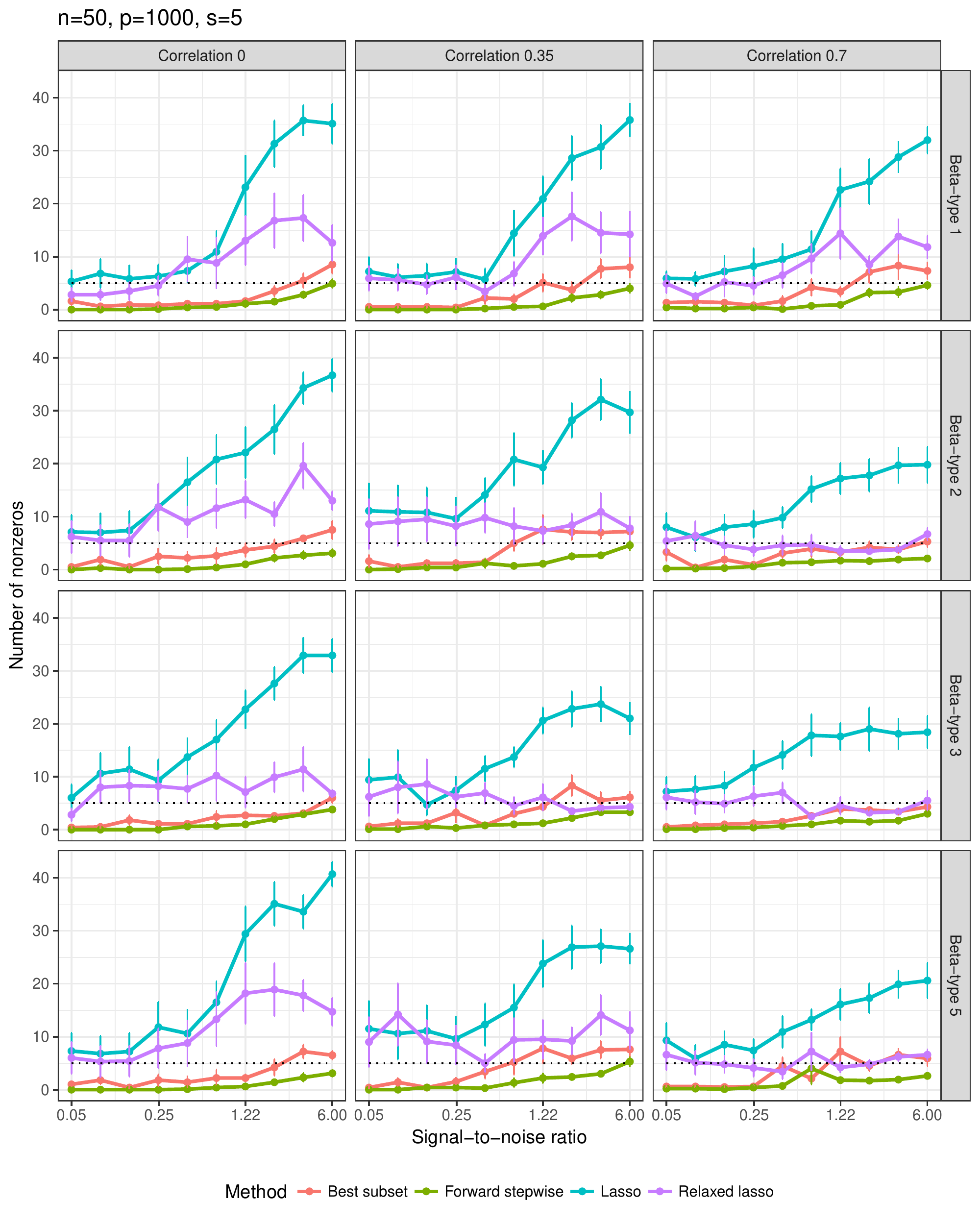}
\end{figure}
\newpage

\subsection{High-10 setting: $n=100$, $p=1000$, $s=10$} 
\subsubsection{Relative risk (to null model)} 
\begin{figure}[!h]
\centering
\includegraphics[width=0.99\textwidth]{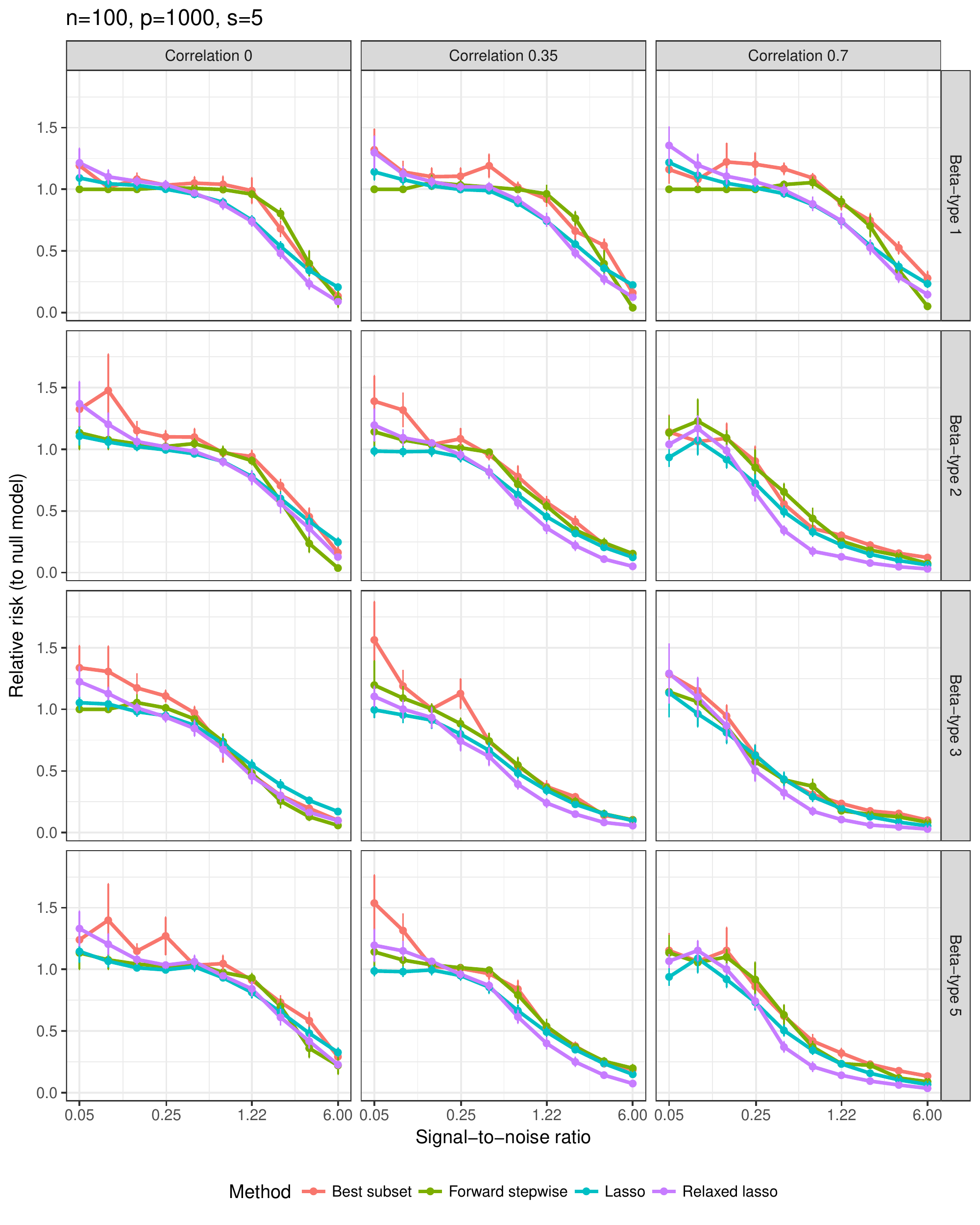}
\end{figure}
\newpage

\subsubsection{Relative test error (to Bayes)}
\begin{figure}[!h]
\centering
\includegraphics[width=0.99\textwidth]{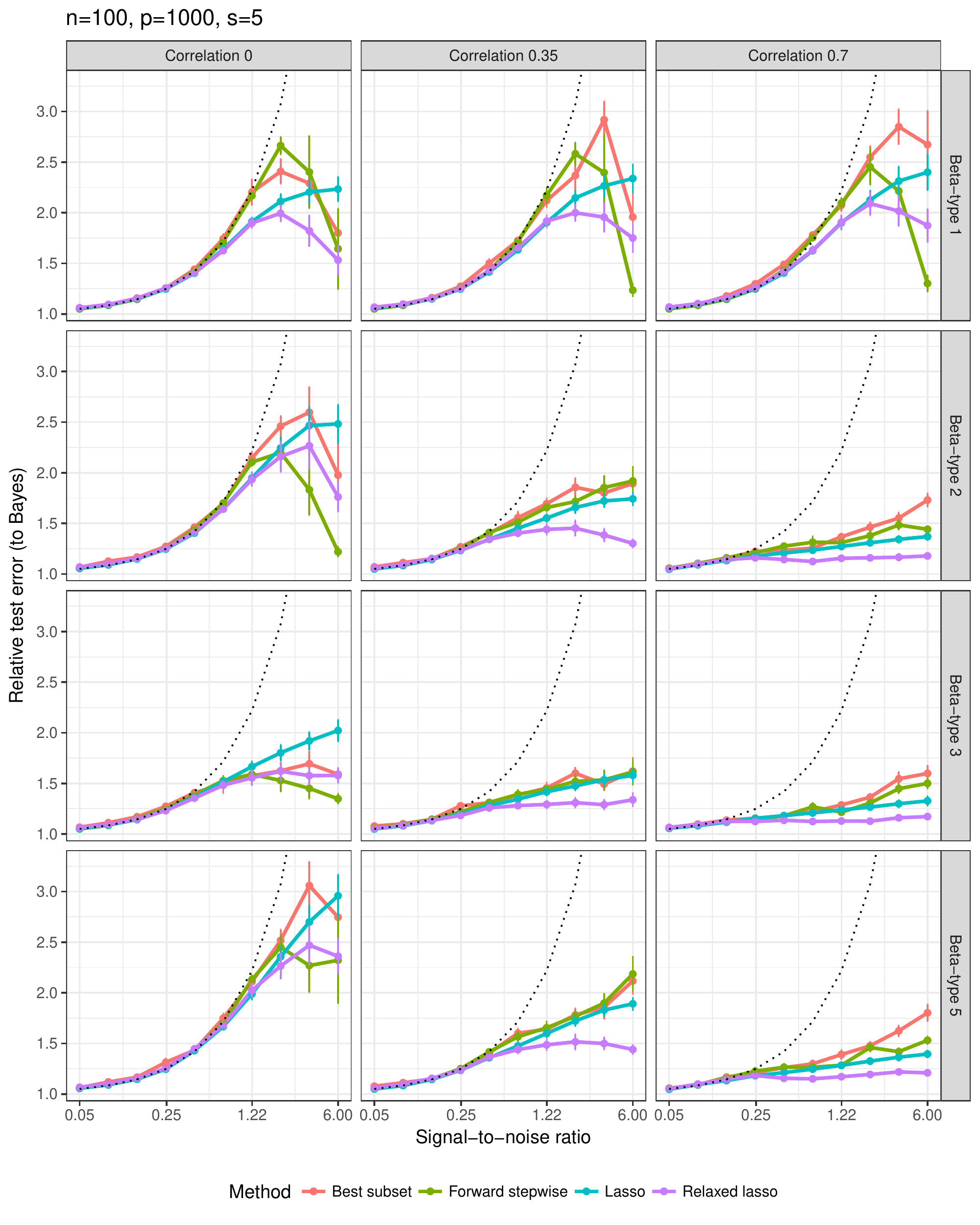}
\end{figure}
\newpage

\subsubsection{Proportion of variance explained}
\begin{figure}[!h]
\centering
\includegraphics[width=0.99\textwidth]{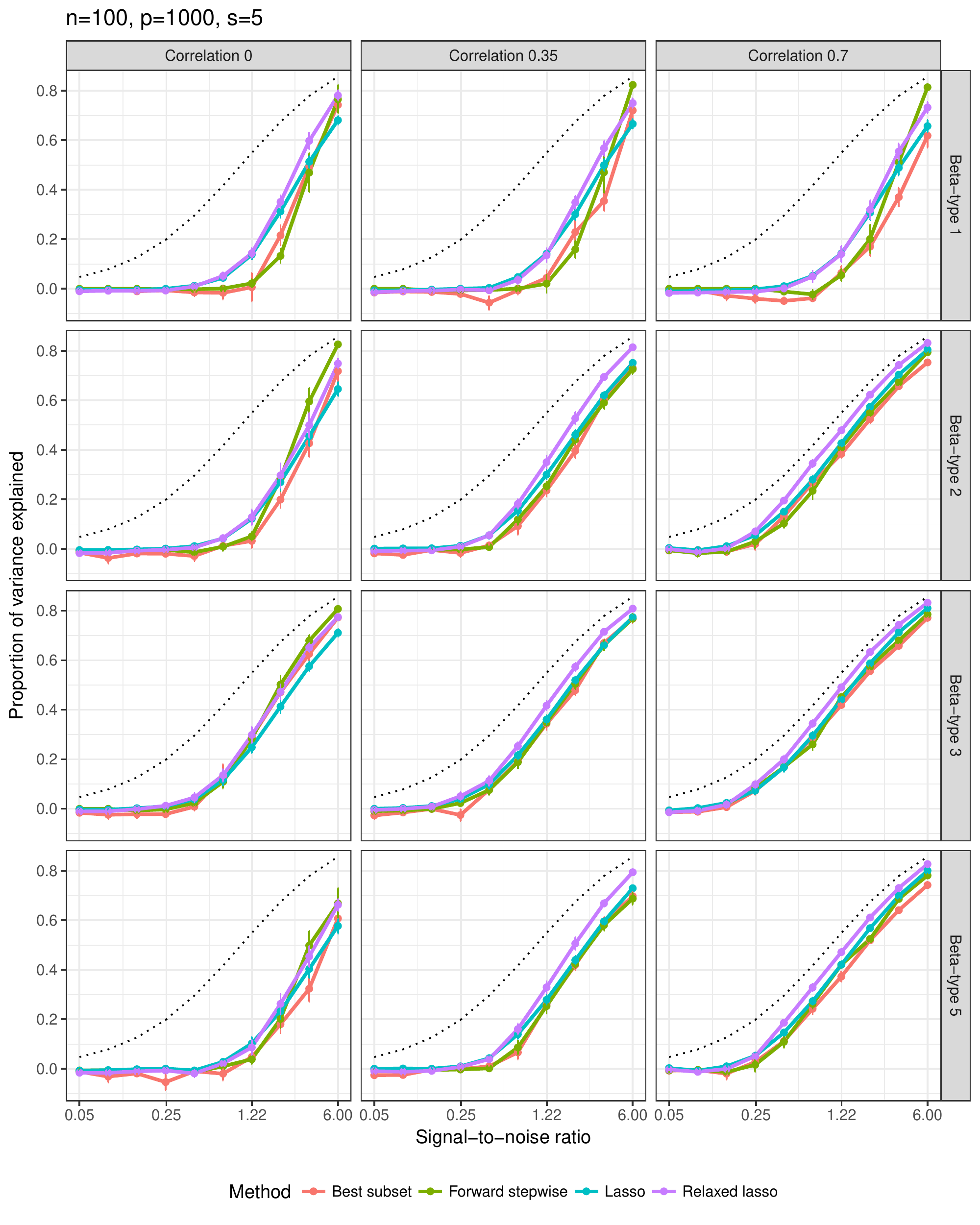}
\end{figure}
\newpage

\subsubsection{Number of nonzero coefficients}
\begin{figure}[!h]
\centering
\includegraphics[width=0.99\textwidth]{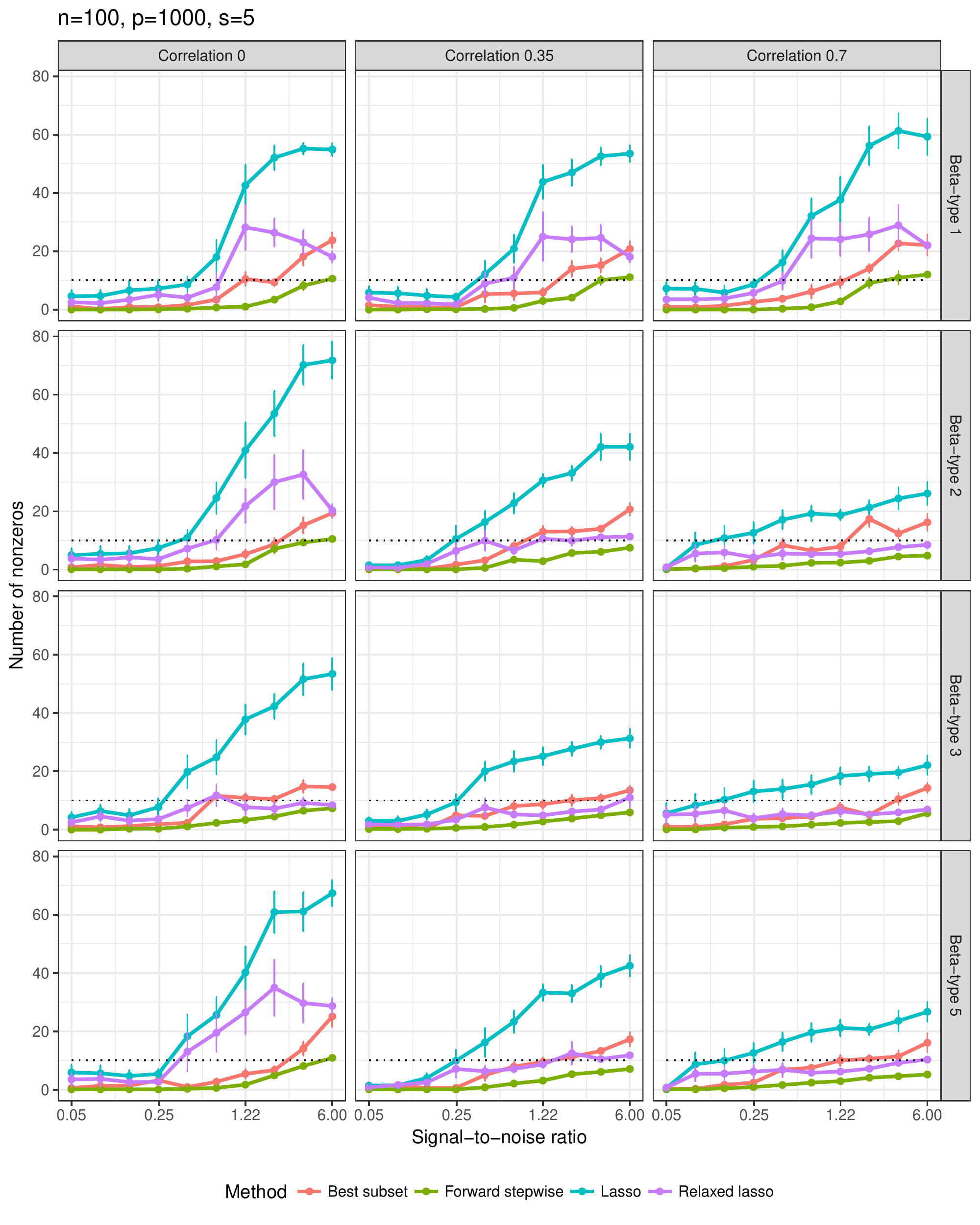}
\end{figure}
\newpage

\section{Oracle tuning}\vspace{-3pt}
\subsection{Low setting: $n=100$, $p=10$, $s=5$} 
\subsubsection{Relative risk (to null model)} 
\begin{figure}[!h]
\centering
\includegraphics[width=0.99\textwidth]{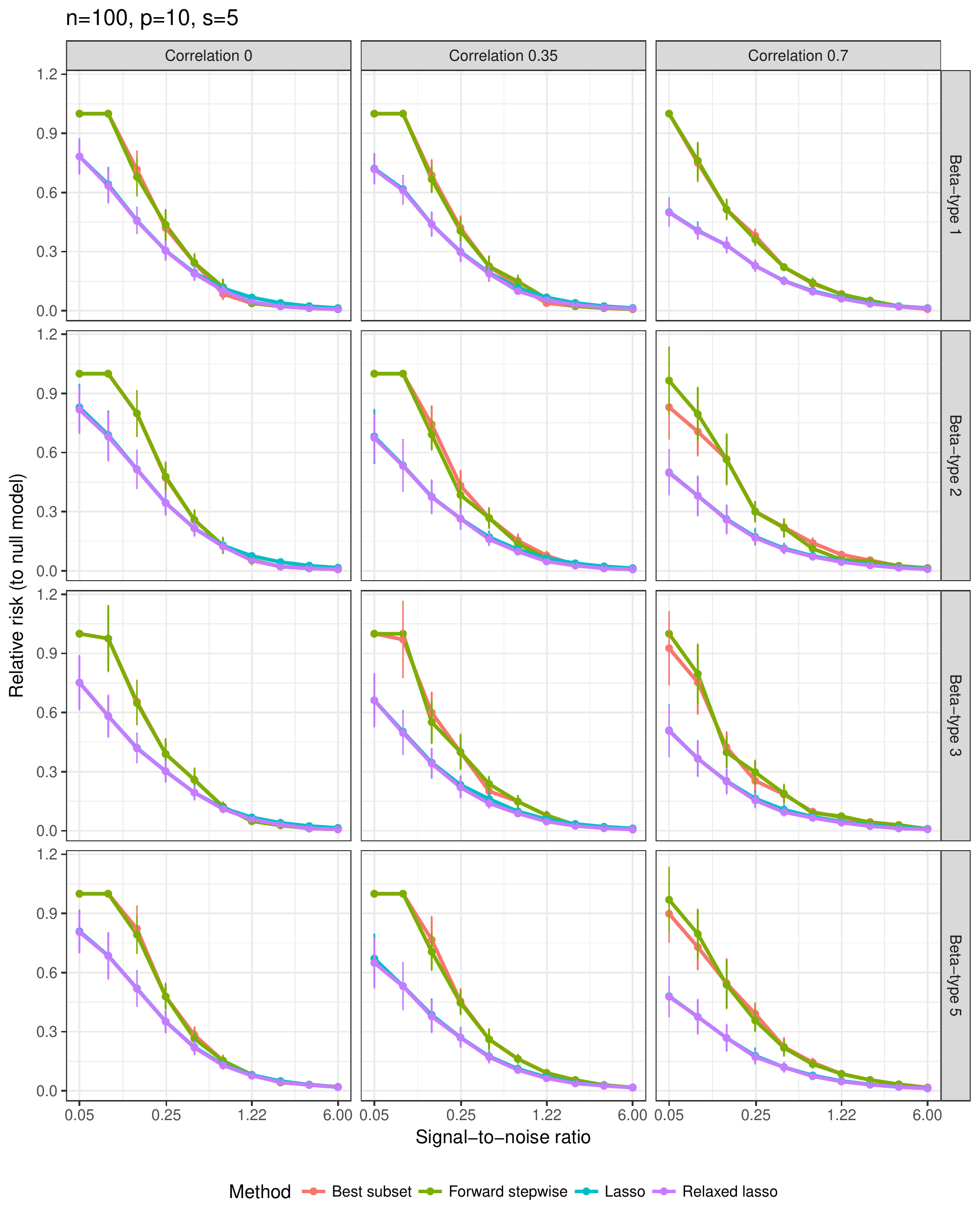}
\end{figure}
\newpage

\subsubsection{Relative test error (to Bayes)}
\begin{figure}[!h]
\centering
\includegraphics[width=0.99\textwidth]{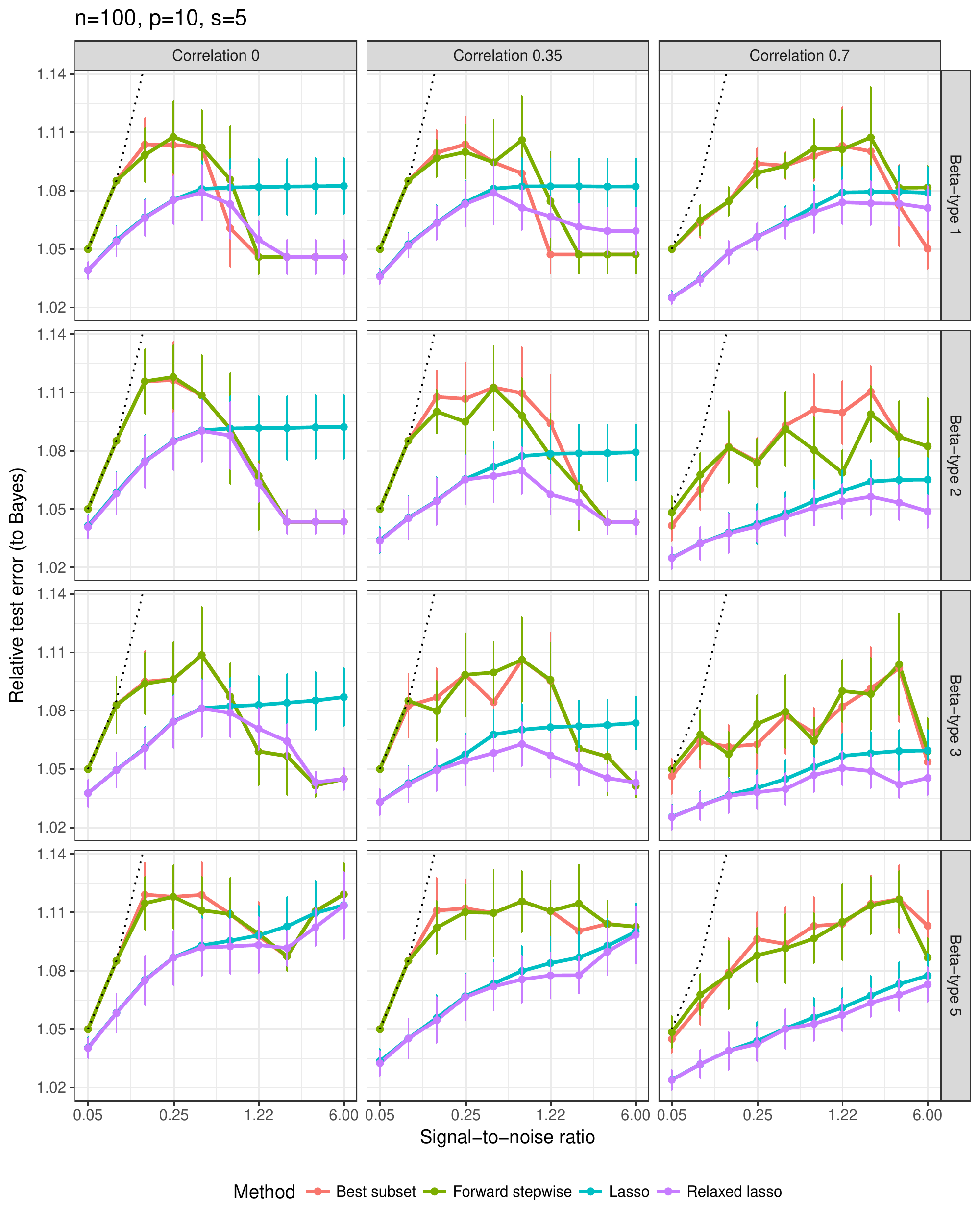}
\end{figure}
\newpage

\subsubsection{Proportion of variance explained}
\begin{figure}[!h]
\centering
\includegraphics[width=0.99\textwidth]{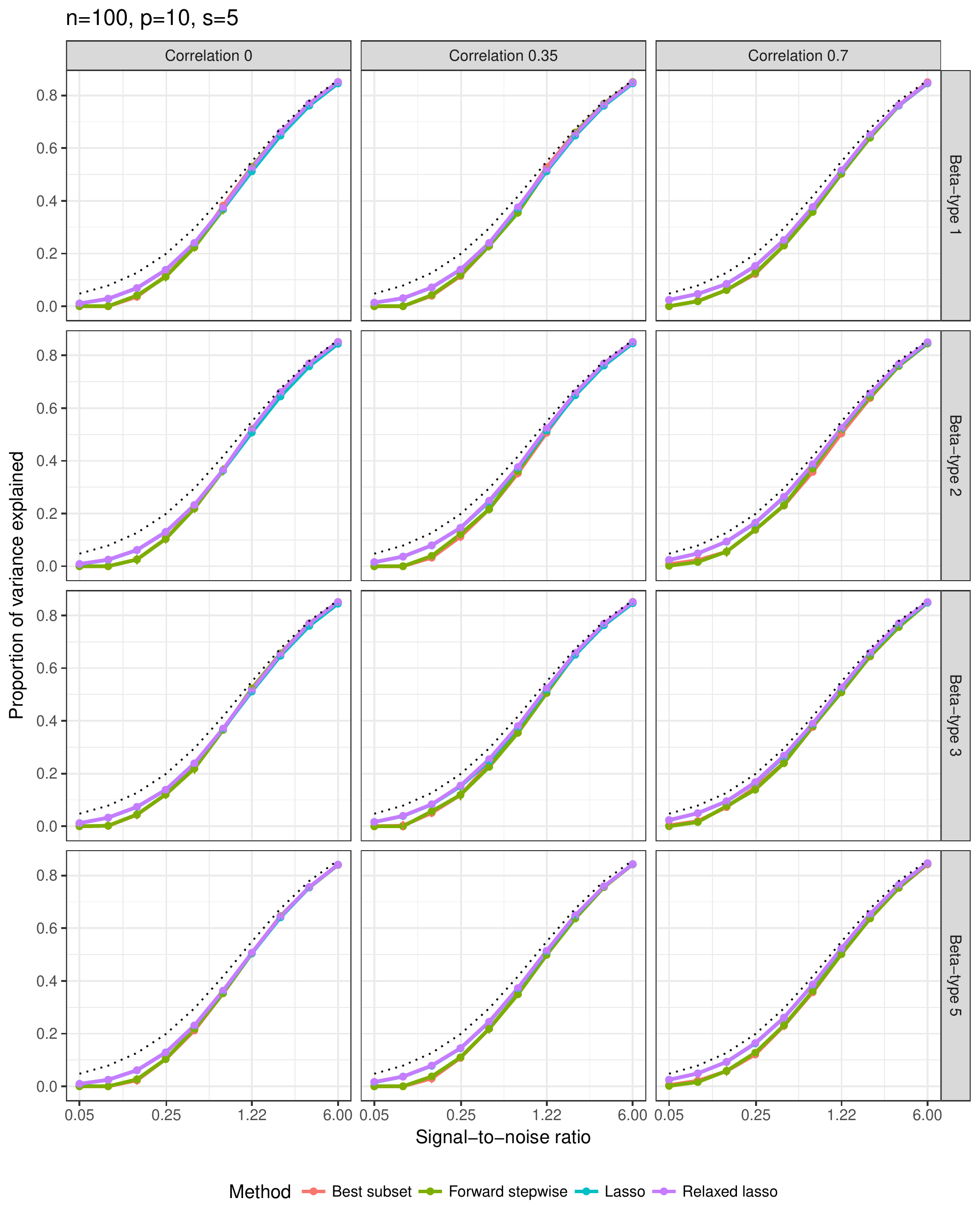}
\end{figure}
\newpage

\subsubsection{Number of nonzero coefficients}
\begin{figure}[!h]
\centering
\includegraphics[width=0.99\textwidth]{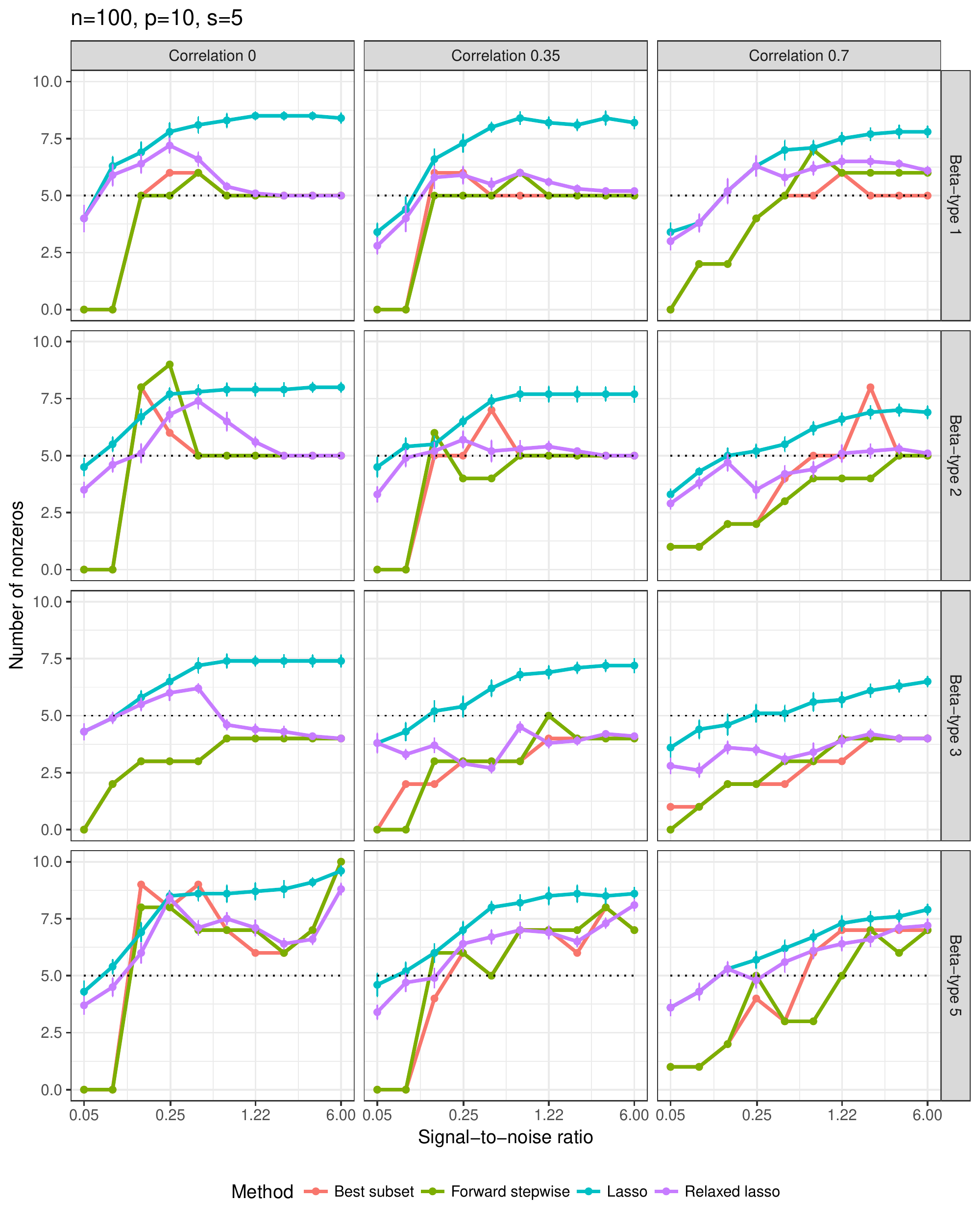}
\end{figure}
\newpage

\subsection{Medium setting: $n=500$, $p=100$, $s=5$} 
\subsubsection{Relative risk (to null model)} 
\begin{figure}[!h]
\centering
\includegraphics[width=0.99\textwidth]{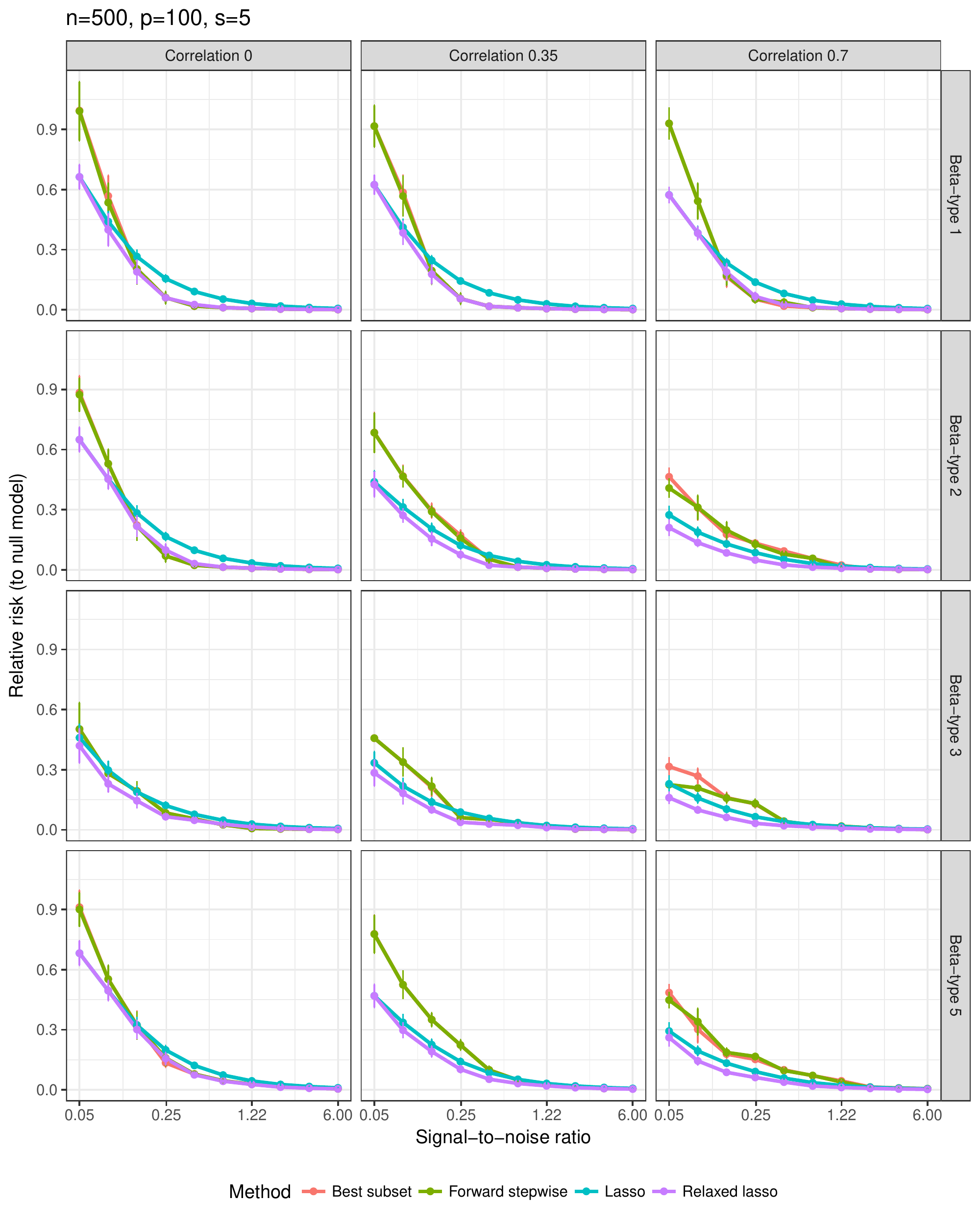}
\end{figure}
\newpage

\subsubsection{Relative test error (to Bayes)}
\begin{figure}[!h]
\centering
\includegraphics[width=0.99\textwidth]{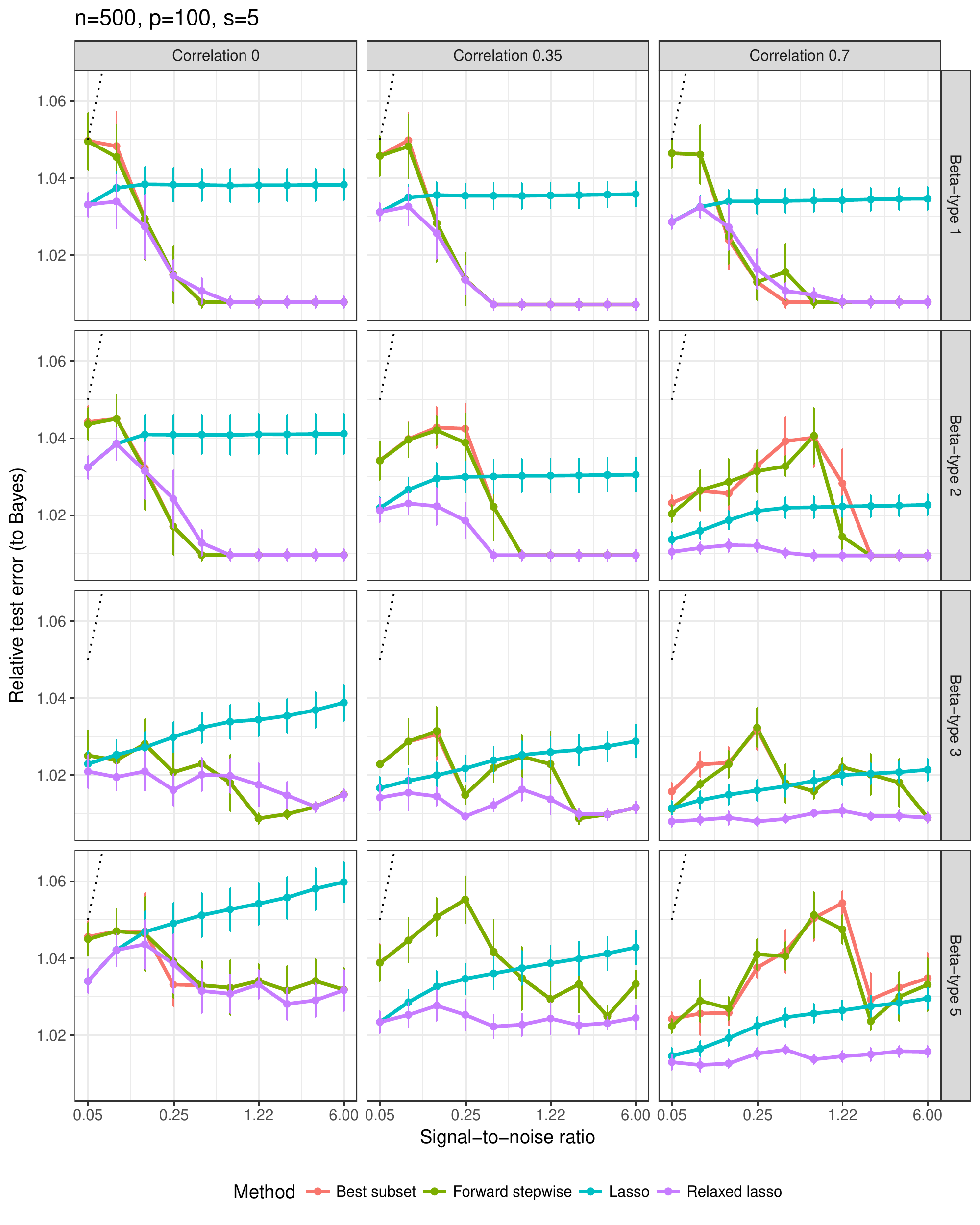}
\end{figure}
\newpage

\subsubsection{Proportion of variance explained}
\begin{figure}[!h]
\centering
\includegraphics[width=0.99\textwidth]{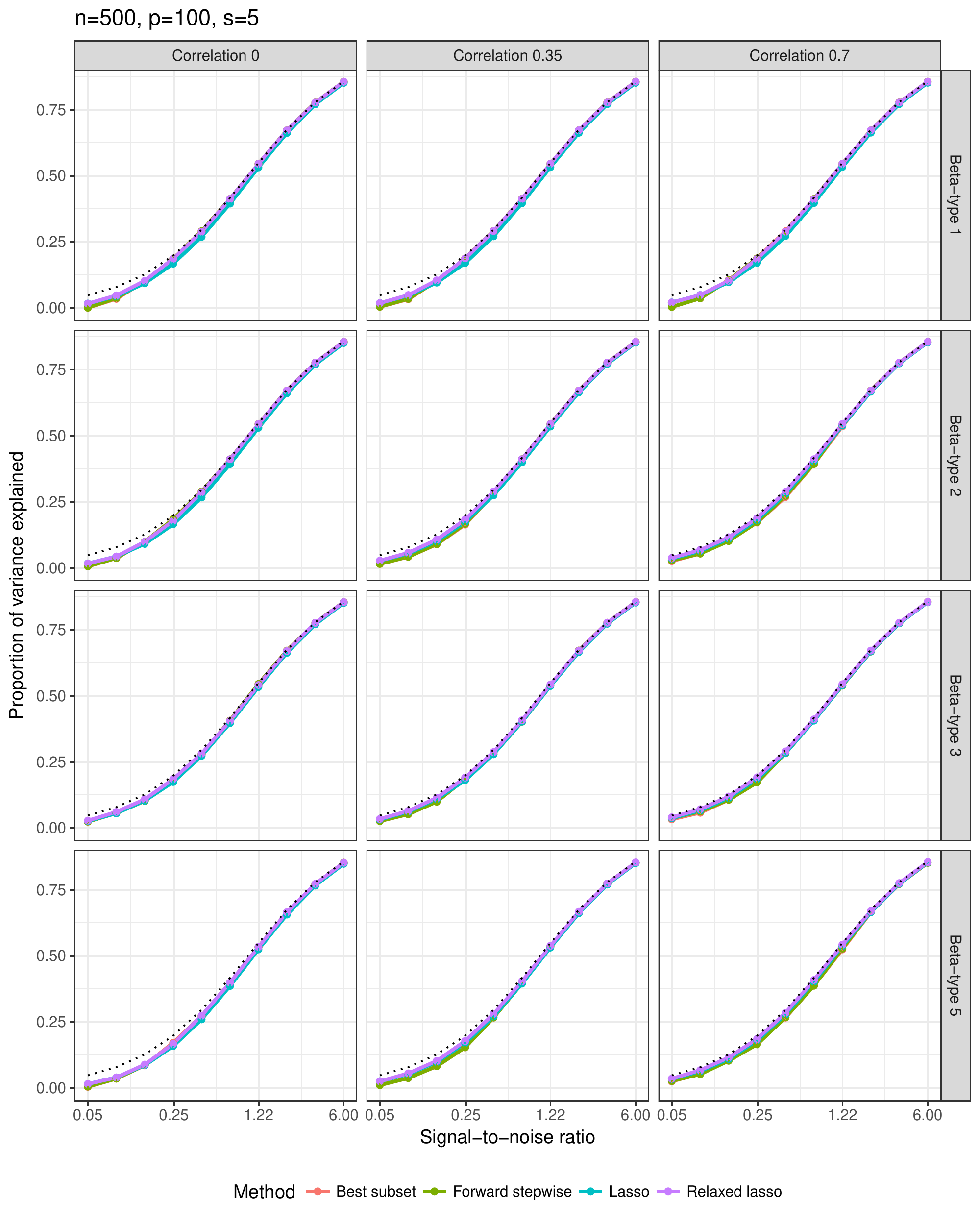}
\end{figure}
\newpage

\subsubsection{Number of nonzero coefficients}
\begin{figure}[!h]
\centering
\includegraphics[width=0.99\textwidth]{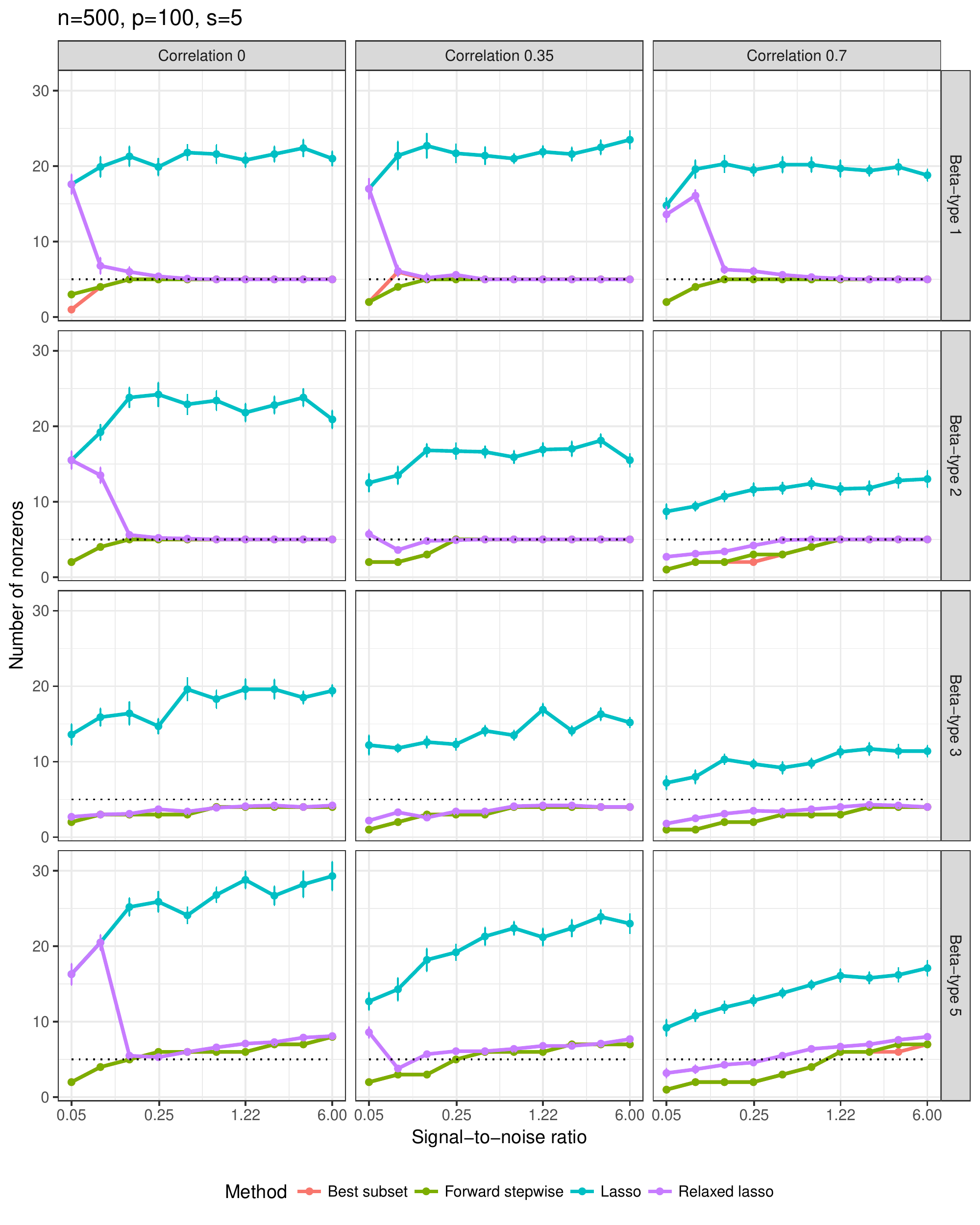}
\end{figure}
\newpage

\subsection{High-5 setting: $n=50$, $p=1000$, $s=5$} 
\subsubsection{Relative risk (to null model)} 
\begin{figure}[!h]
\centering
\includegraphics[width=0.99\textwidth]{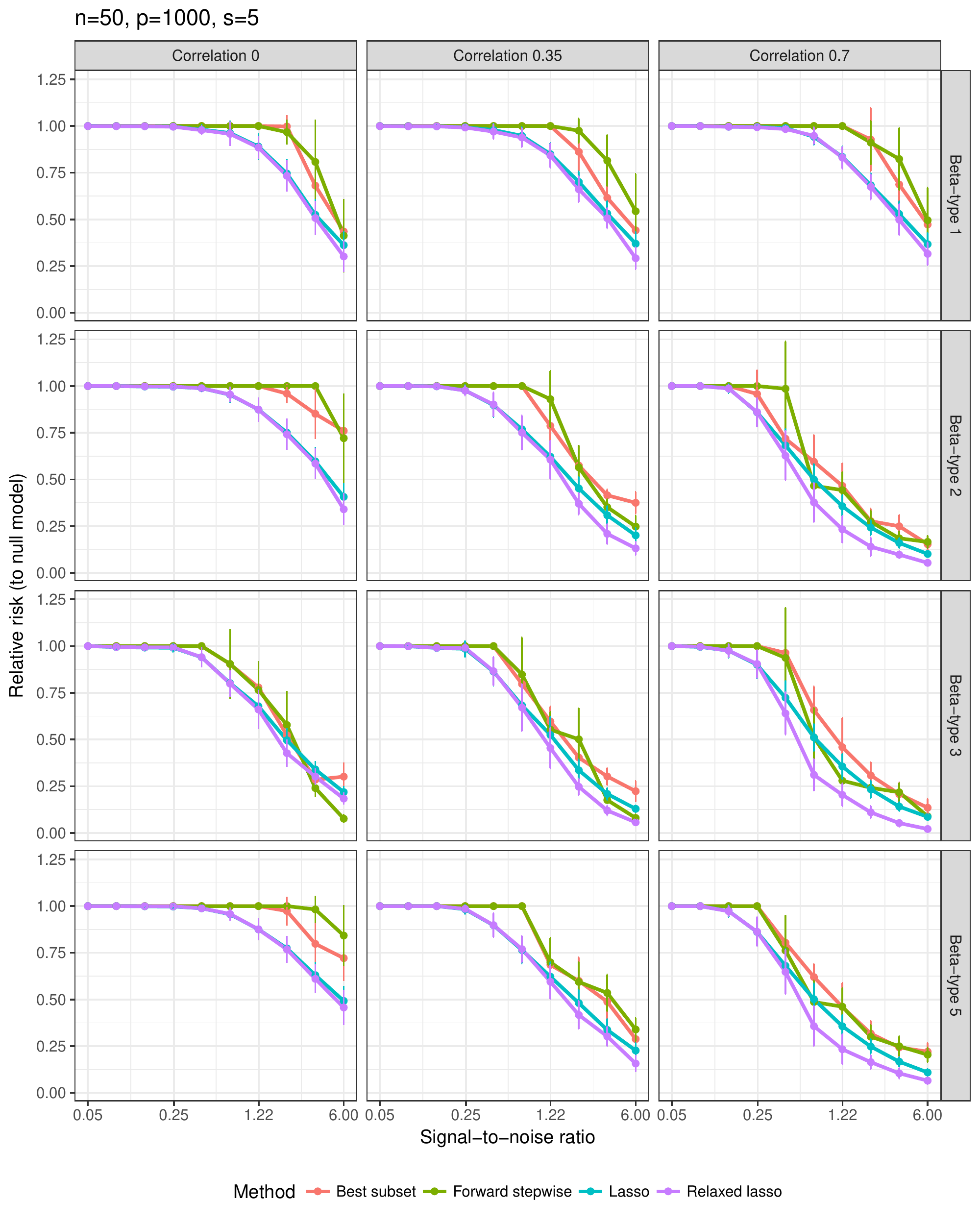}
\end{figure}
\newpage

\subsubsection{Relative test error (to Bayes)}
\begin{figure}[!h]
\centering
\includegraphics[width=0.99\textwidth]{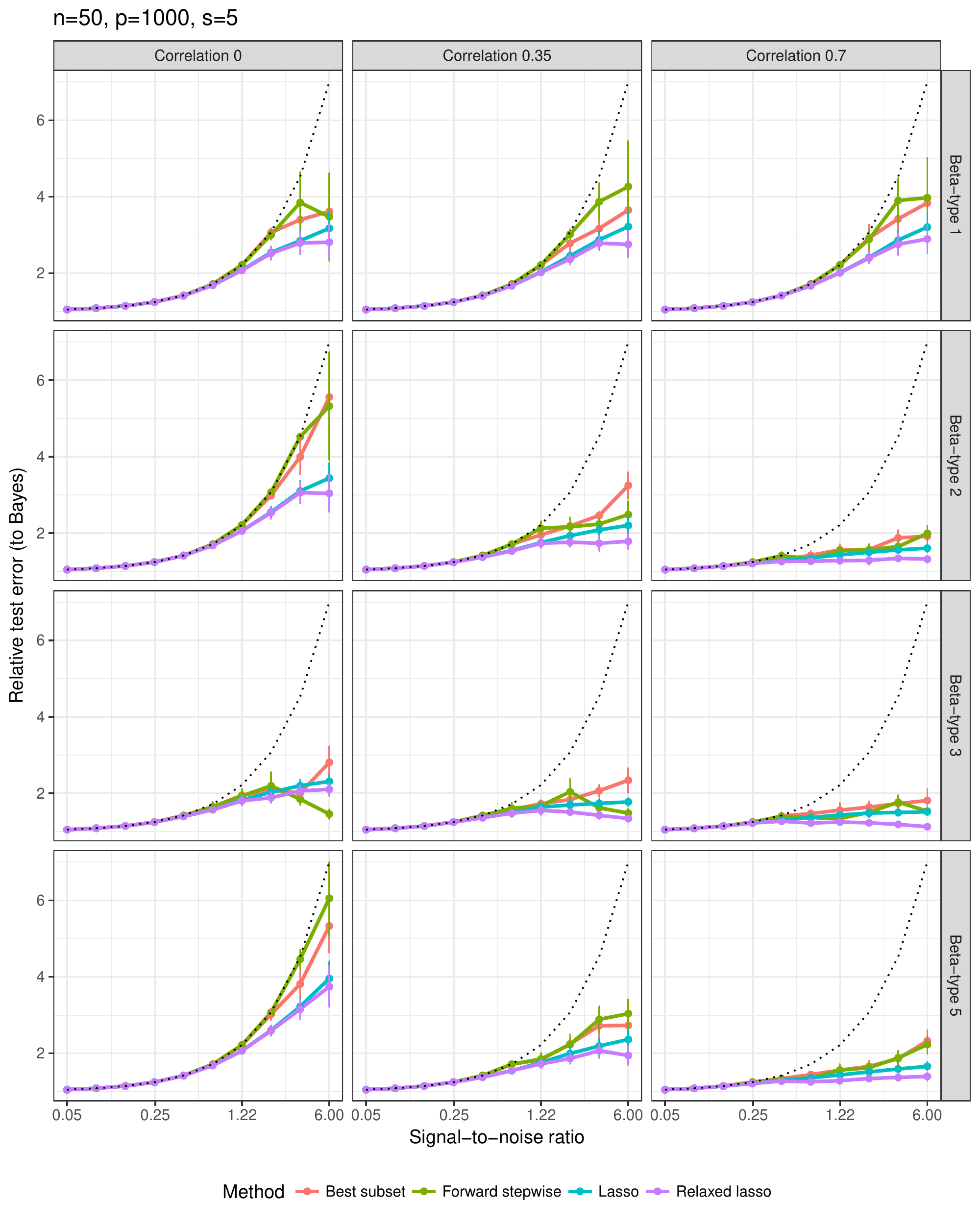}
\end{figure}
\newpage

\subsubsection{Proportion of variance explained}
\begin{figure}[!h]
\centering
\includegraphics[width=0.99\textwidth]{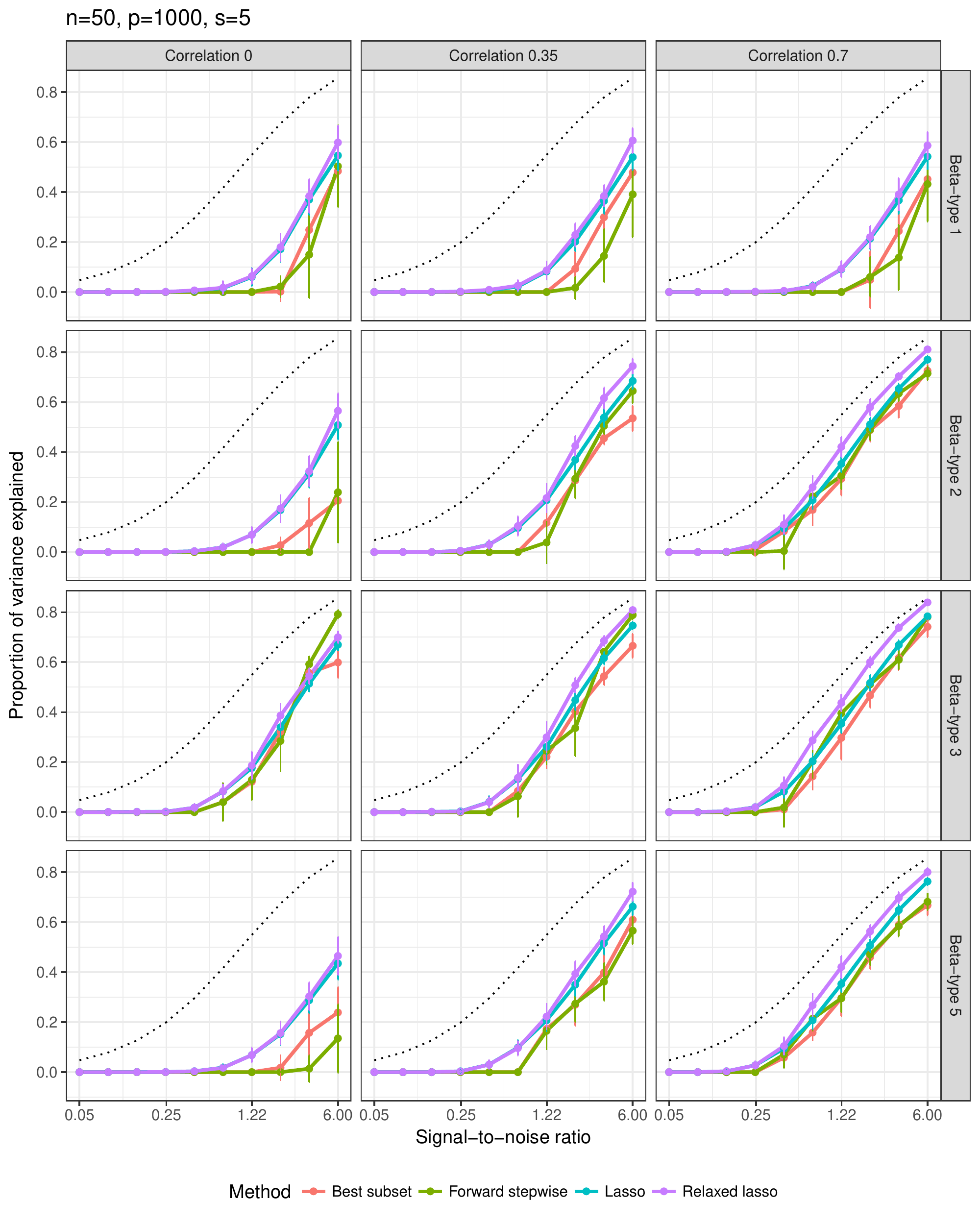}
\end{figure}
\newpage

\subsubsection{Number of nonzero coefficients}
\begin{figure}[!h]
\centering
\includegraphics[width=0.99\textwidth]{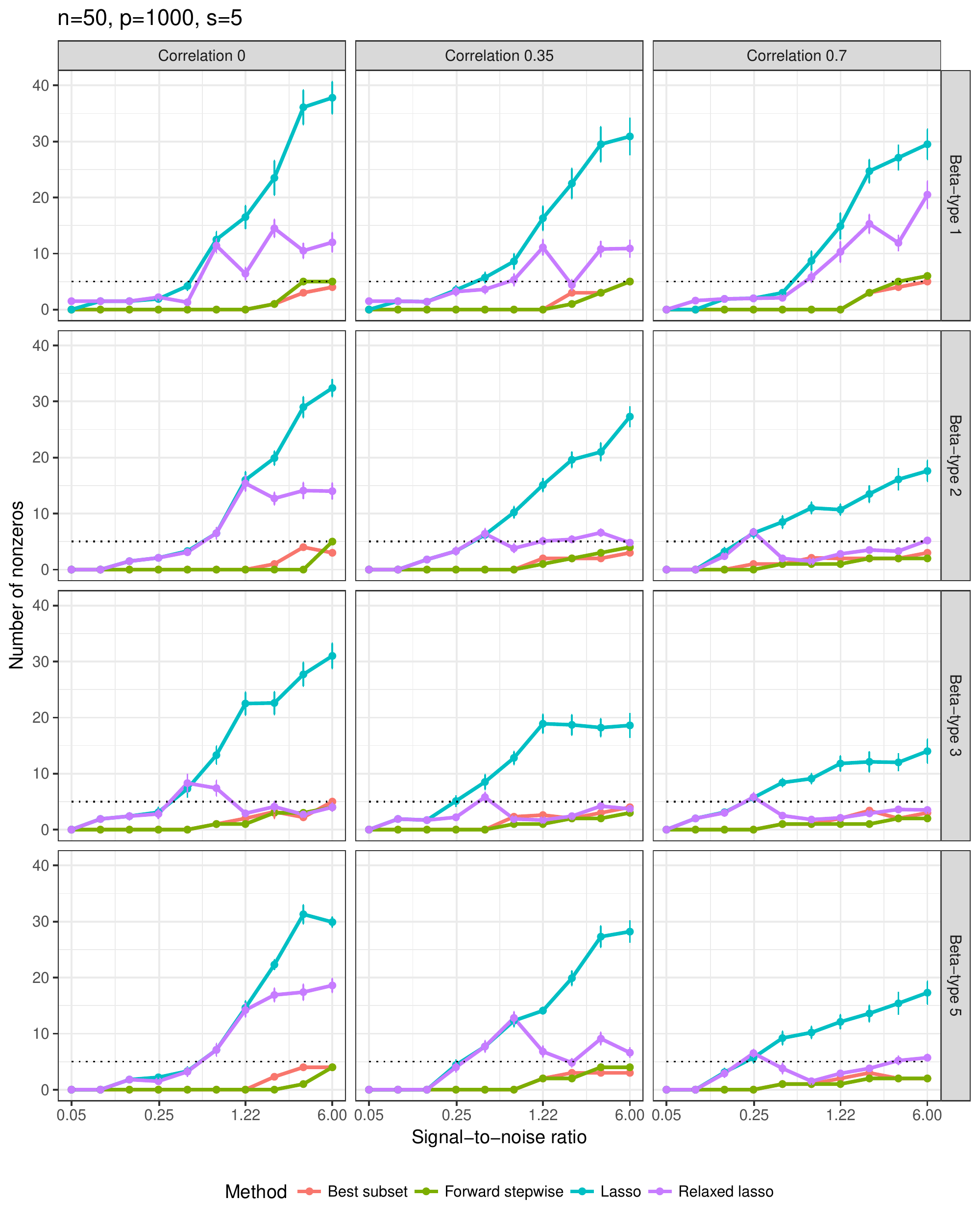}
\end{figure}
\newpage

\subsection{High-10 setting: $n=100$, $p=1000$, $s=10$} 
\subsubsection{Relative risk (to null model)} 
\begin{figure}[!h]
\centering
\includegraphics[width=0.99\textwidth]{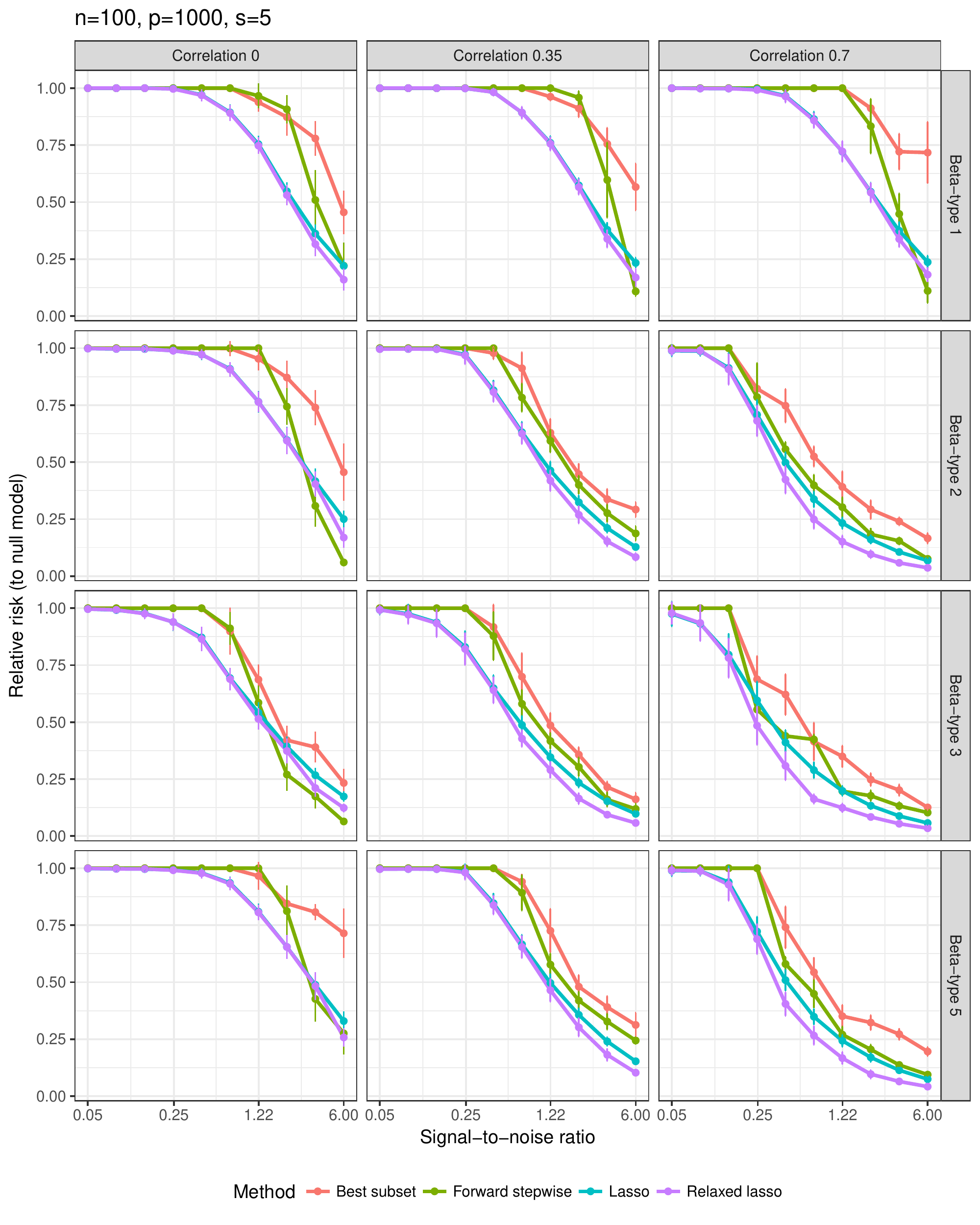}
\end{figure}
\newpage

\subsubsection{Relative test error (to Bayes)}
\begin{figure}[!h]
\centering
\includegraphics[width=0.99\textwidth]{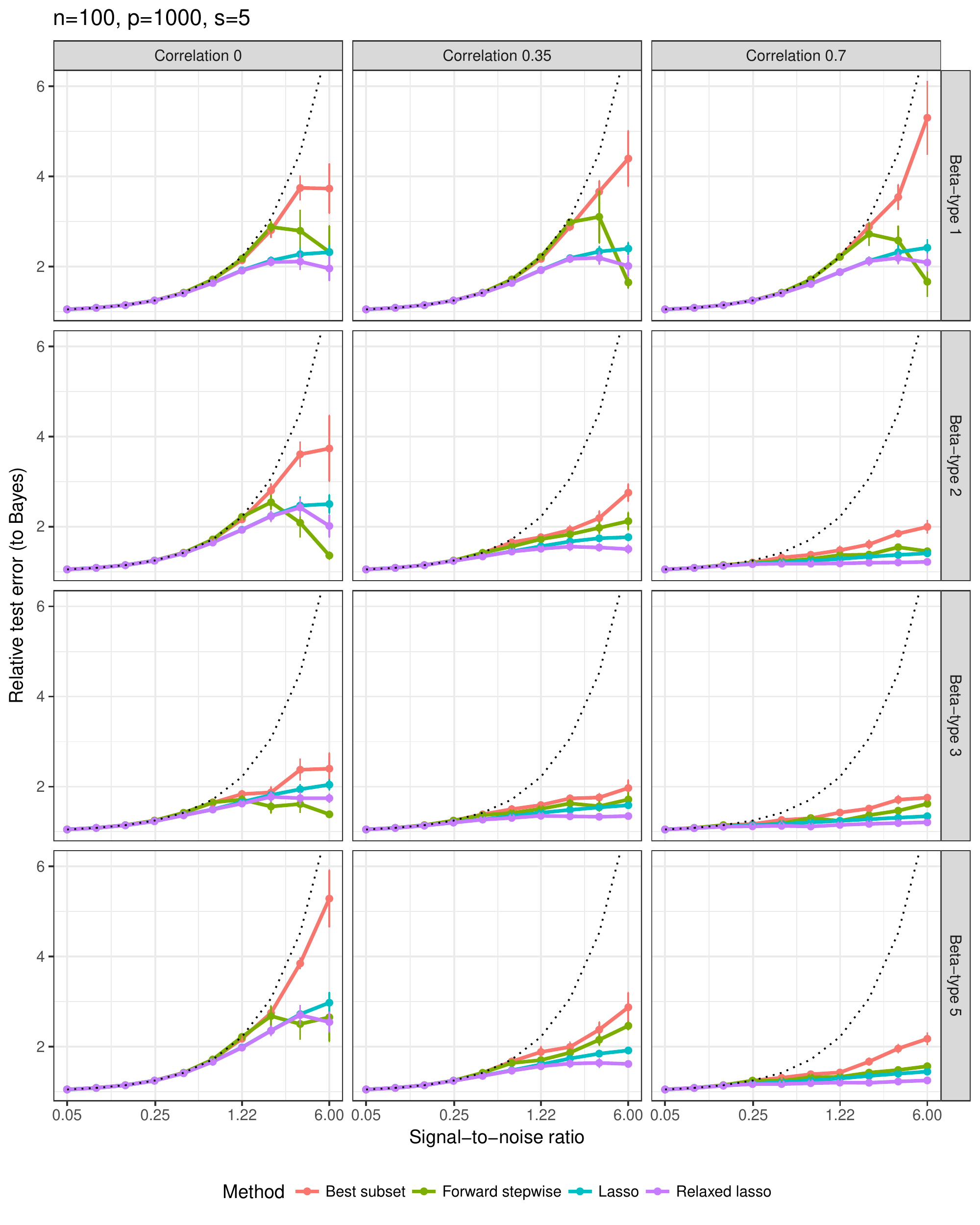}
\end{figure}
\newpage

\subsubsection{Proportion of variance explained}
\begin{figure}[!h]
\centering
\includegraphics[width=0.99\textwidth]{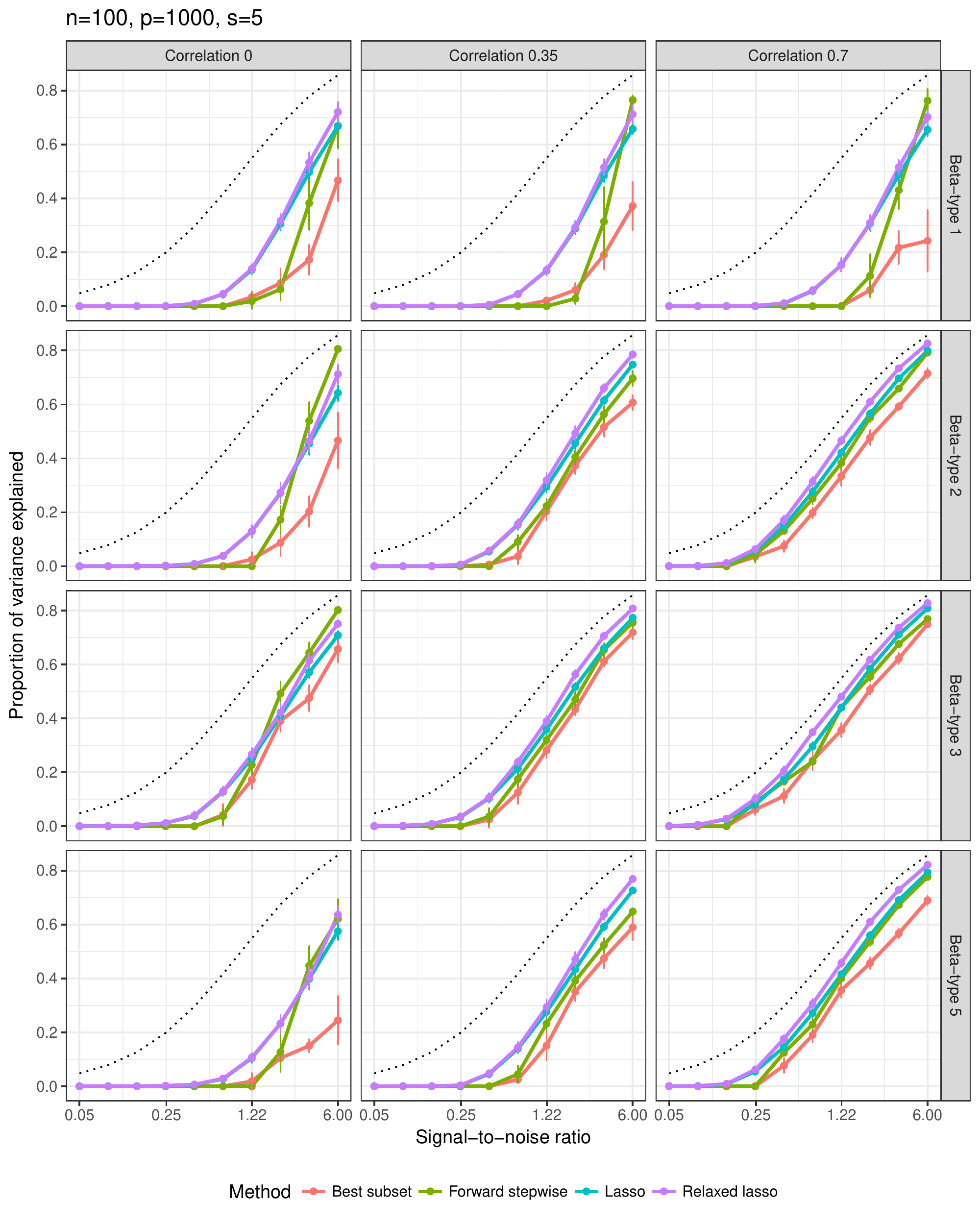}
\end{figure}
\newpage

\subsubsection{Number of nonzero coefficients}
\begin{figure}[!h]
\centering
\includegraphics[width=0.99\textwidth]{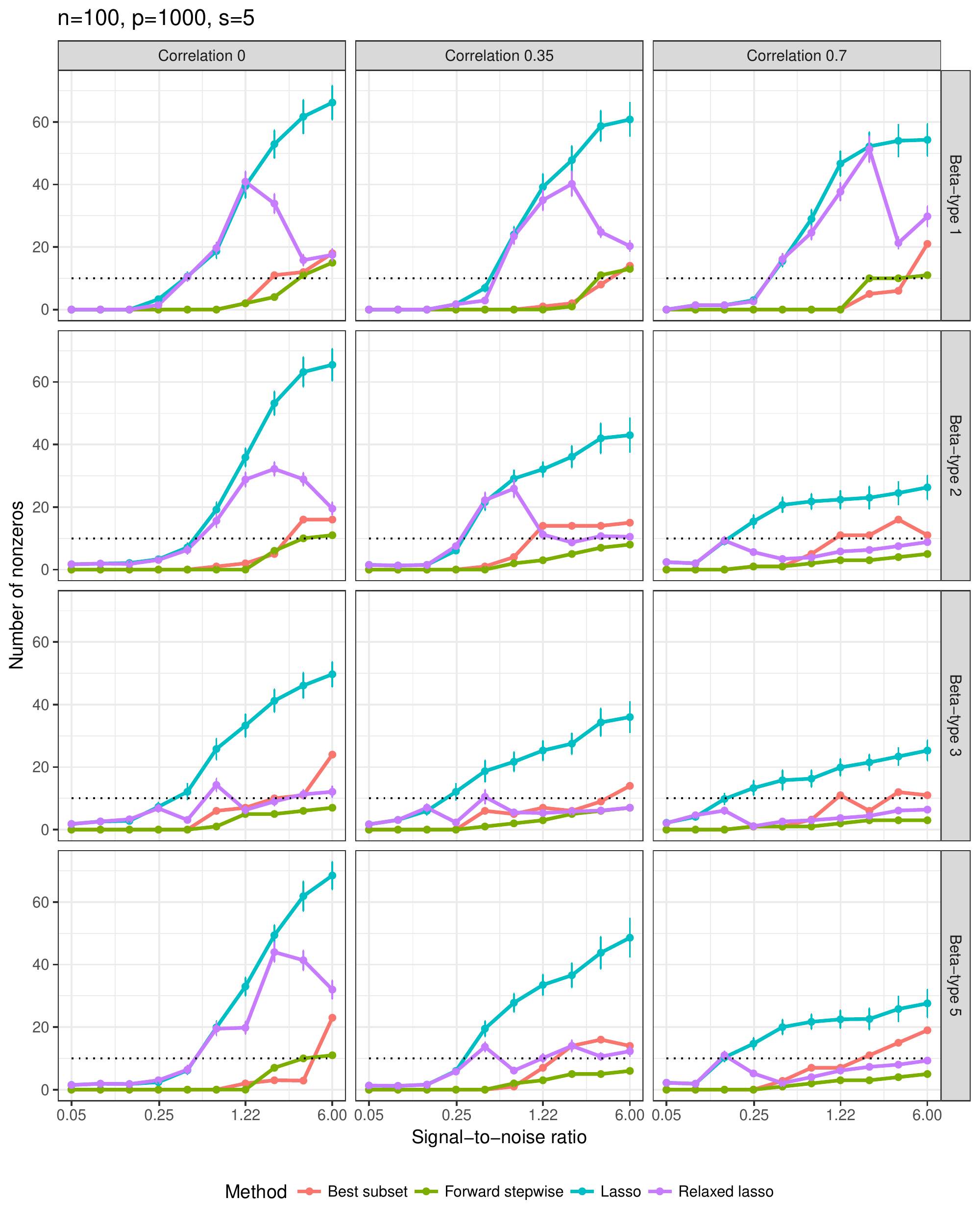}
\end{figure}
\end{document}